\documentclass[a4paper,12pt]{amsart}
\usepackage[foot]{amsaddr}
\usepackage{amssymb}
\usepackage{siunitx}
\usepackage{amsfonts}
\usepackage{booktabs}
\usepackage{amsmath,array}
\usepackage{caption}
\usepackage{subcaption}  
\usepackage{tabularx}
\usepackage{graphicx}
\usepackage[margin=2.7cm]{geometry}
\usepackage{bm}  
\usepackage{hyperref} 
\usepackage[final]{changes}

\begin{document}

\title[ROMs for the incompressible NS equations on collocated grids]{Reduced order models for the incompressible Navier-Stokes equations on collocated grids using a `discretize-then-project' approach}

\author{S. Kelbij Star\textsuperscript{1,2*}}
\address{\textsuperscript{1}SCK CEN, Institute for Advanced Nuclear Systems, Boeretang 200, 2400 Mol, Belgium.}
\address{\textsuperscript{2}Ghent University, Department of Electromechanical, Systems and Metal Engineering, Sint-Pietersnieuwstraat 41, B-9000 Ghent, Belgium.}
\thanks{\textsuperscript{*}Corresponding Author.}
\email{kelbij.star@sckcen.be}

\author{Benjamin Sanderse\textsuperscript{3}}
\email{B.Sanderse@cwi.nl}

\author{Giovanni Stabile\textsuperscript{4}}
\email{gstabile@sissa.it}

\author{Gianluigi Rozza\textsuperscript{4}}
\email{grozza@sissa.it}

\author{Joris Degroote\textsuperscript{2}}
\email{Joris.Degroote@UGent.be}

\address{\textsuperscript{3}Centrum Wiskunde \& Informatica, Scientific Computing, Science Park 123, 1098 XG Amsterdam, the Netherlands.}
\address{\textsuperscript{4}SISSA, International School for Advanced Studies, Mathematics Area, mathLab, via Bonomea 265, 34136 Trieste, Italy.}

\maketitle

\begin{abstract}
A novel reduced order model (ROM) for incompressible flows is developed by performing a Galerkin projection based on a fully (space and time) discrete full order model (FOM) formulation. This `discretize-then-project' approach requires no pressure stabilization technique (even though the pressure term is present in the ROM) nor a boundary control technique (to impose the boundary conditions at the ROM level). These are two main advantages compared to existing approaches. The fully discrete FOM is obtained by a finite volume discretization of the incompressible Navier-Stokes equations on a collocated grid, with a forward Euler time discretization. Two variants of the time discretization method, the inconsistent and consistent flux method, have been investigated. The latter leads to divergence-free velocity fields, also on the ROM level, whereas the velocity fields are only approximately divergence-free in the former method. For both methods, accurate results have been obtained for test cases with different types of boundary conditions: a lid-driven cavity and an open-cavity (with an inlet and outlet). The ROM obtained with the consistent flux method, having divergence-free velocity fields, is slightly more accurate but also slightly more expensive to solve compared to the inconsistent flux method. The speedup ratio of the ROM and FOM computation times is the highest for the open cavity test case with the inconsistent flux method.
\end{abstract}

\section{Introduction}\label{sec:introduction}
Computational Fluid Dynamics (CFD) simulations are essential in many engineering fields, among which aerospace, automotive, civil, naval and nuclear engineering. However, these methods are highly demanding in terms of CPU time and storage, especially for the simulation of turbulent flows, complex geometries, multi-physics phenomena and other types of complex flows. This is even more substantial for parametric (physical or geometrical) problems, such as in flow control, (design) optimization or in (almost) real time modeling for applications that require on-the-spot decision making. This has motivated the development of reduced order modeling techniques that reduce the number of degrees of freedom of the high fidelity models and in that way the computational cost.  

There exist many types of reduced order modeling methods that can be categorized in different ways~\cite{benner2015survey,rozzaadvances}. We make a distinction between methods that are projection-based and those that are not, such as truncation-based methods~\cite{willcox2002balanced,rowley2005model}, goal-oriented methods~\cite{bui2007goal} and low degree-of-freedom models that are based on input-output data~\cite{ravindran2000reduced}. 

The basic principle of the projection-based methods is to retain the essential physics and dynamics of a high fidelity model by projecting the Partial Differential Equations (PDEs) describing the fluid problem onto a low dimensional space, called the reduced basis (RB) space~\cite{rozza2007reduced,veroy2003reduced}. The result is a physics-based model that is reduced in size~\cite{hesthaven2016certified}. Examples of methods to determine the reduced basis are greedy algorithms~\cite{benner2015survey,prud2002reliable}, the dynamic mode decomposition~\cite{schmid2010dynamic,kutz2016dynamic,tissot2014model} and the popular Proper Orthogonal Decomposition (POD) technique~\cite{lumley1981coherent,sirovich1987turbulence}. \added{Many non-linear reduced order modeling methods are, in addition, equipped with a hyper reduction technique, such as gappy POD or a discrete empirical interpolation method, to reduce the complexity and cost associated with solving the non-linear term(s) of the reduced order model (ROM)}~\cite{galbally2010non,chaturantabut2010nonlinear,amsallem2012nonlinear}.

A classical projection-based method is the POD-Galerkin projection approach for which the reduced basis space is spanned by POD modes~\cite{quarteroni2015reduced}. These modes are obtained by applying POD on a set of high fidelity solutions\added{, which are }called snapshots~\cite{hesthaven2016certified, quarteroni2015reduced}. The POD technique is commonly used for incompressible flows due to its optimal convergence property and its applicability to non-linear systems~\cite{bergmann2009enablers,berkooz1993proper}. The projection-based methods are mostly used in combination with a finite element (FE) method as \added{the }full order model~\cite{chakir2009two}. However, POD-Galerkin methods for finite volume (FV) approximations have gained more and more attention in the past years~\cite{Lassila,lorenzi2016pod,Stabile2017CAF,carlberg2018conservative,haasdonk2008reduced} due to the frequent use of the FV method in industry as well as in academics~\cite{Eymard,petrova2012finite,syrakos2017critical}. The FV method owes its popularity to its robustness~\cite{Eymard} and its local and global conservation properties~\cite{versteeg2007,Fletcher}. 

FV discretization methods for the incompressible Navier-Stokes (NS) equations, which describe the fluid dynamics problem, are mainly applied on two kinds of grids: staggered and collocated~\cite{ferziger2002computational}. FV schemes on staggered grids are known to intrinsically conserve mass, momentum and kinetic energy in space and time on Cartesian grids~\cite{harlow1965numerical,vasilyev2000high}. Another favorable property of staggered grids is that the pressure-velocity coupling is inherently enforced, i.e.\ preventing odd-even decoupling of the pressure~\cite{ferziger2002computational,patankar1980numerical}. On the other hand, the collocated grid arrangement offers significant advantages over the staggered grid approach. First of all, the code implementation is generally simpler (easier bookkeeping)~\cite{piller2004finite}. In addition, the collocated grid shortens the computational time and reduces the required memory storage compared to staggered grids on complex solution domains~\cite{peric1988comparison,zang1994non}. Therefore, collocated grids are widely used by popular commercial codes such as ANSYS Fluent~\cite{fluent201114} and STAR-CCM+~\cite{cd2017star} and the open source code OpenFOAM~\cite{guide2015openfoam}, whose libraries we use in this work. 

Despite the potential and the increasing popularity of FV--based POD-Galerkin reduced order models for all sorts of applications, they tend to have issues with accuracy and can exhibit numerical instabilities~\cite{Stabile2017CAF,kalashnikova2011stable, iollo2000stability,balajewicz2013low,balajewicz2016minimal}. \replaced{There are two main sources of instability in the numerical discretization of the incompressible Navier-Stokes equations in classical fluid dynamics: convection-dominated flows and pressure-velocity coupling~\cite{ferziger2002computational}. These sources of instabilities can have a detrimental effect on the reduced order models. Other sources of instabilities can also be present at the reduced order level, such as the mode truncation instability~\cite{balajewicz2016minimal}. In this work, we only focus on the challenge related to the pressure-velocity coupling.}{Challenges regarding pressure-velocity coupling and satisfying the boundary conditions at ROM level make it difficult to generalize the ROM methods such that they can be applied to any problem. An additional difficulty is to make the ROM parametric.}

Several works on POD-Galerkin reduced order models have shown that the pressure gradient term disappears from the reduced set of momentum equations when the reduced basis for the velocity field is (discretely) divergence-free~\cite{holmes_lumley,ma2002low,sanderse2020non}. However, it is (in contrast to a staggered grid) not straightforward to derive a stable `velocity-only' ROM on a collocated grid, since the compatibility relation between divergence and gradient operators is not satisfied~\cite{versteeg2007,ferziger2002computational}. Typically, a combination of Rhie-Chow interpolation at the level of the full order model (FOM)~\cite{rhie1983numerical} and pressure stabilization on the ROM level is required to obtain stable solutions. 

\replaced{Popular techniques that aim at obtaining accurate velocity and pressure ROM approximations, for both finite element and finite volume-based reduced order modeling, are the supremizer enrichment of the velocity space in order to meet the inf-sup condition~\cite{Stabile2017CAF,ballarin2015supremizer} or exploitation of a pressure Poisson equation during the projection stage~\cite{Stabile2017CAF,Akhtar,noack2005need,caiazzo2014numerical}. The advantage of the supremizer enrichment approach is that it eliminates the numerical instabilities in the pressure approximation that are often generated by ROMs that do not satisfy the inf-sup condition. However, the disadvantage of pressure recovery via the momentum equation through the use of a supremizer stabilized velocity basis is that it is hard to determine how many supremizer modes need to be included in the reduced basis~\cite{Stabile2017CAF,ballarin2015supremizer,kean2020error}. The other popular approach makes use of the available velocity approximation to solve a pressure Poisson equation for the pressure in the ROM. A disadvantage of this approach is that it is often not clear how to treat the boundary conditions in the pressure Poisson equation~\cite{Stabile2017CAF,gresho1998incompressible,liu2010stable}. Moreover, even with these techniques, the ROM velocity and pressure fields are often about one or two orders less accurate than the fields obtained by projecting the full order solutions onto the POD basis spaces. This is even the case for non-parametric laminar flow cases, such as the lid driven cavity flow problem~\cite{Stabile2017CAF,ballarin2015supremizer,kean2020error,busto2020pod}.}{Possible stabilization techniques are the supremizer enrichment of the velocity space in order to meet the inf-sup
	condition~\cite{ballarin2015supremizer} or exploitation of a pressure Poisson equation (PPE) during the projection stage~\cite{Akhtar,noack2005need,caiazzo2014numerical}. However, even with these
	techniques, the ROM velocity and pressure fields are about one or two orders less accurate than the fields obtained by projecting
	the full order solutions onto the POD basis spaces. This is even the case for non-parametric laminar flow cases, such as the lid
	driven cavity flow problem~\cite{Stabile2017CAF}.}

\added{Another pressure recovery technique is to approximate the pressure term in the momentum equations using the pressure POD modes in combination with the coefficients for the approximated velocity~\cite{bergmann2009enablers,lorenzi2016pod}. The advantage of this method is that only the momentum equations have to be solved. However, the same number of velocity and pressure modes need to be included in the reduced basis spaces~\cite{caiazzo2014numerical}. Furthermore, techniques that have recently been developed in the context of finite elements are the local projection stabilization technique~\cite{rubino2020numerical,novo2021error} and replacing the incompressibility condition in the Navier-Stokes equations with an artificial compression condition~\cite{decaria2020artificial}. }

\replaced{Another common challenge of projection-based ROMs is satisfying the boundary conditions at the reduced order level. Boundary control strategies are often applied to enforce the boundary conditions in the ROM. Two common methods are the penalty method~\cite{lorenzi2016pod,graham1999optimal1,kalashnikova2012efficient,Sirisup} and the lifting function method~\cite{graham1999optimal1,stabile2017CAIM,ullmann2014pod,fick2018stabilized}. The disadvantage of these methods is that they often require parameter tuning. The penalty method relies on a penalty factor that has to be tuned with a sensitivity analysis or numerical experimentation. Also, it may be hard to find suitable lifting functions that will lead to an accurate ROM and extensive testing of the ROM for different functions can be needed.}{Furthermore, a `discretize-then-project' approach~\cite{lee2020model}, i.e. projecting the fully discrete system, simplifies the treatment of the velocity boundary conditions. A recent study on ROMs on a staggered grid~\cite{sanderse2020non} demonstrated that the boundary conditions of the discrete FOM can be inherited by the ROM via the projection of the boundary vectors. With this approach, no additional
	boundary control method, such as the commonly applied penalty function~\cite{graham1999optimal1,lorenzi2016pod,kalashnikova2012efficient,Sirisup} or lifting function methods~\cite{graham1999optimal1,stabile2017CAIM,ullmann2014pod,fick2018stabilized, bui2007goal}, is needed to handle the BCs at the ROM level.}

\added{The challenges related to projection, pressure stabilization and the boundary conditions at the ROM level make it difficult to generalize the ROM methods such that they can be applied to any problem.} In this work, we develop an efficient ROM for the incompressible NS equations on collocated grids that does not require a pressure stabilization nor an additional method to impose the boundary conditions at the ROM level. \added{We base our approach on reduced order modeling approaches that operate at the discrete level~\cite{carlberg2013gnat,lee2020model} and the recent progression on ROMs on staggered grids~\cite{sanderse2020non}. We derive the reduced order model via projection of the fully discrete system, i.e.\ we project the discrete FOM operators and boundary vectors onto the POD basis spaces. This `discretize-then-project' is not the same as the `discrete projection' approach for which a semi-discrete representation of the full order model is projected onto the POD modes in a discrete inner product~\cite{kalashnikova2012efficient}. Also the projection itself is not discrete~\cite{placzek2011nonlinear}.}

By using this approach, the ROM inherits the boundary conditions of the discrete FOM via the projection of the boundary vectors~\cite{sanderse2020non}, which simplifies the treatment of the boundary conditions~\cite{kalashnikova2012efficient}. In that way, no additional boundary control method is needed to impose the BCs at the ROM level. This approach is easier to implement and more generic than the other boundary control methods that often require parameter tuning. \added{Moreover, projecting the fully discrete FOM operators induces a model consistency between the FOM and the ROM, meaning that all reduced matrices and tensors of the reduced order model match with the linear and non-linear terms of the full order model.} To ease the derivations, we employ explicit time integration methods instead of implicit ones at the FOM and the ROM level~\cite{baiges2013explicit}. Furthermore, we evaluate whether the velocity fields are divergence-free and the necessity of pressure in the ROM formulation\deleted{ to develop a stable ROM}.

\deleted{To better understand how to deal with the challenges related to projection, pressure stabilization and the boundary conditions at ROM level, one needs to have a deep understanding of the underlying full order models.}

\added{The `discretize-then-project' approach developed in this work is an intrusive reduced order modeling method as it is necessary to have access to the solver's discretization and solution algorithm to project the discrete operators. Therefore, we use ITHACA-FV~\cite{ITHACA}, a free and open source code for reduced order modeling applications that makes use of the libraries of OpenFOAM~\cite{guide2015openfoam}, for the development of the reduced order models. Nevertheless, the `discretize-then-project' approach could also be implemented by software developers in commercial software packages for reduced order modeling applications.}

This paper is organized as follows: First, we discuss the incompressible Navier-Stokes equations at the continuous level in Section~\ref{sec:NS}. In Section~\ref{sec:FV}, we discuss the spatial and temporal discretization of the governing equations on a collocated grid for two different approaches for the computation of the convective face fluxes. In Section~\ref{sec:POD_Galerkin}, we apply the POD-Galerkin method at the fully discrete level and show the construction of the ROMs in the online phase. In Section~\ref{sec:setup} the set-up of two numerical test cases, a lid driven cavity flow and an open cavity (with an inlet and an outlet) flow problem, are given and the results are provided and discussed in Sections~\ref{sec:results} and~\ref{sec:discussion}, respectively. Finally, conclusions are drawn in Section~\ref{sec:conclusion} and an outlook for further developments is provided. 

\section{The incompressible Navier--Stokes equations}\label{sec:NS}
\replaced{We take as the governing equations to describe the fluid dynamics problem on a geometrical domain $\Omega$, which coincides with the region of flow, to be the unsteady incompressible Navier-Stokes equations.}{The governing equations to describe the fluid dynamics problem on a geometrical domain $\Omega$, which coincides with the region of flow, are given by the unsteady incompressible Navier--Stokes equations.} For a Newtonian flow with constant fluid density $\rho$ and kinematic viscosity $\nu$ and without gravity and body forces, the general equations of mass and momentum conservation are given, respectively, by
\begin{equation} \label{eq:continuity}
\nabla \cdot \boldsymbol{u} = 0 \hspace{0.2cm}\mbox{in  } \hspace{0.2cm}\Omega,
\end{equation}
\begin{equation} \label{eq:NS}
\frac{\partial \boldsymbol{u}}{\partial t} = - \nabla \cdot \left(\boldsymbol{u} \otimes \boldsymbol{u}\right) + \nu \nabla \cdot ( \nabla \boldsymbol{u}) - \nabla p \hspace{0.2cm}\mbox{in  } \hspace{0.2cm}\Omega,
\end{equation}
\noindent where $\boldsymbol{u}$ = $\boldsymbol{u}(\boldsymbol{x},t)$ represents the vectorial velocity field that is evaluated at $\boldsymbol{x} \in \Omega \subset \mathbb{R}^{d}$ with $d$ = 2 or 3. Furthermore, $p = p(\boldsymbol{x},t)$ is the normalized scalar pressure field, which is divided by the constant fluid density $\rho$, and $t$ denotes time. The right hand side of the momentum equations (Equation~\ref{eq:NS}) contains a convection, diffusion and pressure gradient term, respectively.

Taking the divergence of both sides of Equation~\ref{eq:NS} and applying the continuity constraint of Equation~\ref{eq:continuity} leads to the pressure Poisson equation:
\begin{equation} \label{eq:PPE_FOM}
\nabla^2  p = - \nabla \cdot \left(\nabla \cdot \left(\boldsymbol{u} \otimes \boldsymbol{u}\right) \right) \hspace{0.2cm}\mbox{in  } \hspace{0.2cm}\Omega.
\end{equation}
This equation ensures that continuity is satisfied and can therefore be used as an alternative for the equation of mass conservation (Equation~\ref{eq:continuity}) by solving for $\boldsymbol{u}$ and $p$. Moreover, it shows that velocity and pressure are coupled in the continuous domain.

The equations are supplemented with the initial condition:
\begin{equation} \label{eq:ic_vel_cont}
\boldsymbol{u}(\boldsymbol{x},0) = \boldsymbol{u}_0(\boldsymbol{x}) \hspace{0.2cm}\mbox{in  } \hspace{0.2cm}\Omega,
\end{equation}
where the initial condition is divergence free, i.e. $\nabla \cdot \boldsymbol{u}_0 = 0$.

\subsection{Boundary conditions}
Boundary conditions are required to make the above problem well-posed. In this work, we consider three types of boundary conditions: wall, inflow and outflow. Correspondingly, we subdivide the boundary into $\partial \Omega = \Gamma_{wall} \cup \Gamma_{in} \cup \Gamma_{out}$. All boundary conditions are assumed to be time-independent.

Viscous fluids adjacent to a solid boundary such as a wall satisfy the no-slip condition, which states that the velocity of the fluid is equal to the velocity of the boundary:
\begin{equation} \label{eq:bc_vel_no_slip}
\boldsymbol{u} = \boldsymbol{u}_{wall}(\boldsymbol{x}) \hspace{0.2cm} \mbox{on  } \hspace{0.1cm}\Gamma_{wall} \hspace{0.2cm}  \mbox{for } \hspace{0.2cm} t \geq 0,
\end{equation}
where $\boldsymbol{u}_{wall}$ is the wall velocity that is assumed to be known. In the case of fixed walls, $\boldsymbol{u}_{wall} = 0$. 

The inflow boundary condition is of the same form as the wall boundary condition:
\begin{equation} \label{eq:bc_vel_in}
\boldsymbol{u} = \boldsymbol{u}_{in}(\boldsymbol{x}) \hspace{0.2cm} \mbox{on  } \hspace{0.2cm}\Gamma_{in}\hspace{0.2cm}  \mbox{for }\hspace{0.2cm}  t \geq 0,
\end{equation}
where $\boldsymbol{u}_{in}$ is the velocity at the inlet boundary $\Gamma_{in}$ that is assumed to be known. 

If the problem contains solely wall/inflow boundary conditions, it is also required that the following compatibility condition, which follows from integrating Equation~\ref{eq:continuity} over $\Omega$, is satisfied~\cite{gresho1998incompressible}:
\begin{equation} \label{eq:only_D}
\int_{\partial \Omega}  \boldsymbol{n}  \cdot \boldsymbol{u}_{bc} d\Gamma =  0  \hspace{0.2cm} \mbox{for } t \geq 0,
\end{equation}
where $\boldsymbol{u}_{bc}$ is either the wall (Equation~\ref{eq:bc_vel_no_slip}) or inlet velocity (Equation~\ref{eq:bc_vel_in}) and $\boldsymbol{n}$ denotes the outward pointing normal vector on the boundary $\partial \Omega$. \replaced{Moreover, the pressure can only be determined up to a constant.}{As a consequence, the pressure can then only be determined up to a constant.} This is remedied by imposing the pressure in a selected point in the domain.

For outflow boundaries, the normal component of the stress tensor is specified:
\begin{equation} \label{eq:bc_vel_out}
\boldsymbol{n}  \cdot \nu \nabla \boldsymbol{u}  - \boldsymbol{n}  p= 0 \hspace{0.2cm} \mbox{on  } \hspace{0.2cm}\Gamma_{out} \hspace{0.2cm}  \mbox{for } t \geq 0.
\end{equation}

If the \replaced{pressure Poisson equation}{PPE} (Equation~\ref{eq:PPE_FOM}) is used rather than the equation for mass conservation (Equation~\ref{eq:continuity}), the following boundary conditions apply in addition to Equations~\ref{eq:bc_vel_no_slip} and~\ref{eq:bc_vel_in} for the wall and inflow boundary conditions, respectively:
\begin{equation} \label{eq:PPE_bc_D}
\boldsymbol{n}  \cdot \nabla p  = \boldsymbol{n}  \cdot \left(- \nabla \cdot \left(\boldsymbol{u} \otimes \boldsymbol{u}\right)\right) \hspace{0.2cm} \mbox{on  } \hspace{0.2cm}\Gamma_{wall}, \Gamma_{in} \hspace{0.2cm}  \mbox{for } \hspace{0.2cm} t \geq 0.
\end{equation}
The boundary conditions for the PPE are equivalent to Equation~\ref{eq:bc_vel_out} in the case of an outflow boundary condition.

\section{Finite volume discretization on collocated grids}\label{sec:FV}
\begin{figure}[h!]
	\centering
	\captionsetup{justification=centering}
	\includegraphics[width=10.0cm]{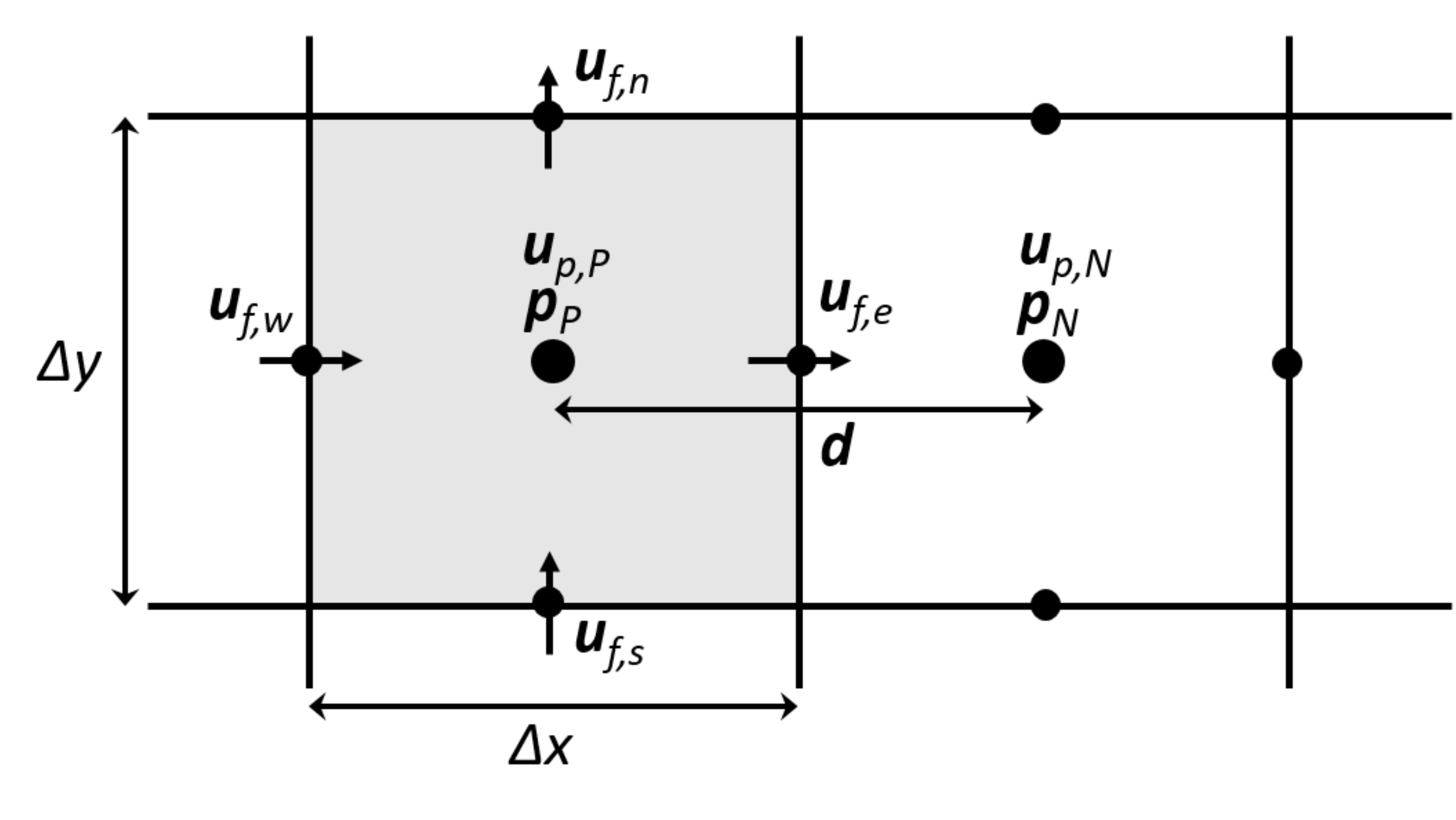}
	\caption{\replaced{Two-dimensional collocated grid with the location of the unknowns of velocity, $\boldsymbol{u}_{p}$, and pressure, $\boldsymbol{p}$, at the center of a cell volume $P$ and neighboring cell volume $N$. $\Delta x$ and $\Delta y$ are the cell lengths and the arrows indicate the mass fluxes, $\boldsymbol{u}_f$, through the cell faces $e$, $w$, $s$ and $n$ of cell $P$. $\boldsymbol{d}$ is the distance vector between the adjacent cell centers $N$ and $P$.}{Two-dimensional collocated grid with the location of the unknowns at the cell center and the faces of a cell volume.}}
	\label{fig:collocated_grid}
\end{figure}

In this section, we discretize the governing partial differential equations, Equations~\ref{eq:continuity} and~\ref{eq:NS}, using the finite volume method on a collocated grid, which is shown in Figure~\ref{fig:collocated_grid}. We present both the spatial and temporal discretization. The fully discretized equations are projected on reduced basis spaces in the next section.

An integral formulation of the governing equations is imposed to all closed cell volumes such that the conservation laws are satisfied locally~\cite{versteeg2007, Fletcher}. The integral form of the conservation equations (Equations~\ref{eq:continuity} and~\ref{eq:NS}) for an arbitrary cell $k$ are given by
\begin{equation} \label{eq:continuity_int}
\int_{\partial(\Omega_h)_k} \boldsymbol{n}  \cdot \boldsymbol{u} dS = 0,
\end{equation}
\begin{equation} \label{eq:NS_int}
\int_{(\Omega_h)_k}\frac{\partial \boldsymbol{u}}{\partial t} d\Omega = - \int_{\partial(\Omega_h)_k}   (\boldsymbol{n}  \cdot   \boldsymbol{u})  \boldsymbol{u}  dS + \nu \int_{\partial(\Omega_h)_k}  \boldsymbol{n} \cdot  \left( \nabla \boldsymbol{u} \right)   dS  - \int_{\partial(\Omega_h)_k} \boldsymbol{n}  p  dS,
\end{equation}
where $(\Omega_h)_k$ is the volume of cell $k$ and $\partial(\Omega_h)_k$ is its boundary. $d\Omega$ is an infinitesimal volume element and $dS$ is an infinitesimal element of surface area.

\subsection{Spatial discretization}\label{sec:SD}
\begin{figure}[h!]
	\centering
	\captionsetup{justification=centering}
	\includegraphics[width=1.\linewidth]{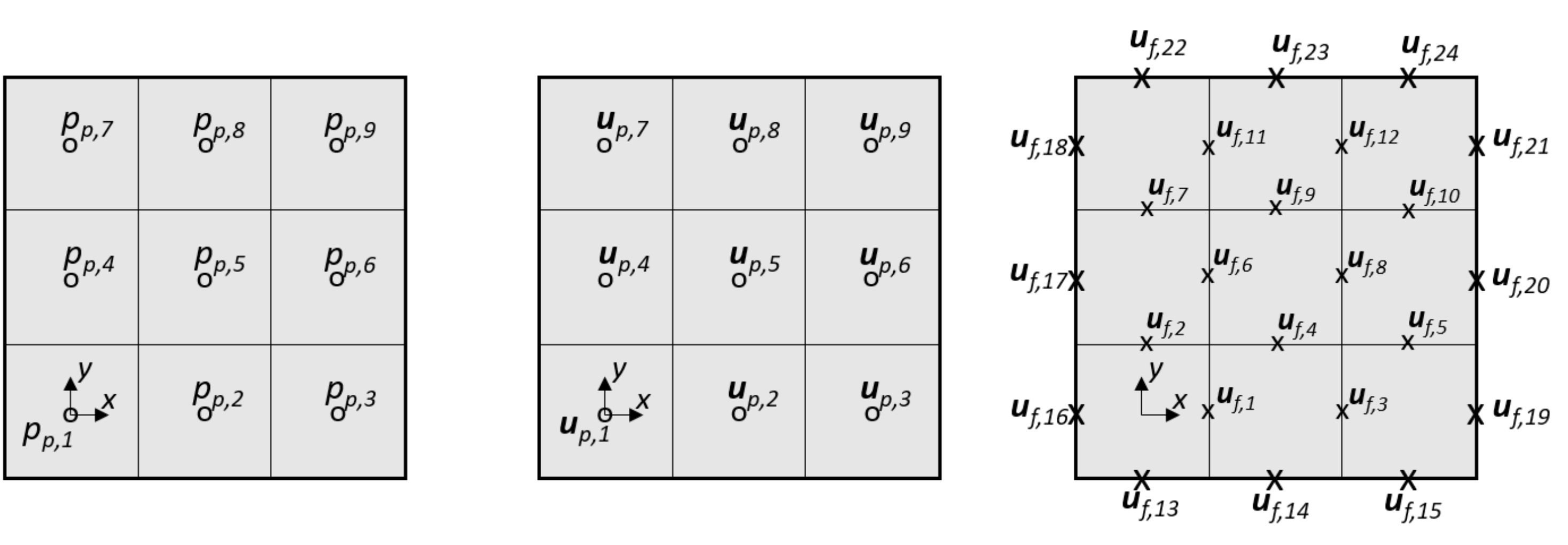}
	\caption{Sketch of a two-dimensional collocated grid with the location of the cell-centered pressure (left), the cell-centered velocity (middle) and the face-centered velocity (right).}
	\label{fig:grid_arrangement_boundary}
\end{figure}

The finite-volume discretization of the governing equations, Equations~\ref{eq:continuity_int} and~\ref{eq:NS_int}, on an arbitrary collocated mesh consisting of $h$ cells can be written in a matrix--vector notation:

\begin{equation} \label{eq:continuity_semi}
\boldsymbol{M} \boldsymbol{u}_f = \boldsymbol{0},
\end{equation}
\begin{equation} \label{eq:NS_semi}
\frac{\text{d} \boldsymbol{u}_{p}}{\text{d}  t}  = - \boldsymbol{C}_p(\boldsymbol{u}_f,\boldsymbol{u}_p) - \boldsymbol{r}_p^C + \nu \boldsymbol{D}_p \boldsymbol{u}_p -  \boldsymbol{G}_p \boldsymbol{p}_p + \nu \boldsymbol{r}_p^D ,
\end{equation}
where $\boldsymbol{p}_p = (p_{p,1}, p_{p,2}, ... ,p_{p,h})^T \in \mathbb{R}^{h}$ is the cell-centered pressure and $\boldsymbol{u}_p \in \mathbb{R}^{dh}$ the cell-centered velocity, which are defined as column vectors containing solely the cell-centered
values. For a three-dimensional problem ($d$ = 3), $\boldsymbol{u}_p$ is arranged as $((\boldsymbol{u}_{p})_1, (\boldsymbol{u}_{p})_2, (\boldsymbol{u}_{p})_3)^T$, where each $(\boldsymbol{u}_{p})_i$ = $((u_{p,1})_i, (u_{p,2})_i, ..., (u_{p,h})_i) $ for $i = 1, 2, 3$.  $(\boldsymbol{u}_f)_i = ((u_{f,1})_i, (u_{f,2})_i, ..., (u_{f,m})_i) \in \mathbb{R}^{dm}$, is the velocity evaluated on the cell faces and $m$ the number of faces. Figure~\ref{fig:grid_arrangement_boundary} depicts the location of the variables on a coarse grid. The face-centered velocity field $\boldsymbol{u}_{f}$ is related to the cell-centered velocity field $\boldsymbol{u}_{p}$ via a linear interpolation operator $\boldsymbol{I}_{p\rightarrow f} \in \mathbb{R}^{dm\times dh}$:
\begin{equation} \label{eq:interpolation}
\boldsymbol{u}_{f} \equiv \boldsymbol{I}_{p\rightarrow f} \boldsymbol{u}_{p} + \boldsymbol{u}_{b},
\end{equation}
where $\boldsymbol{u}_b \in \mathbb{R}^{dm}$ is a vector that contains only the velocity values that are defined at boundary faces of the domain. For the two-dimensional example given in Figure~\ref{fig:grid_arrangement_boundary}, $(\boldsymbol{u}_b)_i = (0, 0, ...,  (u_{f,13})_{i}, ..., (u_{f,24})_{i})$ for $i = 1, 2$ and 13, ..., 24 are the indices of the faces at the boundary of the domain. An alternative option to relate $\boldsymbol{u}_{f}$ to $\boldsymbol{u}_{p}$ will be discussed in Section~\ref{sec:Cflux}. Furthermore, matrix $\boldsymbol{M} \in \mathbb{R}^{h\times dm}$ is the face-to-center discrete divergence operator, $\boldsymbol{D}_p \in \mathbb{R}^{dh\times dh}$ represents the discrete cell-centered Laplacian operator associated with the diffusion term, $\boldsymbol{C}_p(\boldsymbol{u}_f, \boldsymbol{u}_p) \in \mathbb{R}^{dh\times dh}$ represents the non-linear convection operator and matrix $\boldsymbol{G}_p \in \mathbb{R}^{dh\times h}$ is the discrete gradient operator. Furthermore, $\boldsymbol{r}_p^C \in \mathbb{R}^{dh}$ and $\boldsymbol{r}_p^D \in \mathbb{R}^{dh}$ are boundary vectors that contain the contributions of the convection and diffusion terms, respectively. All operators are scaled with the finite volume sizes.

We now detail the discretization of each term of the equations in integral form (Equations~\ref{eq:continuity_int} and~\ref{eq:NS_int}) for an arbitrary cell $k$, i.e. we give the details of the operators that are present in Equations~\ref{eq:continuity_semi} and \ref{eq:NS_semi}. 

The discretization of the continuity equation (Equation~\ref{eq:continuity_int}) yields
\begin{equation}\label{eq:div}
\int_{\partial(\Omega_h)_k}  \boldsymbol{n}  \cdot \boldsymbol{u}  dS = \sum_i^{N_f} \int_{S_{f,i}} \boldsymbol{n}  \cdot \boldsymbol{u}  dS \approx  \sum_{i=1}^{N_f} \boldsymbol{S}_{f,i} \cdot \boldsymbol{u}_{f,i} =  \sum_{i=1}^{N_f} \phi_{f,i} = 0,
\end{equation}
\noindent where $N_f$ is the total number of faces bordering the cell $k$ and $\boldsymbol{S}_f = \boldsymbol{n} S_f$ is the outward-pointing face area vector with $S_f$ the area of the particular face. Hence, the face-to-center discrete divergence operator $\boldsymbol{M}$ of Equation~\ref{eq:continuity_semi} consists of the outward pointing face areas associated with all faces of the discrete domain. However, Equation~\ref{eq:div} shows that the divergence-free constraint is applied to the face flux, $\boldsymbol{\phi}_f = \boldsymbol{S}_{f} \cdot \boldsymbol{u}_{f} $, and not to the cell-centered velocity $\boldsymbol{u}_{p}$. Therefore, we also need to introduce the center-to-center discrete divergence operator $\boldsymbol{M}_p \in \mathbb{R}^{h\times dh}$:
\begin{equation}\label{eq:div_hat}
\boldsymbol{M}_p   \equiv \boldsymbol{M} \boldsymbol{I}_{p\rightarrow f}.
\end{equation}
Hence, the semi-discretized continuity Equation~\ref{eq:continuity_semi} can also be written as

\begin{equation} \label{eq:continuity_semi_hat}
\boldsymbol{M} \boldsymbol{u}_f = \boldsymbol{M} \boldsymbol{I}_{p\rightarrow f} \boldsymbol{u}_p  + \boldsymbol{M} \boldsymbol{u}_b = \boldsymbol{M}_p \boldsymbol{u}_p + \boldsymbol{r}_p^M =  \boldsymbol{0},
\end{equation}
where the boundary vector $\boldsymbol{r}_p^{M} \in \mathbb{R}^{h}$ is given by
\begin{equation} \label{eq:r_p^M}
\boldsymbol{r}_p^{M} \equiv \boldsymbol{M} \boldsymbol{u}_b,
\end{equation}
which contains the contributions of the boundary conditions associated with the continuity equation.

The discretization of the pressure gradient term yields
\begin{equation}\label{eq:gradP}
\int_{\partial(\Omega_h)_k}\boldsymbol{n} p  \mathrm{d}S  = \sum_{i=1}^{N_f} \int_{S_{f,i}} \boldsymbol{n} p \mathrm{d}S \approx \sum_{i=1}^{N_f}\boldsymbol{S}_{f,i} p_{f,i},
\end{equation}
where, the face-centered pressure field $\boldsymbol{p}_{f}$ is related to the cell-centered pressure field $\boldsymbol{p}_{p}$ via a linear interpolation operator $\boldsymbol{\Pi}_{p\rightarrow f} \in \mathbb{R}^{m\times h}$:
\begin{equation} \label{eq:interpolation_scalar}
\boldsymbol{p}_{f} \equiv \boldsymbol{\Pi}_{p\rightarrow f} \boldsymbol{p}_{p}.
\end{equation}
Hence, the discretization of the pressure gradient $\boldsymbol{G}_p$ consists of the face area vectors multiplied by the interpolation factors that are contained in $\boldsymbol{\Pi}_{p\rightarrow f}$.

\begin{figure}[h!]
	\centering
	\includegraphics[width=8.0cm]{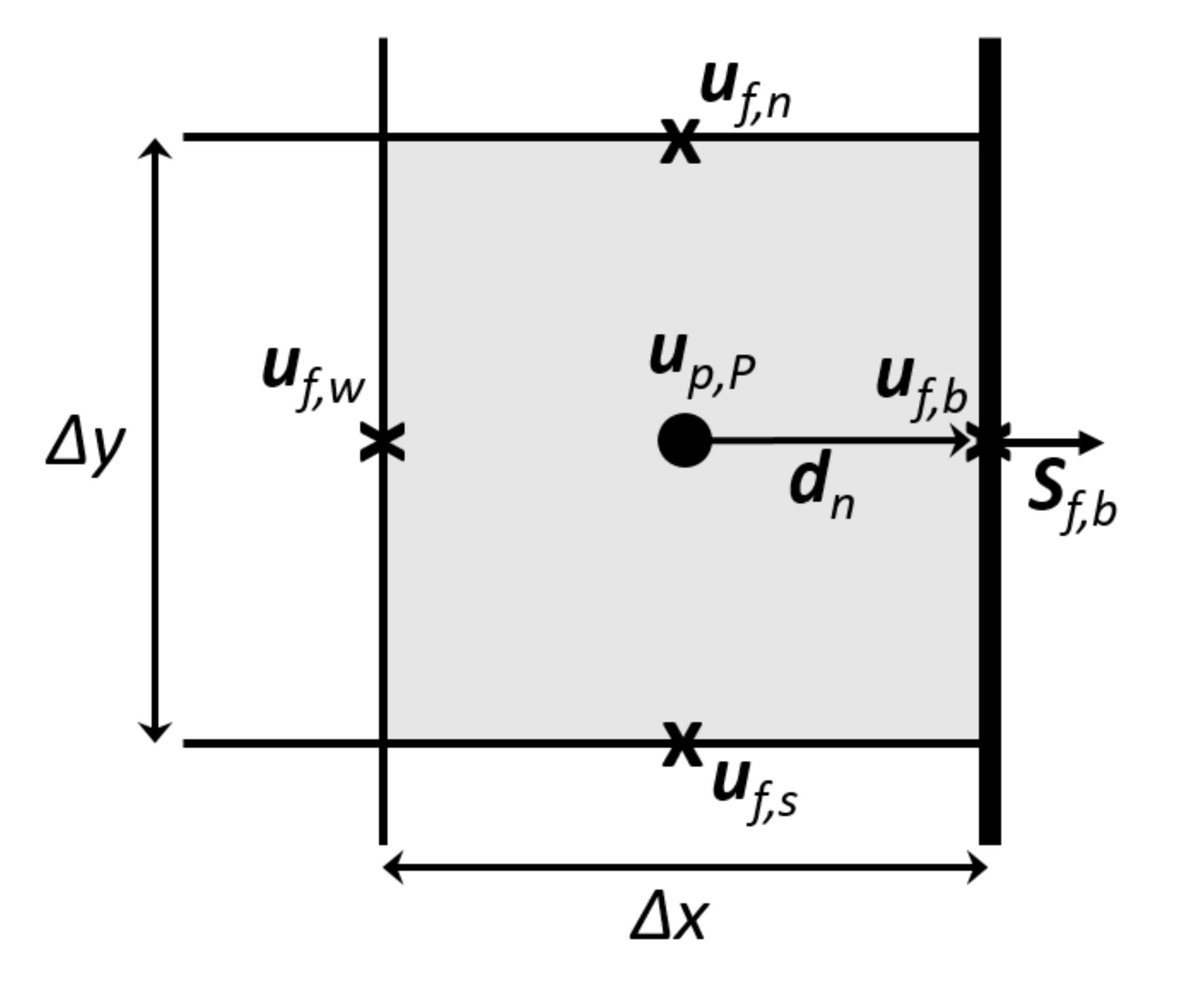}
	\caption{Two-dimensional collocated grid with the location of the velocity at the cell center and the faces of a cell volume of a cell near the boundary face of the domain.}
	\label{fig:collocated_grid_bc}
\end{figure}

Furthermore, the discretization of the diffusion term of the momentum equations for orthogonal meshes yields
\begin{equation}\label{eq:D_FOM}
\int_{\partial(\Omega_h)_k} \!  \boldsymbol{n} \cdot  \nabla \boldsymbol{u}  \, \mathrm{d}S =  \sum_{i=1}^{N_f} \int_{S_{f,i}}  \!  \boldsymbol{n} \cdot  \nabla \boldsymbol{u}  \, \mathrm{d}S  \approx 
\sum_{i=1}^{N_f} |\boldsymbol{S}_{f,i}| \frac{\boldsymbol{u}_{p,N} - \boldsymbol{u}_{p,P}}{|\boldsymbol{d}|},
\end{equation}
where $\boldsymbol{d}$ is the distance vector between any adjacent cell centers $N$ and $P$ to a particular face as shown in Figure~\ref{fig:collocated_grid}. Hence, the discrete diffusion operator $\boldsymbol{D}_p$ consists of coefficients associated with the face area vectors and the reciprocal of center-to-center distances. If the cell is neighboring a face, $b$, that is coinciding with the boundary of the computational domain, as shown in Figure~\ref{fig:collocated_grid_bc}, the discretization associated to that face changes to:
\begin{equation}\label{eq:D_FOMbc}
|\boldsymbol{S}_{f,b}| \frac{ \boldsymbol{u}_{f,b} - \boldsymbol{u}_{p,P}}{|\boldsymbol{d_n}|} ,
\end{equation}
which is split in two terms:
\begin{equation}\label{eq:D_FOMbc2}
\underbrace{|\boldsymbol{S}_{f,b}| \frac{ \boldsymbol{0} - \boldsymbol{u}_{p,P}}{|\boldsymbol{d_n}|}}_{\boldsymbol{D}_p}+ \underbrace{|\boldsymbol{S}_{f,b}| \frac{ \boldsymbol{u}_{f,b} - \boldsymbol{0}}{|\boldsymbol{d_n}|}}_{\boldsymbol{r}_p^D},
\end{equation}
where $\boldsymbol{u}_{f,b}$ is the value of velocity at the boundary face $b$, $\boldsymbol{d_n}$ is the distance vector between the face at the boundary of the domain and the center of the cell and $\boldsymbol{S}_{f,b}$ is the face area vector of $b$. The first term of Equation~\ref{eq:D_FOMbc2} is contained in the discrete diffusion operator $\boldsymbol{D}_p$, while the second term is contained in the boundary vector $\boldsymbol{r}_p^D \in \mathbb{R}^{dh}$.

Finally, the discretization of the convection term yields
\begin{equation}\label{eq:C_FOM}
\int_{\partial(\Omega_h)_k} \left( \boldsymbol{n} \cdot \boldsymbol{u}  \right) \boldsymbol{u}  dS = \sum_{i=1}^{N_f} \int_{S_{f,i}}  \!  \left( \boldsymbol{n} \cdot \boldsymbol{u}  \right) \boldsymbol{u}  \, \mathrm{d}S  \approx   \sum_{i=1}^{N_f} \left(\boldsymbol{S}_{f,i} \cdot \boldsymbol{u}_{f,i} \right) \boldsymbol{u}_{f,i}   =  \sum_{i=1}^{N_f}  \boldsymbol{\phi}_{f,i} \boldsymbol{u}_{f,i}.
\end{equation}
This shows that the convection operator $\boldsymbol{C}_p(\boldsymbol{u}_f,\boldsymbol{u}_p)$ is a non-linear operator that depends on the face fluxes $\boldsymbol{\phi}_{f}$. In the case that the cell has a face that corresponds to the boundary of the domain, as shown in Figure~\ref{fig:collocated_grid_bc}, the term:
\begin{equation}\label{eq:C_bc}
\underbrace{\left(\boldsymbol{S}_{f,b} \cdot \boldsymbol{u}_{f,b} \right) \boldsymbol{u}_{f,b}}_{\boldsymbol{r}_p^C} 
\end{equation}
is contained in the boundary vector $\boldsymbol{r}_p^C \in \mathbb{R}^{dh}$ instead of the matrix associated with the convection operator.
In this work, the non-linearity of the discretized convection term is quadratic, because $\boldsymbol{u}_f$ is obtained via linear interpolation of $\boldsymbol{u}_p$. Hence, we can redefine the convection operator in terms of a matrix-vector product:
\begin{equation}\label{eq:C_split}
\tilde{\boldsymbol{C}}_p(\boldsymbol{u}_f)\boldsymbol{u}_p \equiv \boldsymbol{C}_p(\boldsymbol{u}_f,\boldsymbol{u}_p).
\end{equation}

Finally, substituting Equation~\ref{eq:continuity_semi_hat} in the continuity equation (Equation~\ref{eq:continuity_semi}) and Equations~\ref{eq:C_split} in the momentum equations (Equation~\ref{eq:NS_semi}) results in the following spatially discretized system of equations:
\begin{equation} \label{eq:continuity_semi_split}
\boldsymbol{M}_p\boldsymbol{u}_p + \boldsymbol{r}_p^M = \boldsymbol{0},
\end{equation} 
\begin{equation} \label{eq:NS_semi_r_p}
\frac{\text{d}\boldsymbol{u}_{p} }{\text{d}  t}  = - \tilde{\boldsymbol{C}}_p(\boldsymbol{u}_f)\boldsymbol{u}_p + \nu \boldsymbol{D}_p \boldsymbol{u}_p - \boldsymbol{G}_p \boldsymbol{p}_p + \boldsymbol{r}_p,
\end{equation}
where $\boldsymbol{r}_p \in \mathbb{R}^{dh} \equiv  -\boldsymbol{r}_p^C + \nu \boldsymbol{r}_p^D$. All operators are scaled with the finite volume sizes.

\subsection{Explicit projection methods (temporal discretization)} \label{sec:TD}
Applying a central discretization stencil to the velocity divergence (Equation~\ref{eq:div}) and the pressure gradient (Equation~\ref{eq:gradP}) together with collocated grids generally results in spurious pressure oscillations~\cite{klaij2015stabilization}. These oscillations occur because the compatibility relation between the divergence and gradient operators is not satisfied (in contrast to a staggered grid)~\cite{versteeg2007,ferziger2002computational}. This so-called checkerboard problem is caused by a wide stencil in the PPE, which yields a pressure--velocity decoupling at adjacent cell centers~\cite{date1993solution}. The typical solution for this problem is to use the Rhie--Chow interpolation~\cite{rhie1983numerical} for the cell-centered face velocities. The PISO (Pressure-Implicit with Splitting of Operators~\cite{issa1986solution}) solver that is standardly available in OpenFOAM corresponds to this method of Rhie and Chow~\cite{komen2020analysis}. However, the use of the Rhie--Chow interpolation is unnecessary even on collocated grids when finite volume projection methods (also called fractional methods and not to be confused with the Galerkin projection method) with explicit time integration methods are employed~\cite{hirsch2007numerical}.

We showed in the previous section that in a collocated setting there exist two different velocity fields, namely, the velocities at the cell faces, $\boldsymbol{u}_f$, and the cell-centered velocities, $\boldsymbol{u}_p$. The cell-centered velocities together with the pressure form the primary solution variables. They can be related to the face-centered velocity via a linear interpolation (Equation~\ref{eq:interpolation}). We call this approach the inconsistent flux method (IFM). The fluxes at the cell centers are only approximately discretely divergence free with this approach, which is shown in the next subsection. We therefore also discuss a second approach for which we have an additional equation for the face velocities. We call this method the consistent flux method (CFM). 

Recently, Komen et al.~\cite{komen2020analysis} analyzed five numerical algorithms in finite volume collocated grid solvers for the incompressible Navier--Stokes equations for a selection of explicit (and implicit) Runge--Kutta schemes. They demonstrated that the temporal order reduces to approximately one also for the higher order schemes (except for the high-order method of Kazemi~\cite{kazemi2015analysis}, which however turns out to be very dissipative). Therefore, and for simplicity reasons as mentioned in the Introduction, we describe both projection methods (IFM and CFM) with the explicit Euler method~\cite{vuorinen2014implementation} (also called Forward Euler) i.e. the original Chorin--Temam algorithm~\cite{canuto2007spectral,chorin1968numerical,temam1968methode}. The extension of our approach to multi-stage (Runge-Kutta) methods is straightforward.

\subsubsection{Inconsistent flux method} \label{sec:ICflux}
We discretize in the time using Forward Euler, which is first order~\cite{sanderse2012accuracy}, the spatially discretized mass and momentum equations including boundary conditions (Equations~\ref{eq:continuity_semi_split} and~\ref{eq:NS_semi_r_p}). Writing them in vector form results in:
\begin{equation} \label{eq:mass_time}
\boldsymbol{M}_p\boldsymbol{u}_p^{n+1} + \boldsymbol{r}_p^M = \boldsymbol{0},
\end{equation} 
\begin{equation} \label{eq:NS_time}
\frac{\boldsymbol{u}_{p}^{n+1} - \boldsymbol{u}_{p}^{n}}{\Delta t} =  -\tilde{\boldsymbol{C}}_p(\boldsymbol{u}_f^n)\boldsymbol{u}_p^n + \nu \boldsymbol{D}_p \boldsymbol{u}_p^n - \boldsymbol{G}_p \boldsymbol{p}_p^{n+1} + \boldsymbol{r}_p , 
\end{equation}
where $\Delta t$ is the time step and $\boldsymbol{u}_p^{n} \approx \boldsymbol{u}_p(t^{n})$ is the solution at the $n^{th}$ time step.

As we showed in the previous section, velocity and pressure are coupled. The projection method computes first an intermediate velocity $\boldsymbol{u}_{p}^{*}$ by ignoring the pressure gradient term in the momentum equations:
\begin{equation} \label{eq:NS_time_intermediate}
\frac{\boldsymbol{u}_{p}^{*} - \boldsymbol{u}_{p}^{n}}{\Delta t} =  -\tilde{\boldsymbol{C}}_p(\boldsymbol{u}_f^n)\boldsymbol{u}_p^n + \nu \boldsymbol{D}_p \boldsymbol{u}_p^n + \boldsymbol{r}_p.
\end{equation}
Only the viscous and convective forces are thus considered in this step. Moreover, $\boldsymbol{u}_{p}^{*}$ is, in general, not divergence free.

Then the projection step follows, where the intermediate velocity field is corrected by the pressure in order to obtain the solution of $\boldsymbol{u}_{p}$ at time step $n+1$:
\begin{equation} \label{eq:NS_PM}
\boldsymbol{u}_{p}^{n+1}= \boldsymbol{u}_{p}^{*} - \Delta t \boldsymbol{G}_p \boldsymbol{p}_p^{n+1}.
\end{equation}

In order to obtain a divergence free velocity field at the next time step, Equation~\ref{eq:mass_time}, we take the divergence of Equation~\ref{eq:NS_PM}:
\begin{equation} \label{eq:NS_time_div}
\boldsymbol{M}_p \boldsymbol{u}_{p}^{n+1} + \boldsymbol{r}_p^M  = \left(\boldsymbol{M}_p \boldsymbol{u}_{p}^{*} + \boldsymbol{r}_p^M \right) - \Delta t \boldsymbol{M}_p \boldsymbol{G}_p \boldsymbol{p}_p^{n+1} = \boldsymbol{0}.
\end{equation}

Rewriting Equation~\ref{eq:NS_time_div} leads to the PPE in fully-discretized form:
\begin{equation} \label{eq:PPE_time_div}
\boldsymbol{L}_p \boldsymbol{p}_p^{n+1} =  \frac{1}{\Delta t} \left(\boldsymbol{M}_p \boldsymbol{u}_p^{*}+\boldsymbol{r}_p^M \right),
\end{equation}
where $\boldsymbol{L}_p \equiv \boldsymbol{M}_p \boldsymbol{G}_p \in \mathbb{R}^{h \times h}$ is a wide stencil Laplacian operator. Basically, this operator is based on interpolating the computed cell-centered pressure gradients to the cell faces. As a result, the pressure is decoupled at neighboring cells~\cite{klaij2015stabilization}. Hence, the pressure solution may contain non-physical spurious modes, which is known as the checkerboard problem.

By taking the pressure gradient directly at the cell faces, the checkerboard problem is avoided. This is similar to the way the diffusion operator is discretized in Equation~\ref{eq:D_FOM}, for which the direct neighboring cells are used without alternately skipping neighboring cells~\cite{versteeg2007}. This approach corresponds to the original interpolation method of Rhie and Chow. Therefore, we use the compact stencil given by the compact Laplacian operator $\boldsymbol{L}_f \in \mathbb{R}^{h \times h}$ instead of $\boldsymbol{L}_p$. $\boldsymbol{L}_f$ is also the standard Laplacian operator used in OpenFOAM~\cite{vuorinen2014implementation}. However, when using $\boldsymbol{L}_f$ instead of $\boldsymbol{L}_p$, the continuity constraint at the cell centers $\boldsymbol{M}_p\boldsymbol{u}_p^{n + 1 }  + \boldsymbol{r}_p^M= \boldsymbol{0}$ is no longer satisfied. 

Finally, the solution of the PPE is used to correct the cell-centered velocity field as done in Equation~\ref{eq:NS_PM}. As a result, the cell-centered velocity fields do not conserve mass and are only approximately divergence free~\cite{morinishi1998fully,felten2006kinetic}. Moreover, the computation of the face velocity lacks the correction by the flux that appears in the PPE and an incomplete flux term remains~\cite{komen2020analysis}.

To make the Galerkin projection procedure that will be introduced in Section~\ref{sec:POD_Galerkin}  straightforward, we rewrite the fully discrete system of equations (Equations~\ref{eq:NS_time_intermediate}--\ref{eq:PPE_time_div}) in such a way that we have only one equation for the pressure and one equation for the cell-centered velocity at the next time step, respectively:

\begin{equation}\label{eq:ifm_p}
\boldsymbol{L}_f \boldsymbol{p}_p^{n+1} = \frac{1}{\Delta t}\left(\boldsymbol{M}_p\boldsymbol{u}_p^n + \boldsymbol{r}_p^M \right) +  \boldsymbol{M}_p \left( - \tilde{\boldsymbol{C}}_p(\boldsymbol{u}_{f}^n)\boldsymbol{u}_p^{n}  + \nu \boldsymbol{D}_p \boldsymbol{u}_p^{n} + \boldsymbol{r}_p \right) ,
\end{equation}
\begin{equation}\label{eq:ifm_u}
\boldsymbol{u}_{p}^{n+1} = \boldsymbol{u}_{p}^n +  \Delta t \left( - \tilde{\boldsymbol{C}}_p(\boldsymbol{u}_{f}^n)\boldsymbol{u}_p^{n}  + \nu \boldsymbol{D}_p \boldsymbol{u}_p^{n} + \boldsymbol{r}_p \right) -\Delta t \boldsymbol{G}_p \boldsymbol{p}_p^{n+1}.
\end{equation}
For the inconsistent flux method, the velocity at the faces $\boldsymbol{u}_f$ are approximated using the interpolation operator $\boldsymbol{I}_{p\rightarrow f}$ of Equation~\ref{eq:interpolation}. Furthermore, the linear system of Equation~\ref{eq:ifm_p} needs to be solved to obtain $\boldsymbol{p}_p^{n+1}$, while Equation~\ref{eq:ifm_u} is fully explicit.

\subsubsection{Consistent flux method} \label{sec:Cflux}
In this method, we use the pressure field obtained by solving the PPE (Equation~\ref{eq:ifm_p}) to also correct the face fluxes. We apply first the linear interpolation operator onto Equation~\ref{eq:ifm_u}:
\begin{equation}\label{eq:cfm_phi2}
\boldsymbol{u}_f^{n+1} = \boldsymbol{I}_{p\rightarrow f} \left[\boldsymbol{u}_p^n +  \Delta t \left( - \tilde{\boldsymbol{C}}_p(\boldsymbol{u}_{f}^n)\boldsymbol{u}_p^{n}  + \nu \boldsymbol{D}_p \boldsymbol{u}_p^{n} + \boldsymbol{r}_p \right) -\Delta t \boldsymbol{G}_p \boldsymbol{p}_p^{n+1}\right],
\end{equation}
which is equivalent to Equation~\ref{eq:interpolation}. However, rather than interpolating the cell-centered pressure gradients (using $\boldsymbol{I}_{p\rightarrow f}\boldsymbol{G}_p$), we directly evaluate the pressure gradients at the faces using a new discrete face gradient operator $\boldsymbol{G}_f \in \mathbb{R}^{dm \times h}$. Therefore, Equation~\ref{eq:cfm_phi2} can be rewritten as:
\begin{equation}\label{eq:cfm_phi}
\boldsymbol{u}_f^{n+1} = \boldsymbol{I}_{p\rightarrow f} \boldsymbol{u}_p^n + \Delta t \boldsymbol{I}_{p\rightarrow f} \left( -\tilde{\boldsymbol{C}}_p(\boldsymbol{u}_f^n)\boldsymbol{u}_p^n + \nu \boldsymbol{D}_p \boldsymbol{u}_p^n  + \boldsymbol{r}_p\right) - \Delta t \boldsymbol{G}_f \boldsymbol{p}_p^{n+1}.
\end{equation}

The spatial discretization of the pressure gradient term, i.e. the last term of Equation~\ref{eq:cfm_phi} on the right hand side, for a cell $k$ is approximated by
\begin{equation}\label{eq:Gf}
\sum_{i=1}^{N_f} \boldsymbol{S}_{f,i} \frac{p_{p,N} - p_{p,P}}{|\boldsymbol{d}|}.
\end{equation}
Hence, the gradient operator $\boldsymbol{G}_f$ consists of coefficients associated with the surface normal vectors and the reciprocal of center-to-center distances. $\boldsymbol{G}_f$ directly uses the cell-centered pressure to calculate the gradient, while $\boldsymbol{G}_p$ (Equation~\ref{eq:gradP}) is based on the linear interpolation of the pressure in the cell centers.

If we then take the divergence of Equation~\ref{eq:cfm_phi} according to \ref{eq:continuity_semi_hat}:
\begin{equation}\label{eq:prove}
\boldsymbol{M} \boldsymbol{u}_f^{n+1} = \left(\boldsymbol{M}_p \boldsymbol{u}_p^n  + \boldsymbol{r}_p^M  \right)+ \boldsymbol{M}_p \left [\Delta t \left( -\tilde{\boldsymbol{C}}_p(\boldsymbol{u}_f^n)\boldsymbol{u}_p^n + \nu \boldsymbol{D}_p \boldsymbol{u}_p^n  + \boldsymbol{r}_p\right) \right]  - \Delta t \boldsymbol{M}\boldsymbol{G}_f \boldsymbol{p}_p^{n+1},
\end{equation}
we see that the combination of the first two terms on the right hand side is equal to the right hand side of Equation~\ref{eq:ifm_p} (multiplied by $\Delta t$). Therefore, substituting the pressure computed with Equation~\ref{eq:ifm_p}, proves that the face velocity fields are discretely divergence free as the right hand side of Equation~\ref{eq:prove} is zero.

The system of equations for the consistent flux method is then formed by Equations~\ref{eq:ifm_p}, \ref{eq:ifm_u} and \ref{eq:cfm_phi}, which are solved in this particular order to obtain the solution for $\boldsymbol{u}_f$, $\boldsymbol{u}_p$ and $\boldsymbol{p}_p$ at  $t^{n+1}$.

\section{POD-Galerkin reduced order models for the explicit projection methods} \label{sec:POD_Galerkin}
We apply the POD-Galerkin method~\cite{sirovich1987turbulence,Lassila} directly on the fully discrete formulations given by Equations~\ref{eq:ifm_p} and~\ref{eq:ifm_u} for the inconsistent flux method and Equations~\ref{eq:ifm_p}, \ref{eq:ifm_u} and~\ref{eq:cfm_phi} for the consistent flux method. Therefore, the full order models and reduced order models are both first order in time (Forward Euler).

We assume that the FOM solutions can be expressed as a linear combination of orthonormal spatial modes multiplied by time-dependent coefficients~\cite{Lassila}. The discrete cell-centered velocity fields, $\boldsymbol{u}_p$, are approximated by
\begin{equation}\label{eq:approx_up}
\boldsymbol{u}_p \approx \boldsymbol{u}_{p,r} =  \boldsymbol{\Phi}\boldsymbol{a},
\end{equation}
\noindent where $\boldsymbol{\Phi} = (\boldsymbol{\varphi}_1, \boldsymbol{\varphi}_2, ..., \boldsymbol{\varphi}_{N_r^u}) \in \mathbb{R}^{dh \times N_r^u}$ is a matrix containing the cell-centered velocity modes $\boldsymbol{\varphi} \in \mathbb{R}^{dh}$.  For a three-dimensional problem ($d$ = 3), $\boldsymbol{\varphi}$ is arranged as $((\boldsymbol{\varphi})_1, (\boldsymbol{\varphi})_2, (\boldsymbol{\varphi})_3)^T$, where each $(\boldsymbol{\varphi})_i$ = $((\varphi_{1})_i, (\varphi_{2})_i, ..., (\varphi_{h})_i) $ for $i = 1, 2, 3$. $\boldsymbol{a}= (a^1, a^2, ..., a^{N_r^u})^T \in \mathbb{R}^{N_r^u} $ are the corresponding time-dependent coefficients with $N_r^u$ the number of velocity modes. The subscript $r$ denotes quantities associated to the ROM.

Similarly, the discrete pressure fields are approximated by
\begin{equation}\label{eq:approx_p}
\boldsymbol{p}_p \approx \boldsymbol{p}_{p,r} = \boldsymbol{X}\boldsymbol{b}, 
\end{equation}
\noindent where $\boldsymbol{X} = (\boldsymbol{\chi}_1, \boldsymbol{\chi}_2, ..., \boldsymbol{\chi}_{N_r^p}) \in \mathbb{R}^{h \times N_r^p}$ is a matrix containing the cell-centered pressure modes $\boldsymbol{\chi} = (\chi_1, \chi_2, ...,\chi_h)^T \in \mathbb{R}^{h}$  and  $\boldsymbol{b}^n = (b^1, b^2, ..., b^{N_r^p})^T\in \mathbb{R}^{N_r^p} $ the corresponding time-dependent coefficients with $N_r^p$ the number of pressure modes. 

Finally, the discrete face velocity fields are approximated by
\begin{equation}\label{eq:approx_phi}
\boldsymbol{u}_f \approx \boldsymbol{u}_{f,r} = \boldsymbol{\Psi}\boldsymbol{c}, 
\end{equation}
\noindent where $\boldsymbol{\Psi} = (\boldsymbol{\psi}_1, \boldsymbol{\psi}_2, ..., \boldsymbol{\psi}_{N_r^u}) \in \mathbb{R}^{dm \times N_r^{u}}$ is a matrix containing the face velocity modes $\boldsymbol{\psi} \in \mathbb{R}^{m}$ and $\boldsymbol{c}(t)= (c^1, c^2, ..., c^{N_r^u})^T\in \mathbb{R}^{N_r^u}$ the corresponding time-dependent coefficients.  For a three-dimensional problem ($d$ = 3), $\boldsymbol{\psi}$ is arranged as $((\boldsymbol{\psi})_1, (\boldsymbol{\psi})_2, (\boldsymbol{\psi})_3)^T$, where each $(\boldsymbol{\psi})_i$ = $((\psi_{1})_i, (\psi_{2})_i, ..., (\psi_{m})_i) $ for $i = 1, 2, 3$.

\subsection{Proper Orthogonal Decomposition}\label{sec:POD}
The optimal POD basis space for the cell-centered velocity, $E_{POD}^{u_p}$ = span($\boldsymbol{\varphi}_1$,$\boldsymbol{\varphi}_2$, ...,$\boldsymbol{\varphi}_{N_r^u}$) is constructed by minimizing the difference between the snapshots, i.e. the discrete solutions at several time instances, and their orthogonal projection onto the reduced basis for the $L_2$-norm:
\begin{equation} \label{eq:min}
E_{POD}^{u_p} = \textrm{arg}\underset{\boldsymbol{\varphi}_1, ..., \boldsymbol{\varphi}_{N_r^u}}{\textrm{min}}\frac{1}{N_s^u} \sum\limits_{n=1}^{N_s^u} \left\Vert \boldsymbol{u}_{p}^n - \sum\limits_{i=1}^{N_r^u} \left( \boldsymbol{u}_{p}^n, \boldsymbol{\varphi}_i \right)_{L_2(\Omega_h)} \boldsymbol{\varphi}_i\right\Vert_{L_2(\Omega_h)}^2,
\end{equation}
subjected to the orthogonality constraint $\boldsymbol{\Phi}^T \replaced{\boldsymbol{V}}{\boldsymbol{\Omega}}\boldsymbol{\Phi}$ = $\boldsymbol{I}$, where $\replaced{\boldsymbol{V}}{\boldsymbol{\Omega}} \in \mathbb{R}^{dh\times dh}$ is a diagonal matrix with the cell-centered control volumes and $\boldsymbol{I}\in \mathbb{R}^{N_r^u \times N_r^u}$ is the identity matrix. $N_s^u$ is the number of velocity snapshots and $N_r^u$ $\leq$ $N_s^u$. ${\left( \cdot,\cdot\right)_{L_2(\Omega_h)}}$ is the discrete $L_2$-inner product of the fields over the whole discrete domain $\Omega_h$. The $L_2$-norm is the preferred norm for discrete numerical schemes~\cite{Stabile2017CAF,BuStaRoCen2018} with
\begin{equation} 
{\left( \boldsymbol{u}_{p}^n,\boldsymbol{\varphi}_i \right)_{L_2(\Omega_h)}} \equiv \sum_{j=1}^{h} \boldsymbol{u}_{p,j}^n \cdot \boldsymbol{\varphi}_{i,j} (\Omega_h)_j.
\end{equation}
The optimal POD basis space for the cell-centered pressure, $E_{POD}^{p}$ = span($\boldsymbol{\chi}_1$,$\boldsymbol{\chi}_2$, ...,$\boldsymbol{\chi}_{N_r^p}$) is constructed in a similar way. 

For the face velocity, $E_{POD}^{u_f}$ = span($\boldsymbol{\psi}_1$,$\boldsymbol{\psi}_2$, ...,$\boldsymbol{\psi}_{N_r^u}$) is constructed as follows:
\begin{equation} \label{eq:min_uf}
E_{POD}^{u_f} = \textrm{arg}\underset{\boldsymbol{\psi}^1, ..., \boldsymbol{\psi}_{N_r^u}}{\textrm{min}}\frac{1}{N_s^u} \sum\limits_{n=1}^{N_s^u} \left\Vert \boldsymbol{u}_{f}^n - \sum\limits_{i=1}^{N_r^u} \left( \boldsymbol{u}_{f}^n, \boldsymbol{\psi}_i \right)_{L_2(\Sigma)} \boldsymbol{\psi}_i\right\Vert_{L_2(\Sigma)}^2,
\end{equation}
where the discrete inner product ${\left( \cdot,\cdot\right)_{L_2(\Sigma)}}$ is defined over all face areas $\Sigma$:
\begin{equation} 
{\left( \boldsymbol{u}_{f}^n,\boldsymbol{\psi}_i \right)_{L_2(\Sigma)}} \equiv \sum_{j=1}^{m} \boldsymbol{u}_{f,j}^n \cdot \boldsymbol{\psi}_{i,j} (S_{f})_j.
\end{equation}

The minimization problem mentioned in Equation~\ref{eq:min} is equivalent to solving the following eigenvalue problem on a set of snapshots:
\begin{equation} \label{eq:ev} 
\boldsymbol{C}^u\boldsymbol{Q}^u=\boldsymbol{Q}^u\boldsymbol{\lambda}^u,
\end{equation}
with
\begin{equation} \label{eq:ev_C} 
C^u_{ij} = {\left( \boldsymbol{u}_{p}^i,\boldsymbol{u}_{p}^j\right)_{L_2(\Omega_h)}} \hspace{0.5cm}
\text{for} \hspace{0.1cm} i,j = 1, ..., N_s^u,
\end{equation}
where $\boldsymbol{C}^u$ $\in \mathbb{R}^{N_s^u \times N_s^u}$  is the correlation matrix of velocity snapshots, $\boldsymbol{Q}^u$ $\in \mathbb{R}^{N_s^u \times N_s^u}$ is a square matrix of eigenvectors and $\boldsymbol{\lambda}^u$ $\in \mathbb{R}^{N_s^u \times N_s^u}$  is a diagonal matrix containing the eigenvalues. The POD modes, $\boldsymbol{\varphi}_i$, are then constructed as follows
\begin{equation} \label{eq:POD}
\boldsymbol{\varphi}_i = \frac{1}{N_s^u\sqrt{\lambda_i^u}} \sum\limits_{n=1}^{N_s^u}\boldsymbol{u}_{p}^n Q^u_{in}\text{\hspace{0.5cm} for \hspace{0.1cm}}i = 1,...,N_r^u.
\end{equation}
The cell-centered velocity modes $\boldsymbol{\varphi}$ are only approximately discretely divergence free like the cell-centered velocity $\boldsymbol{u}_p$. As a consequence, it is necessary to include pressure in the ROM formulations\deleted{ to develop a stable ROM}.

The most energetic (dominant) POD modes are selected based on the decay of the eigenvalues $\lambda_i^u$. The procedure is the same for obtaining the pressure modes and the face velocity modes using the appropriate inner products.

\subsection{Galerkin projection for the inconsistent flux method}\label{sec:GalerkinICflux}
The approximations of the discrete velocity and pressure fields (Equations~\ref{eq:approx_up} and~\ref{eq:approx_p}) are substituted into the FOM of the inconsistent flux method (Equations~\ref{eq:ifm_p} and~\ref{eq:ifm_u}). The PPE (Equation~\ref{eq:ifm_p}) is then projected onto the reduced basis spanned by the pressure modes by left-multiplying with $\boldsymbol{X}^T\replaced{\boldsymbol{V}}{\boldsymbol{\Omega}}$:
\begin{equation}\label{eq:ROM_p_ifm}
\begin{split}
\boldsymbol{X}^T\boldsymbol{V} \boldsymbol{L}_f \boldsymbol{X}\boldsymbol{b}^{n+1}  = &\frac{1}{\Delta t} \left( \boldsymbol{X}^T\boldsymbol{V} \boldsymbol{M}_p\boldsymbol{\Phi}\boldsymbol{a}^{n} + \boldsymbol{X}^T \boldsymbol{V} \boldsymbol{r}_p^M\right)\\&
+ \boldsymbol{X}^T\boldsymbol{V} \boldsymbol{M}_p \left( - \tilde{\boldsymbol{C}}_p(\boldsymbol{I}_{p\rightarrow f}\boldsymbol{\Phi}\boldsymbol{a}^{n})\boldsymbol{\Phi}\boldsymbol{a}^{n} + \nu \boldsymbol{D}_p \boldsymbol{\Phi}\boldsymbol{a}^{n} + \boldsymbol{r}_p \right).
\end{split}
\end{equation}
Rewriting leads to the following ROM formulation for the equation for pressure:
\begin{equation} \label{eq:continuity_semi_ROM_in}
\boldsymbol{L}_r \boldsymbol{b}^{n+1} =  \frac{1}{\Delta t} \left( \boldsymbol{M}_r\boldsymbol{a}^n + \boldsymbol{q}^M_r \right)  -  \hat{\boldsymbol{A}}_r(\boldsymbol{a}^{n})\boldsymbol{a}^n + \nu \boldsymbol{B}_r \boldsymbol{a}^{n} + \boldsymbol{q}_r,
\end{equation}
where the reduced matrices associated with the linear terms, $\boldsymbol{L}_r = \boldsymbol{X}^T\replaced{\boldsymbol{V}}{\boldsymbol{\Omega}} \boldsymbol{L}_f \boldsymbol{X} \in \mathbb{R}^{N_r^p \times N_r^p}$, $\boldsymbol{M}_r = \boldsymbol{X}^T\replaced{\boldsymbol{V}}{\boldsymbol{\Omega}} \boldsymbol{M}_p\boldsymbol{\Phi} \in \mathbb{R}^{N_r^p \times N_r^u}$ and $\boldsymbol{B}_r = \boldsymbol{X}^T\replaced{\boldsymbol{V}}{\boldsymbol{\Omega}} \boldsymbol{M}_p \boldsymbol{D}_p\boldsymbol{\Phi} \in \mathbb{R}^{N_r^p \times N_r^u}$ and the reduced vector $\boldsymbol{q}_r = \boldsymbol{X}^T\replaced{\boldsymbol{V}}{\boldsymbol{\Omega}} \boldsymbol{M}_p\boldsymbol{r}_p \in \mathbb{R}^{N_r^p}$, can all be determined during the offline stage. The non-linear convection term $\hat{\boldsymbol{A}}_r(\boldsymbol{a}) \in \mathbb{R}^{N_r^p \times N_r^u \times N_r^u}$ is also precomputed during the offline stage and is stored as a third order tensor. Therefore, $\hat{\boldsymbol{A}}_r$ consists of $N_r^u$ components $\hat{\boldsymbol{A}}_{r,i} \in \mathbb{R}^{N_r^p \times N_r^u}$ and is constructed as:
\begin{equation} \label{eq:A_r}
\hat{\boldsymbol{A}}_{r,i} = \boldsymbol{X}^T \boldsymbol{V}\boldsymbol{M}_p \tilde{\boldsymbol{C}}_p(\boldsymbol{I}_{p\rightarrow f} \boldsymbol{\Phi}_i) \boldsymbol{\Phi}.
\end{equation}
During the online stage, the term $\hat{\boldsymbol{A}}_r(\boldsymbol{a}^{n})\boldsymbol{a}^n$ of Equation~\ref{eq:continuity_semi_ROM_in} is evaluated as
\begin{equation} \label{eq:A_r_online}
\sum_{i=1}^{N_r^u} (\boldsymbol{a}^{n})^T \hat{\boldsymbol{A}}_{r,i} \boldsymbol{a}^n.
\end{equation}
This only holds when the interpolation operator $\boldsymbol{I}_{p\rightarrow f}$ is linear, e.g. the convection term is quadratic and discretized with a linear discretization scheme.

Similarly, the discrete momentum equations~\ref{eq:ifm_u} are projected onto the reduced basis spanned by the velocity modes by left-multiplying with $\boldsymbol{\Phi}^T\replaced{\boldsymbol{V}}{\boldsymbol{\Omega}}$:
\begin{equation}
\boldsymbol{\Phi}^T\boldsymbol{V} \boldsymbol{\Phi} \boldsymbol{a}^{n+1} =  \boldsymbol{\Phi}^T\boldsymbol{V} \boldsymbol{\Phi} \boldsymbol{a}^n +  \Delta t \boldsymbol{\Phi}^T\boldsymbol{V}\left( - \tilde{\boldsymbol{C}}_p(\boldsymbol{I}_{p\rightarrow f}\boldsymbol{\Phi}\boldsymbol{a}^{n})\boldsymbol{\Phi}\boldsymbol{a}^{n} + \nu \boldsymbol{D}_p \boldsymbol{\Phi}\boldsymbol{a}^{n} + \boldsymbol{r}_p \right) -\Delta t\boldsymbol{\Phi}^T \boldsymbol{V} \boldsymbol{G}_p \boldsymbol{X}\boldsymbol{b}^{n+1} .
\end{equation}

Rewriting this is in matrix-vector notation leads to the following ROM formulation for the momentum equations:
\begin{equation} \label{eq:NS_semi_ROM_in}
\boldsymbol{a}^{n+1} = \boldsymbol{a}^n + \Delta t \left(-\hat{\boldsymbol{C}}_r( \boldsymbol{a}^n) \boldsymbol{a}^n  + \nu \boldsymbol{D}_r  \boldsymbol{a}^n + \boldsymbol{r}_r\right) - \Delta t \hat{\boldsymbol{G}}_r \boldsymbol{b}^{n+1}  ,
\end{equation}
where the reduced matrices $\boldsymbol{D}_r = \boldsymbol{\Phi}^T\replaced{\boldsymbol{V}}{\boldsymbol{\Omega}} \boldsymbol{D}_p\boldsymbol{\Phi} \in \mathbb{R}^{N_r^u \times N_r^u}$ and $\hat{\boldsymbol{G}}_r = \boldsymbol{\Phi}^T \replaced{\boldsymbol{V}}{\boldsymbol{\Omega}} \boldsymbol{G}_p \boldsymbol{X} \in \mathbb{R}^{N_r^u \times N_r^p}$ and the reduced vector $\boldsymbol{r}_r = \boldsymbol{\Phi}^T\replaced{\boldsymbol{V}}{\boldsymbol{\Omega}}\boldsymbol{r}_p \in \mathbb{R}^{N_r^u}$ can all be determined during the offline stage. The equation is simplified by $\boldsymbol{\Phi}^T\replaced{\boldsymbol{V}}{\boldsymbol{\Omega}} \boldsymbol{\Phi}$ = $\boldsymbol{I}$. Similar to Equation~\ref{eq:A_r}, the non-linear convection term $\hat{\boldsymbol{C}}_r(\boldsymbol{a}) \in \mathbb{R}^{N_r^u \times N_r^u \times N_r^u}$ is precomputed during the offline stage and stored as a third order tensor.

During the online stage, the linear system of Equation~\ref{eq:continuity_semi_ROM_in} can be solved for the pressure coefficients $\boldsymbol{b}^{n+1}$ as all terms of the right hand side depend solely on the solutions at time step $t^n$. This vector of coefficients is then used to calculate the velocity coefficients $\boldsymbol{a}^{n+1}$ at the new time step $t^{n+1}$ from Equation~\ref{eq:NS_semi_ROM_in}. The boundary conditions are incorporated in the ROM (Equations~\ref{eq:ROM_p_ifm} and~\ref{eq:NS_semi_ROM_in}) as the boundary vector $\boldsymbol{r}_p$ is also projected onto the reduced \replaced{basis spaces}{bases}. Therefore, no additional boundary control method is needed.

In many POD-Galerkin ROMs it is assumed that the POD velocity modes satisfy the strong divergence free constraint and that the pressure only enters the ROM on the boundary~\cite{Lassila,lorenzi2016pod}. Then, the pressure gradient term completely vanishes in the case of enclosed flow. This is not true for the inconsistent flux method as the discrete cell-centered velocity field is only approximately discretely divergence free. Therefore the divergence free constraint is also not fully satisfied neither at the FOM nor at the ROM level.

\subsection{Galerkin projection for the consistent flux method}\label{sec:GalerkinCflux}
We obtain the ROM for the consistent flux method by following the same Galerkin projection procedure for the inconsistent flux method as described in the previous subsection. The approximations of the discrete cell-centered velocity, face-centered velocity and pressure fields (Equations~\ref{eq:approx_up}, ~\ref{eq:approx_p} and ~\ref{eq:approx_phi}) are substituted into the FOM of the consistent flux method (Equations~\ref{eq:ifm_p},~\ref{eq:ifm_u} and ~\ref{eq:cfm_phi}). This results in the following reduced system of equations in matrix-vector notation:
\begin{equation} \label{eq:CFM_ROM_p}
\boldsymbol{L}_r \boldsymbol{b}^{n+1} =  \frac{1}{\Delta t} \left( \boldsymbol{M}_r\boldsymbol{a}^n + \boldsymbol{q}^M_r \right) -  \boldsymbol{A}_r(\boldsymbol{c}^{n})\boldsymbol{a}^n + \nu \boldsymbol{B}_r \boldsymbol{a}^{n} + \boldsymbol{q}_r,
\end{equation}
\begin{equation} \label{eq:CFM_ROM_u}
\boldsymbol{a}^{n+1} = \boldsymbol{a}^n + \Delta t\left(-\boldsymbol{C}_r( \boldsymbol{c}^n ) \boldsymbol{a}^n  + \nu \boldsymbol{D}_r  \boldsymbol{a}^n + \boldsymbol{r}_r  \right) - \Delta t\hat{\boldsymbol{G}}_r \boldsymbol{b}^{n+1} ,
\end{equation}
\begin{equation}\label{eq:cfm_ROM_phi3}
\boldsymbol{W}_r\boldsymbol{c}^{n+1} = \boldsymbol{N}_r \boldsymbol{a}^n + \Delta t\left(- \boldsymbol{K}_r(\boldsymbol{c}^{n})\boldsymbol{a}^n + \nu \boldsymbol{P}_r \boldsymbol{a}^n  + \boldsymbol{s}_r \right)- \Delta t\boldsymbol{G}_r \boldsymbol{b}^{n+1},
\end{equation}
with $\boldsymbol{W}_r = \boldsymbol{\Psi}^T\boldsymbol{\Sigma}\boldsymbol{\Psi} \in \mathbb{R}^{N_r^u \times N_r^u}$, \replaced{$\boldsymbol{N}_r$}{$\boldsymbol{V}_r$}$ = \boldsymbol{\Psi}^T\boldsymbol{\Sigma} \boldsymbol{I}_{p\rightarrow f} \boldsymbol{\Phi} \in \mathbb{R}^{N_r^u \times N_r^u}$, $\boldsymbol{P}_r = \boldsymbol{\Psi}^T\boldsymbol{\Sigma} \boldsymbol{D}_p\boldsymbol{\Phi} \in \mathbb{R}^{N_r^u \times N_r^u}$, $\boldsymbol{G}_r = \boldsymbol{\Psi}^T\boldsymbol{\Sigma} \boldsymbol{G}_f\boldsymbol{X} \in \mathbb{R}^{N_r^u \times N_r^p}$ and the reduced vector $\boldsymbol{s}_r = \boldsymbol{\Psi}^T\boldsymbol{\Sigma}\boldsymbol{r}_p \in \mathbb{R}^{N_r^u}$. The matrix $\boldsymbol{\Sigma} \in \mathbb{R}^{dm \times dm}$ contains the face areas of the cells. The reduced convection terms $\boldsymbol{A}_r(\boldsymbol{a}^n) \in \mathbb{R}^{N_r^p \times N_r^u \times N_r^u}$, $\boldsymbol{C}_r(\boldsymbol{a}^n) \in \mathbb{R}^{N_r^u \times N_r^u \times N_r^u}$ and $\boldsymbol{K}_r(\boldsymbol{a}^n) \in \mathbb{R}^{N_r^u \times N_r^u \times N_r^u}$ are determined, respectively, by
\begin{equation}\label{eq:A_r_CFM}
\boldsymbol{A}_{r,i} = \boldsymbol{X}^T \boldsymbol{V}\boldsymbol{M}_p \tilde{\boldsymbol{C}}_p( \boldsymbol{\Psi}_i) \boldsymbol{\Phi},
\end{equation}
\begin{equation}\label{eq:C_r}
\boldsymbol{C}_{r,i} = \boldsymbol{\Phi}^T \boldsymbol{V}\tilde{\boldsymbol{C}}_p( \boldsymbol{\Psi}_i) \boldsymbol{\Phi},
\end{equation}
\begin{equation}\label{eq:K_r}
\boldsymbol{K}_{r,i} = \boldsymbol{\Psi}^T \boldsymbol{\Sigma} \tilde{\boldsymbol{C}}_p( \boldsymbol{\Psi}_i) \boldsymbol{\Phi}.
\end{equation}

As the face-centered velocity fields are discretely divergence free, also the POD flux modes are discretely divergence free~\cite{caiazzo2014numerical}. Therefore, the pressure gradient term of Equation~\ref{eq:cfm_ROM_phi3} completely vanishes in the case of enclosed flow~\cite{Lassila}. 

The reduced system of the CFM (Equations~\ref{eq:CFM_ROM_p}--\ref{eq:cfm_ROM_phi3}) differs from the reduced system of the IFM (Equations~\ref{eq:continuity_semi_ROM_in}--\ref{eq:NS_semi_ROM_in}) in three ways. First of all, the reduced equation for the coefficients of the face-centered velocity is added to the CFM-ROM in the same way that the CFM-FOM also has an additional equation for $\boldsymbol{u}_f$ at the new time step. Secondly, the convection terms of Equations~\ref{eq:CFM_ROM_p}--\ref{eq:cfm_ROM_phi3} depend on the face-centered velocity coefficients $\boldsymbol{c}$ instead of the cell-centered velocity coefficients $\boldsymbol{a}$. Thirdly, more reduced matrices need to be precomputed during the offline stage, which results in additional storage and CPU costs compared to the IFM.

\section{Numerical set-up} \label{sec:setup}
In this section the numerical set-up of two cases is described. The first test case is the classical lid driven cavity benchmark, which is a closed flow problem. The second test case consists of an open cavity flow problem featuring an inlet and outlet boundary. This is an important test case for testing the projection of the boundary vectors. Both cases are modeled on a two-dimensional domain. Full order simulations are carried out for both the consistent and inconsistent flux method that have been implemented in ITHACA-FV~\cite{ITHACA}, which is an open source C$^{++}$ library based on OpenFOAM~\cite{Jasak}. The libraries of OpenFOAM 6 are used in this work. 
\added{Since we are focusing on the pressure-velocity coupling challenge and not on instabilities in convection-dominated flows, the simulations are carried out for low Reynolds numbers, i.e.\ the flows are considered laminar.} For the full order simulations, the spatial discretization is performed using central differencing schemes. For the open cavity, an upwind discretization scheme is used for the convective term due to a higher Peclet number of this case and to test the methods for different numerical schemes.

We focus on the non-parametric case. Therefore, the same boundary conditions are applied in the ROM as in the FOM for which the snapshots are collected. The time step, the total simulation time and the Reynolds number are also identical for the FOM and the ROM. 

\subsection{Lid-driven cavity flow problem}\label{sec:setupLDC}
\begin{figure}[h!]
	\centering
	\captionsetup{justification=centering}
	\includegraphics[width=7.0cm]{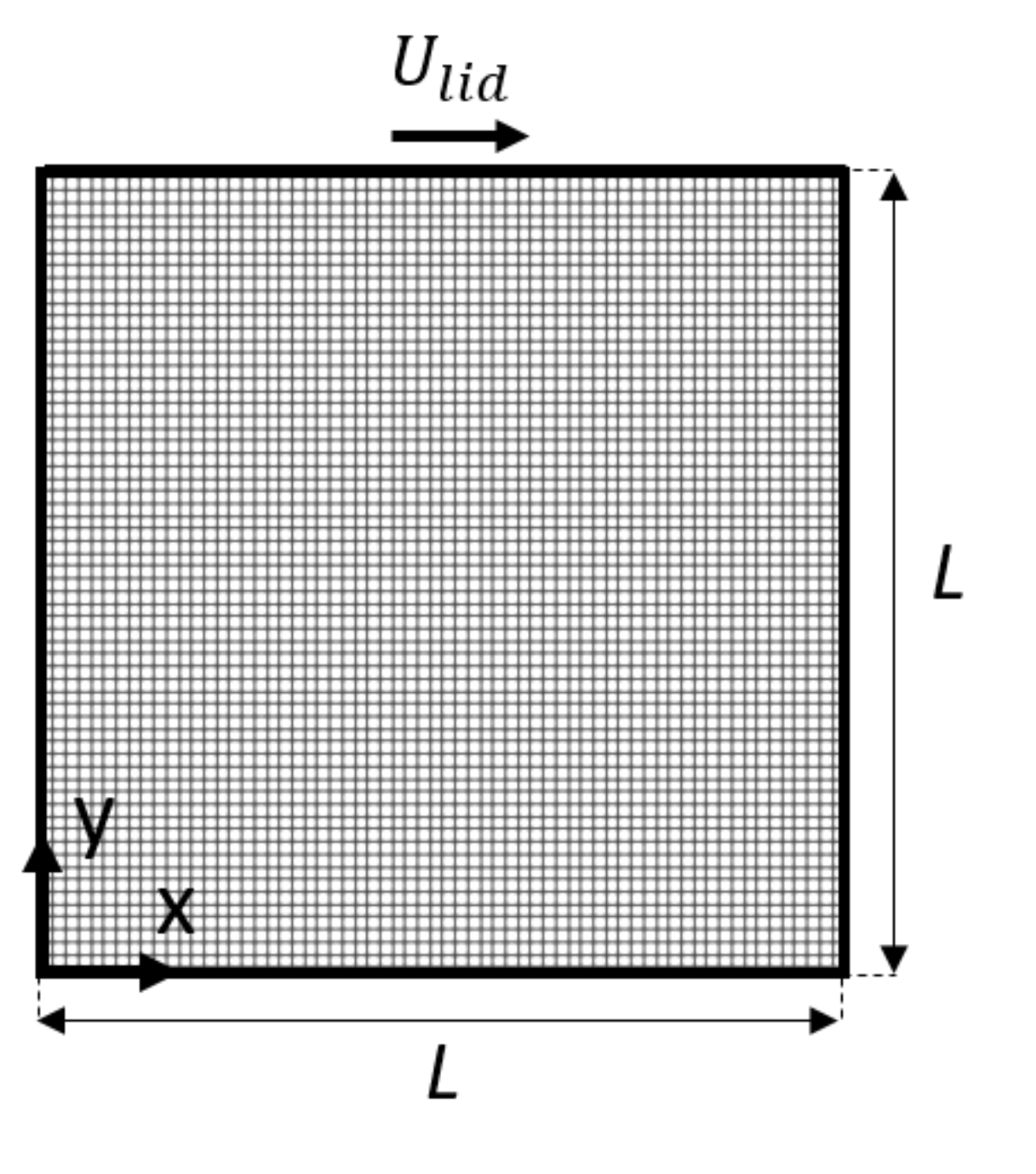}
	\caption{Sketch of the geometry and mesh of the 2D square cavity with a moving top lid.}
	\label{fig:LID_setup}
\end{figure}

Figure~\ref{fig:LID_setup} depicts a sketch of the geometry of the two-dimensional lid driven cavity problem. The length of the square cavity, $L$, equals 1.0 m. A (64 $\times$ 64) structured mesh with quadrilateral cells is constructed on the domain. A tangential uniform velocity $U_{lid}$ = 1.0 m/s is prescribed at the top wall and non-slip conditions are applied to the other walls. The Reynolds number based on the velocity of the lid and the cavity characteristic length is 100 and the flow is considered laminar. The pressure reference value is set to 0 Pa at coordinate (0,0) at the lower left corner of the cavity. The initial condition for the cell-centered velocity is a zero field: $\boldsymbol{u}_0$ = $\boldsymbol{0}$. 
Simulation are run with a constant time step of $\Delta t$ = 5 $\cdot$ \num{e-3} s and for a total simulation time, $T$, of 1.0 s. 

\newpage

\subsection{Open cavity flow problem}\label{sec:setupOC}
\begin{figure}[h!]
	\centering
	\captionsetup{justification=centering}
	\includegraphics[width=16.0cm]{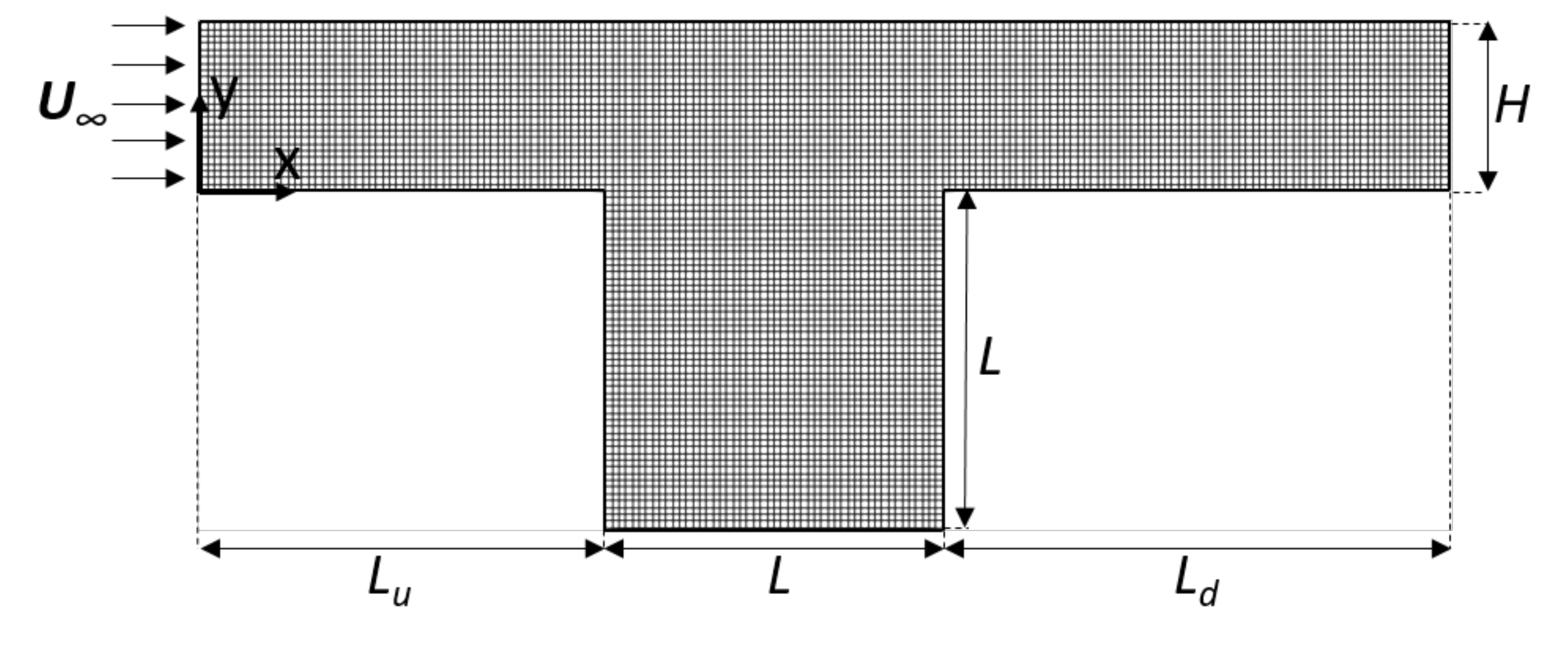}
	\caption{Sketch of the geometry and mesh of the 2D open cavity.}
	\label{fig:OC_setup}
\end{figure}

The second test case consists of a two-dimensional square cavity problem with an inlet and outlet along the top~\cite{poussot2015parametric,barbagallo2009closed,sipp2007global}. Figure~\ref{fig:OC_setup} depicts a sketch of the geometry. The height of the cavity equals its length $L$ = 1.0 m. The fluid enters from the left of the domain at a uniform velocity $U_\infty$ = 1.0 m/s. The inlet is located $L_u = 1.2 L$ upstream of the cavity and the exit $L_d = 1.5 L$ downstream of the cavity. The outflow boundary condition of Equation~\ref{eq:bc_vel_out} is considered at the outlet. The no-slip boundary condition is applied to all walls. The pressure reference value is set to 0 Pa at coordinate (0,0).
The computational domain is divided into 7125 quadrilateral cells. The Reynolds number based on the free-stream velocity $U_\infty$ and the cavity characteristic length $L$ is 200.

The initial condition for the cell-centered velocity is determined by solving a potential flow problem subjected to the problem's boundary conditions is given by
\begin{align} \label{eq:potential_flow}
\begin{cases}
\nabla \cdot \boldsymbol{u}_0 = 0   &\mbox{in  } \Omega,  \\
\nabla^2 p = 0 &\mbox{in  } \Omega.
\end{cases}
\end{align}
The total simulation time is $T = 2.0$ s with a time step $\Delta t = 2.5\times10^{-3}$ s. Snapshots of the flow fields are collected every time step. Table~\ref{tab:compdetails} summarizes the computational details for the open cavity flow problem.

\begin{table}[h!]
	\centering
	\renewcommand{\arraystretch}{1.2}
	\caption{Computational details for the lid driven cavity and open cavity flow problems.}
	\begin{tabular}{lll}
		\toprule	
		\textbf{Variables} & \textbf{Lid driven cavity} & \textbf{Open cavity} \\
		\midrule
		Number of cells & 4096 & 7125  \\
		Cavity length $L$  & 1.0 m & 1.0 m \\
		$U_{lid}$, $U_{\infty}$  & 1.0 m/s&1.0 m/s \\
		Viscosity $\nu$ & 0.01 m$^2$/s& 0.005 m$^2$/s\\
		Reynolds number & 100  & 200\\ 
		Simulation time $T$  & 1.0 s & 2.0 s\\
		Time step $\Delta t$  & 0.005 s & 0.0025 s\\
		Spatial scheme convection  & Linear (central differencing) & Linear (upwind) \\
		Temporal scheme  & Forward Euler& Forward Euler\\		
		\bottomrule
	\end{tabular}\label{tab:compdetails}
\end{table}

\newpage 

\section{Results} \label{sec:results}
In this section, we show the full order and reduced order results of two test cases: the lid driven cavity flow problem and the open cavity flow problem. These open and closed flow test cases are excellent test cases to demonstrate the difference in the treatment of the (non-homogeneous) boundary conditions: In the case of the closed cavity, an tangential boundary condition is applied on the top wall of the cavity, while an inflow and outflow boundary condition are applied for the open flow problem.

One of the main goals of this work is to \added{accurately} reproduce the FOM results with our developed reduced order models\deleted{ in a stable and accurate way}. Therefore, rather than validating the models against experimental results and/or other numerical models, we directly compare the ROM results with the corresponding FOM results. 

We analyze and compare the FOM and ROM results of the inconsistent flux method and the consistent flux method. The main difference between the two projection methods is that mass conserving face fluxes are obtained with the CFM, while the fluxes are only approximately discretely divergence free in the case of the IFM. Therefore, we compare the summation of the local continuity errors for every cell at all time instances as they give an indication of how well the continuity equation is satisfied in the simulations. The local time step continuity error is calculated, according to the definition used by OpenFOAM~\cite{greenshields2015openfoam}, as follows for the FOM fields:
\begin{equation}\label{local_continuity}
\epsilon_{local} (t^n) =  \sum_{k=1}^{h} \frac{\Delta t}{(\Omega_{h})_k} \left| \left[\sum_{i=1}^{N_f}  \phi_{f,i} (t^n) \right]_k \right|.
\end{equation}
Similarly, the local time step continuity error can be determined for the POD velocity modes and the fields obtained with the ROMs.

Furthermore, we compute the relative error of the cell-centered fields at each time step to show the performance of the proposed methods. For this we consider the following three types of fields at a time instance $t^n$: the full order fields $\boldsymbol{u}_p^n$ and $\boldsymbol{p}_p^n$, the projected fields $\hat{\boldsymbol{u}}_{p,r}^n$ = $\boldsymbol{\Phi}\boldsymbol{\Phi}^T\replaced{\boldsymbol{V}}{\boldsymbol{\Omega}}\boldsymbol{u}_p^n$ and $\hat{\boldsymbol{p}}_{p,r}^n = \boldsymbol{X}\boldsymbol{X}^T\replaced{\boldsymbol{V}}{\boldsymbol{\Omega}}\boldsymbol{p}_p^n$, which are obtained by the $L_2$-projection of the snapshots onto the POD bases and lastly, the predicted fields $\boldsymbol{u}_{p,r}^n$ and $\boldsymbol{p}_{p,r}^n$ obtained by solving the ROMs. For every time instance, $t^n$, the relative basis projection error is given by
\begin{equation}\label{eq:l2_projection}
\|\hat{\epsilon}^u\|_{L_2(\Omega_h)}(t^n) = \frac{\|\boldsymbol{u}_p^n -\hat{\boldsymbol{u}}_{p,r}^n\|_{L_{2}(\Omega_h)}}{\|\boldsymbol{u}^n_p \|_{L_{2}(\Omega_h)}},
\end{equation}
\noindent and the prediction error is determined by
\begin{equation}\label{eq:l2_prediction}
\|\epsilon^u\|_{L_2(\Omega_h)}(t^n) = \frac{\|\boldsymbol{u}_p^n -\boldsymbol{u}_{p,r}^n\|_{L_{2}(\Omega_h)}}{\|\boldsymbol{u}^n_p \|_{L_{2}(\Omega_h)}}.
\end{equation}
Similarly, $\|\hat{\epsilon}^p\|_{L_2(\Omega_h)}(t^n)$ and $\|\epsilon^p\|_{L_2(\Omega_h)}(t^n)$ are computed for the pressure fields. For each of the cases and methods we compare the relative prediction error with the basis projection error, which is the `best possible' error at every time instance. 

Finally, we determine the speedup in computational time, which is defined as the FOM CPU time divided by the ROM CPU time.

\subsection{Lid driven cavity flow problem}\label{sec:resultsLID}
\begin{figure}[h!]
	\centering
	\begin{subfigure}{.5\linewidth}
		\centering
		\includegraphics[width=1.\linewidth]{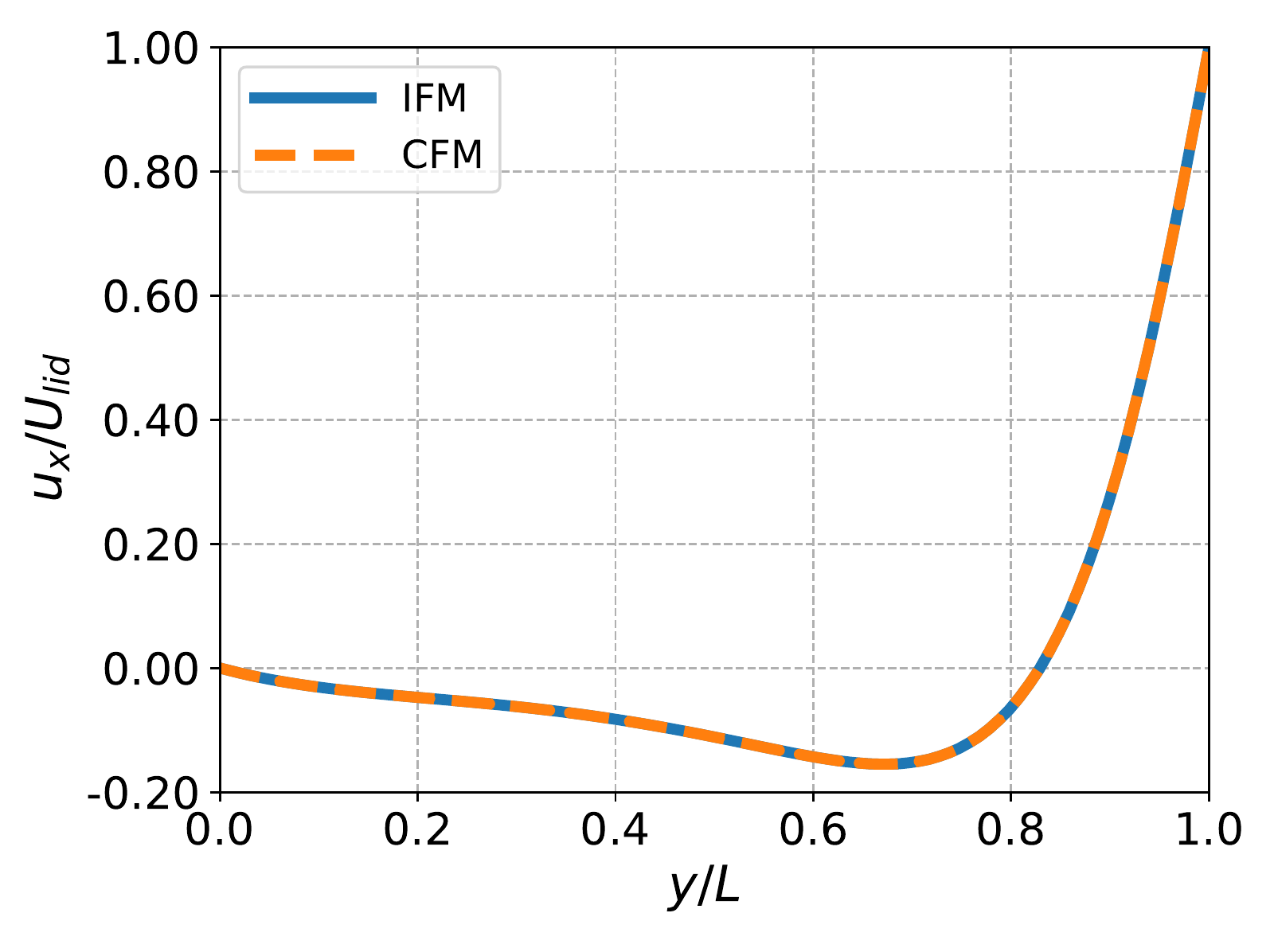}
	\end{subfigure}%
	\begin{subfigure}{.5\linewidth}
		\centering
		\includegraphics[width=1.\linewidth]{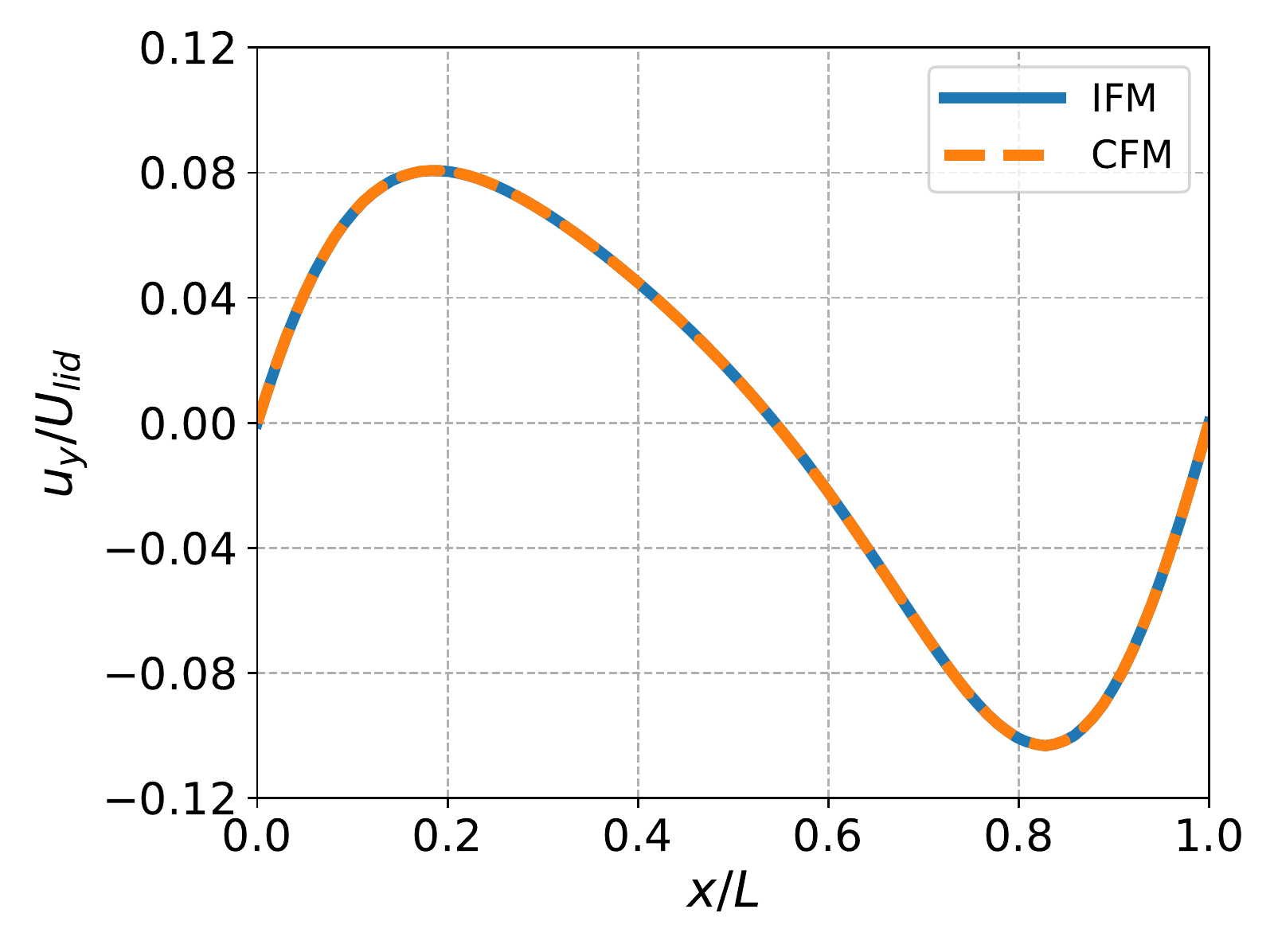}
	\end{subfigure}
	\caption{Velocity profiles for the lid driven cavity flow case at final simulation time: (left) normalized velocity component in the $x$-direction at $x/L$ = 0.5; (right) normalized velocity component in the $y$-direction at $y/L$ = 0.5. }
	\label{fig:u_profiles_LDC}
\end{figure}

\begin{figure}[h!]
	\centering
	\begin{subfigure}{.5\linewidth}
		\centering
		\includegraphics[width=1.\linewidth]{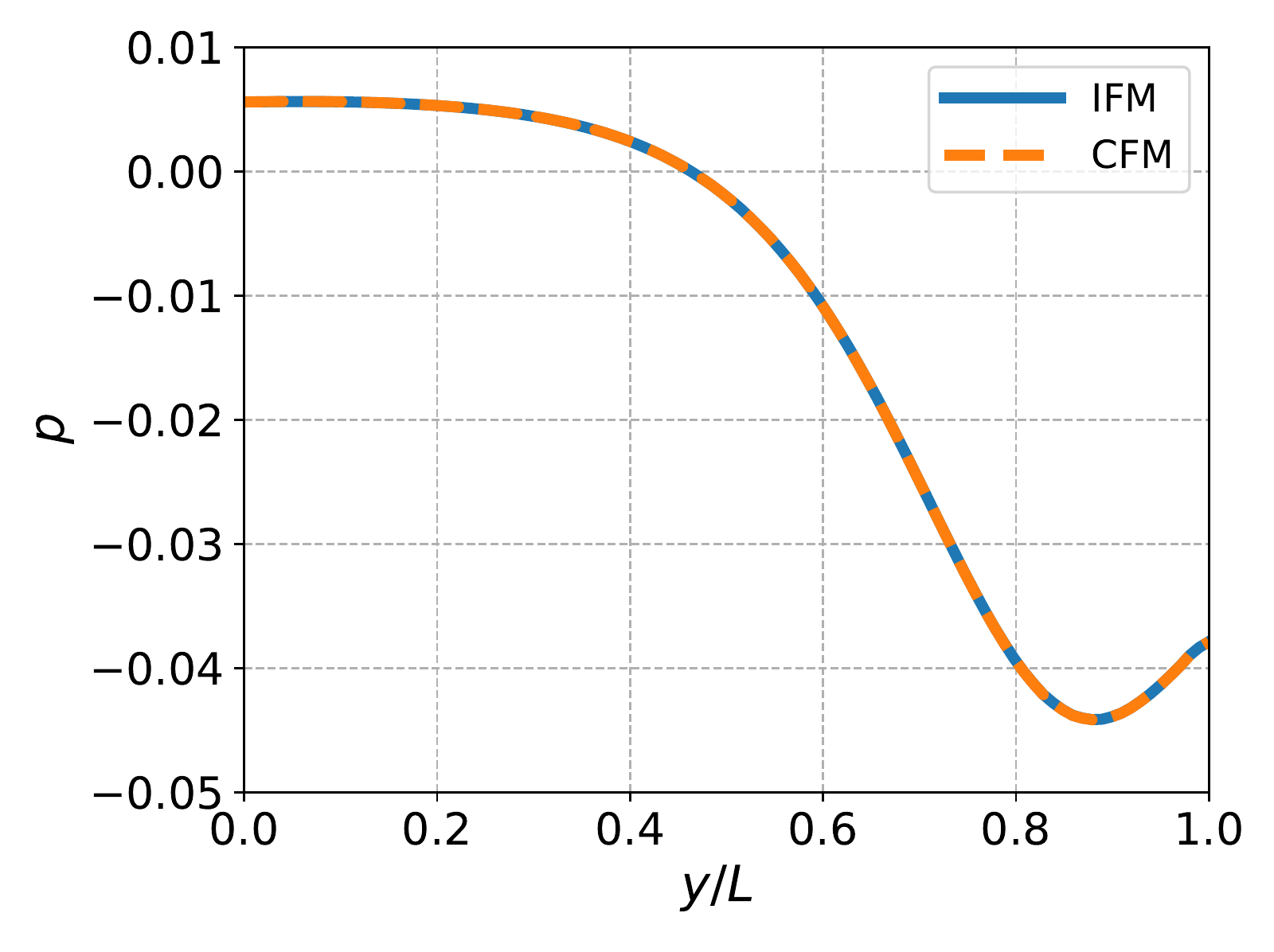}
	\end{subfigure}%
	\begin{subfigure}{.5\linewidth}
		\centering
		\includegraphics[width=1.\linewidth]{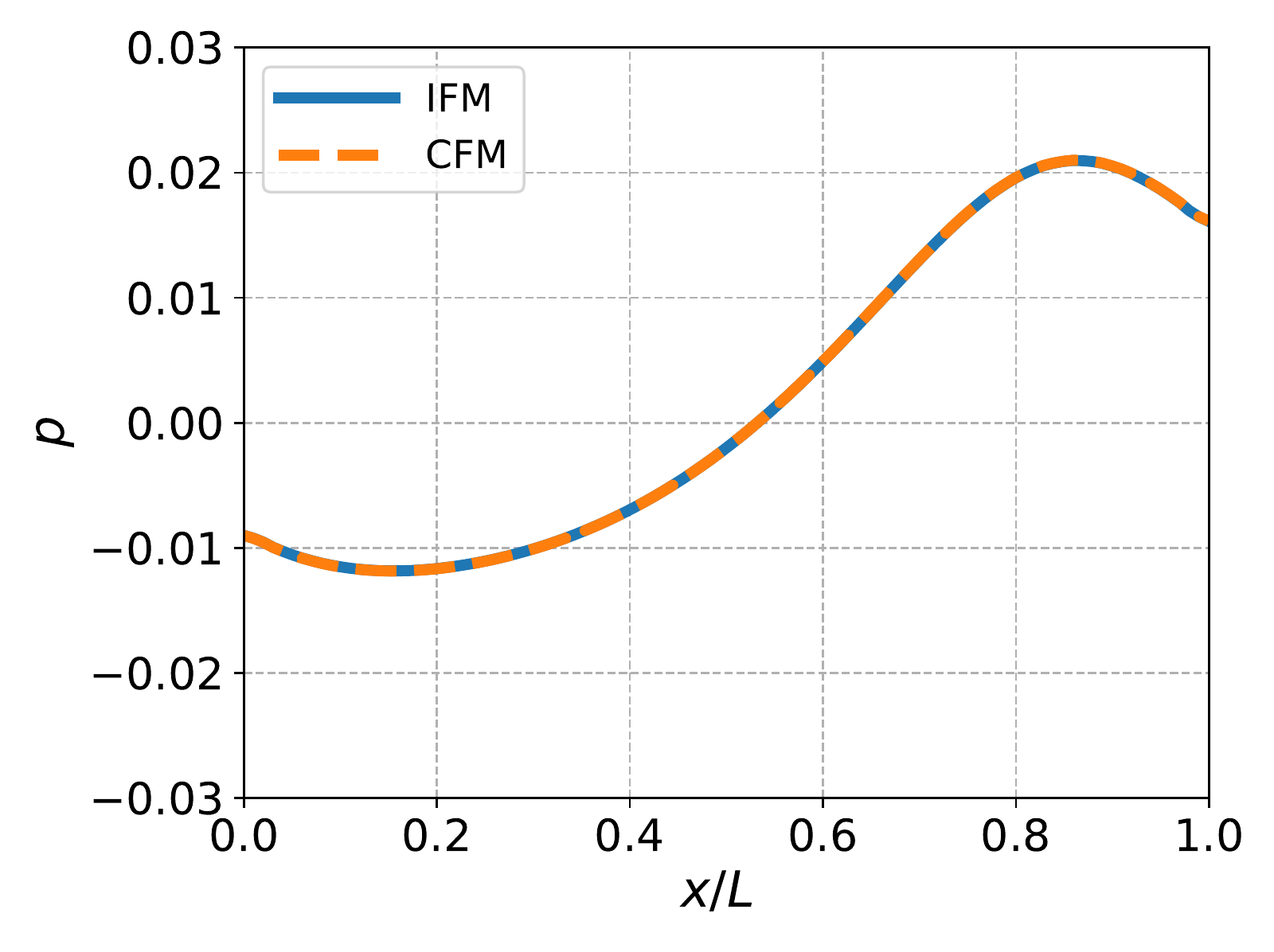}
	\end{subfigure}
	\caption{Normalized pressure profiles for the lid driven cavity flow case at final simulation time: (left) at $x/L$ = 0.5; (right) at $y/L$ = 0.5. }
	\label{fig:p_profiles_LDC}
\end{figure}

Full order simulations are performed for the lid driven cavity problem according to Section~\ref{sec:setupLDC}. The velocity and pressure profiles at the centerlines of the cavity at final simulation time are shown in Figures~\ref{fig:u_profiles_LDC} and~\ref{fig:p_profiles_LDC}, respectively. These figures show that the full order solutions obtain with the inconsistent flux method are close to the consistent-flux solutions.

The local continuity errors (Equation~\ref{local_continuity}) of the IFM-FOM is of the order \num{e-6}, while it is of the order \num{e-16} in the case of the CFM-FOM. Nevertheless, this difference can be considered negligible in this particular case as the Figures~\ref{fig:u_profiles_LDC} and~\ref{fig:p_profiles_LDC} show that the methods perform equally.

\begin{figure}[h!]
	\centering
	\begin{subfigure}{.5\linewidth}
		\centering
		\includegraphics[width=1.\linewidth]{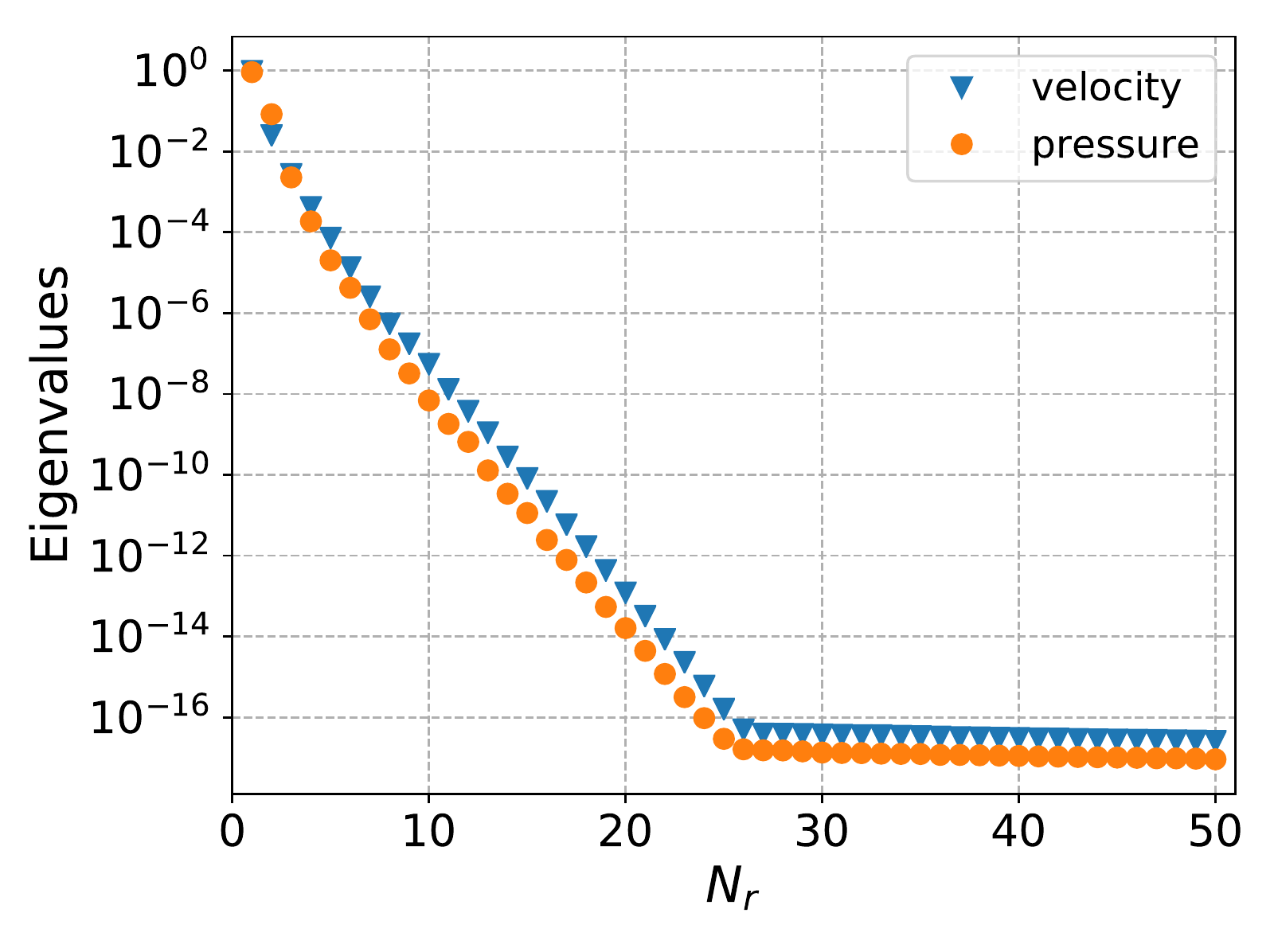}
	\end{subfigure}%
	\begin{subfigure}{.5\linewidth}
		\centering
		\includegraphics[width=1.\linewidth]{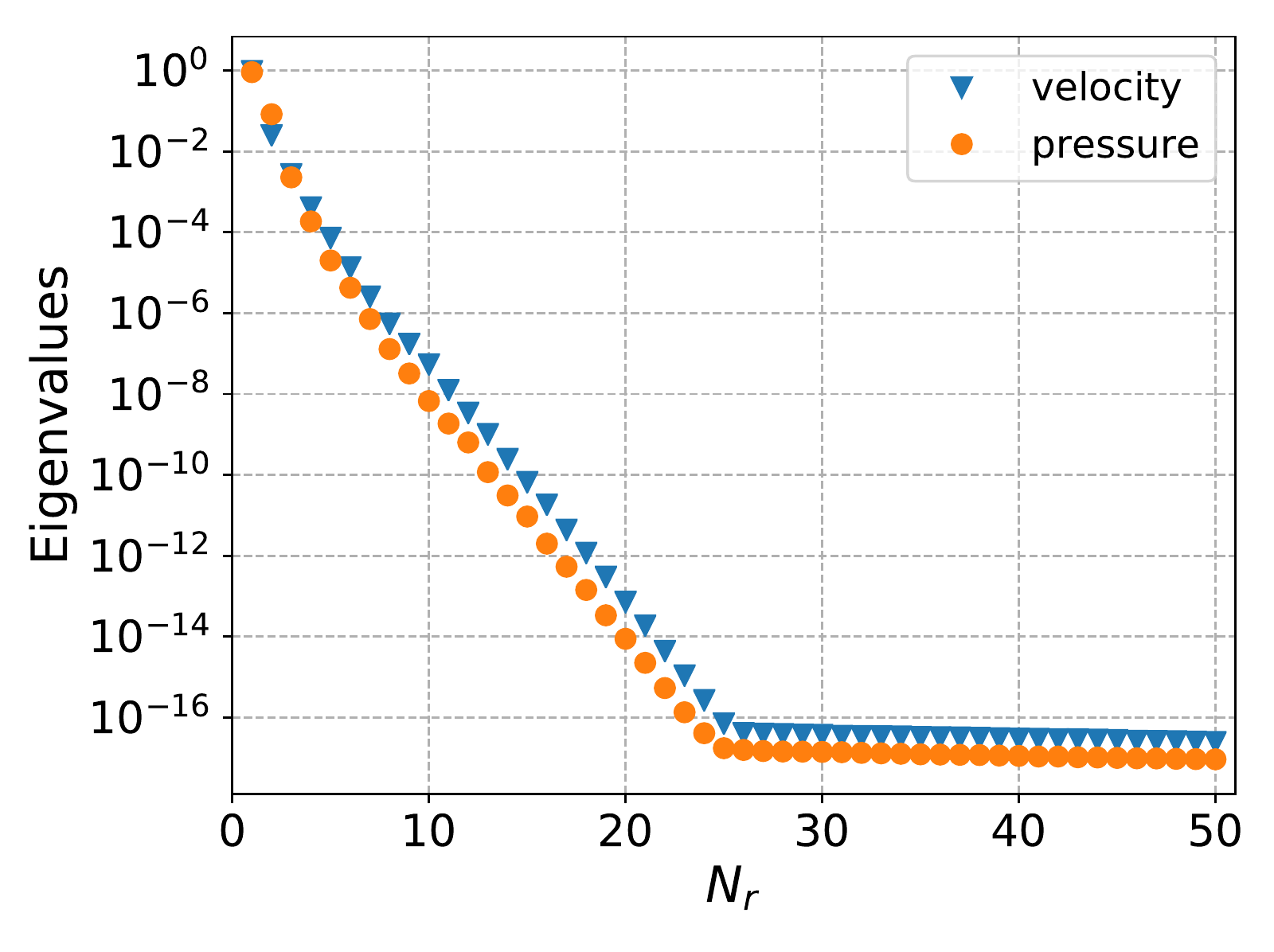}
	\end{subfigure}
	\caption{Eigenvalues as function of the number of modes for the lid driven cavity flow case: (left) inconsistent flux method; (right) consistent flux method.}
	\label{fig:ev_LDC}
\end{figure}

The POD eigenvalues of the cell-centered velocity and pressure modes are shown in Figure~\ref{fig:ev_LDC}. The eigenvalues are approximately the same for both projection methods. For both velocity and pressure, the values decay rapidly for increasing number of modes. Therefore, the problem is suited for dimension reduction. A plateau is reached at about 25 modes due to the machine precision. As the slope of eigenvalue decay is almost the same for pressure and velocity, we take an equal number of modes $N_r$ for the reduced pressure basis and reduced velocity basis: $N_r = N_r^u = N_r^p$. 

We study the effect of increasing the number of modes on the accuracy of the cell-centered velocity field, while using the full snapshot set as basis for the POD. We take $N_r$ = 2; 5; 10; 15; 20. 

\begin{figure}[h!]
	\centering
	\begin{subfigure}{.5\linewidth}
		\centering
		\includegraphics[width=1.\linewidth]{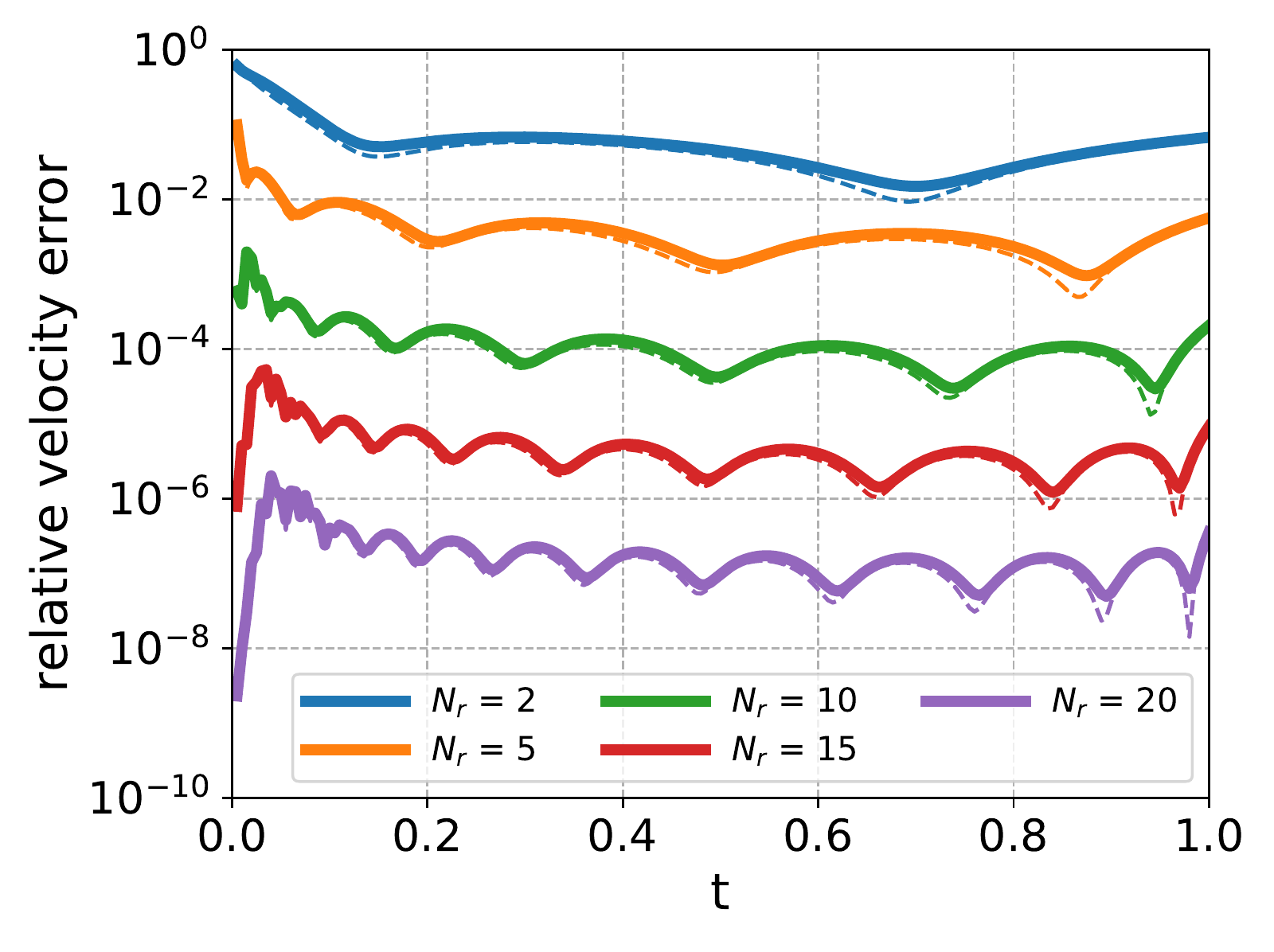}
	\end{subfigure}%
	\begin{subfigure}{.5\linewidth}
		\centering
		\includegraphics[width=1.\linewidth]{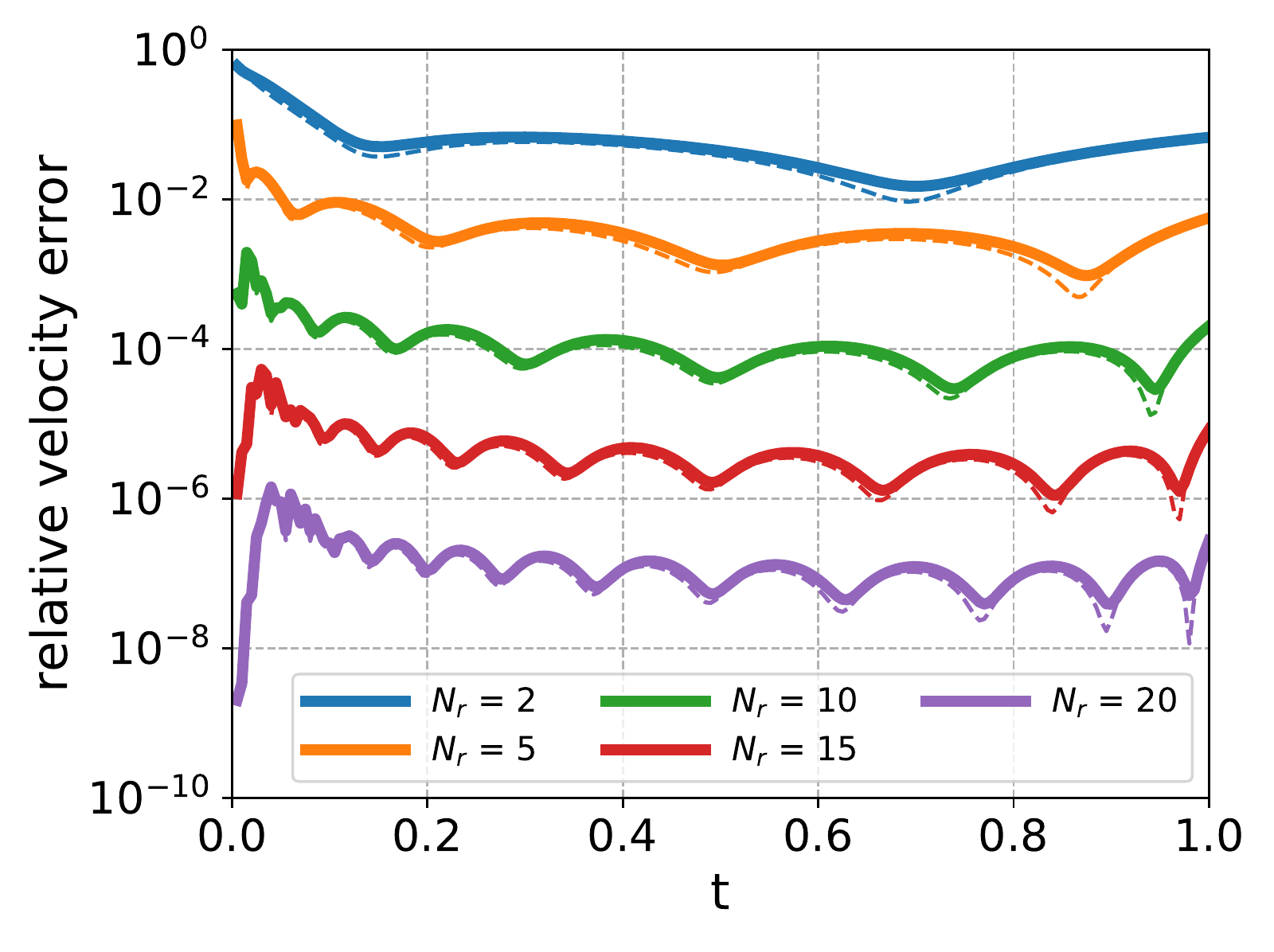}
	\end{subfigure}
	\caption{Relative cell-centered velocity error as a function of time for different number of modes for the lid driven cavity flow case: (left) inconsistent flux method; (right) consistent flux method. Dashed lines: basis projection error (projecting snapshots onto truncated basis).}
	\label{fig:L2error_U_LDC}
\end{figure}

\begin{figure}[h!]
	\centering
	\begin{subfigure}{.5\linewidth}
		\centering
		\includegraphics[width=1.\linewidth]{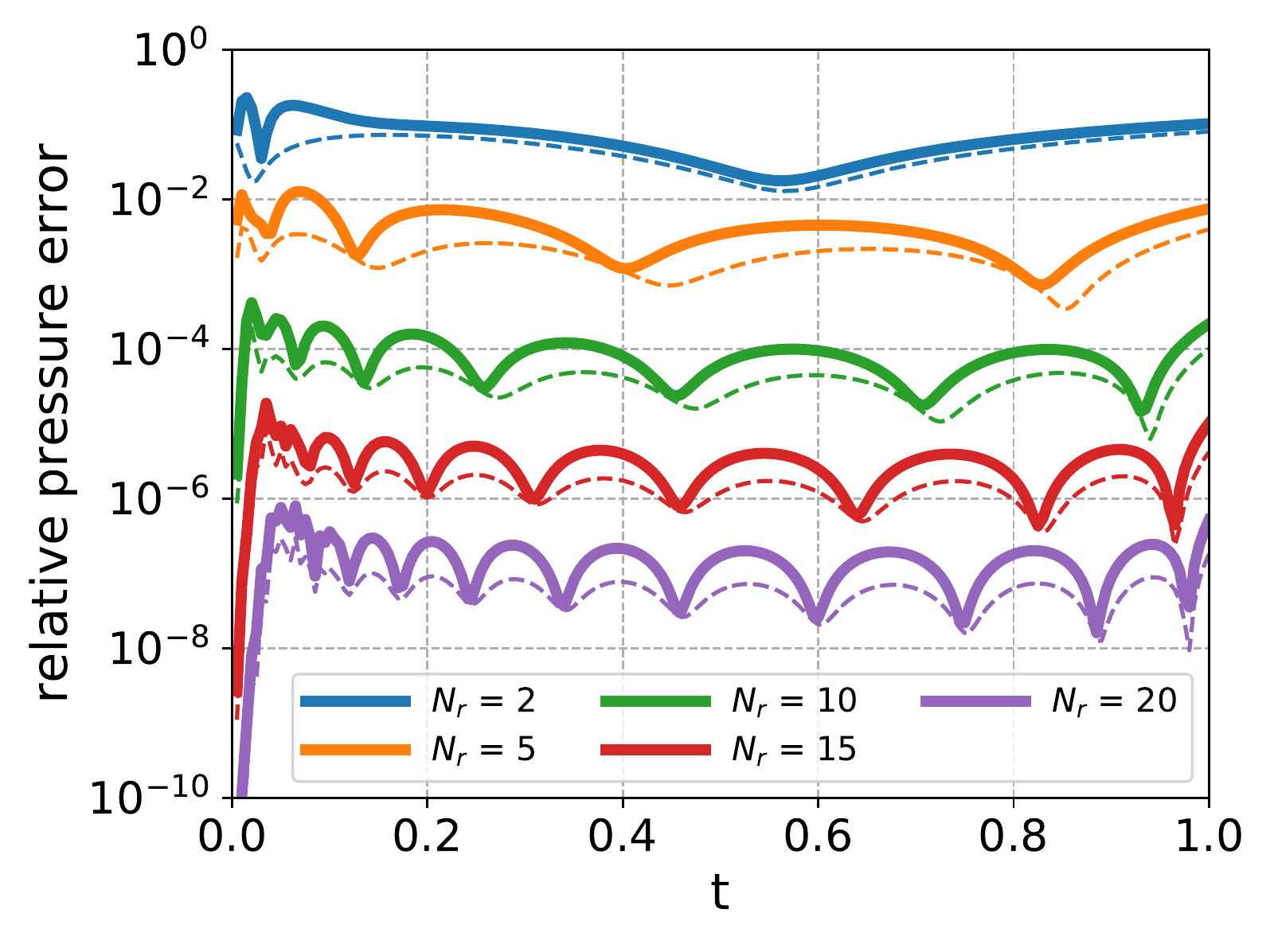}
	\end{subfigure}%
	\begin{subfigure}{.5\linewidth}
		\centering
		\includegraphics[width=1.\linewidth]{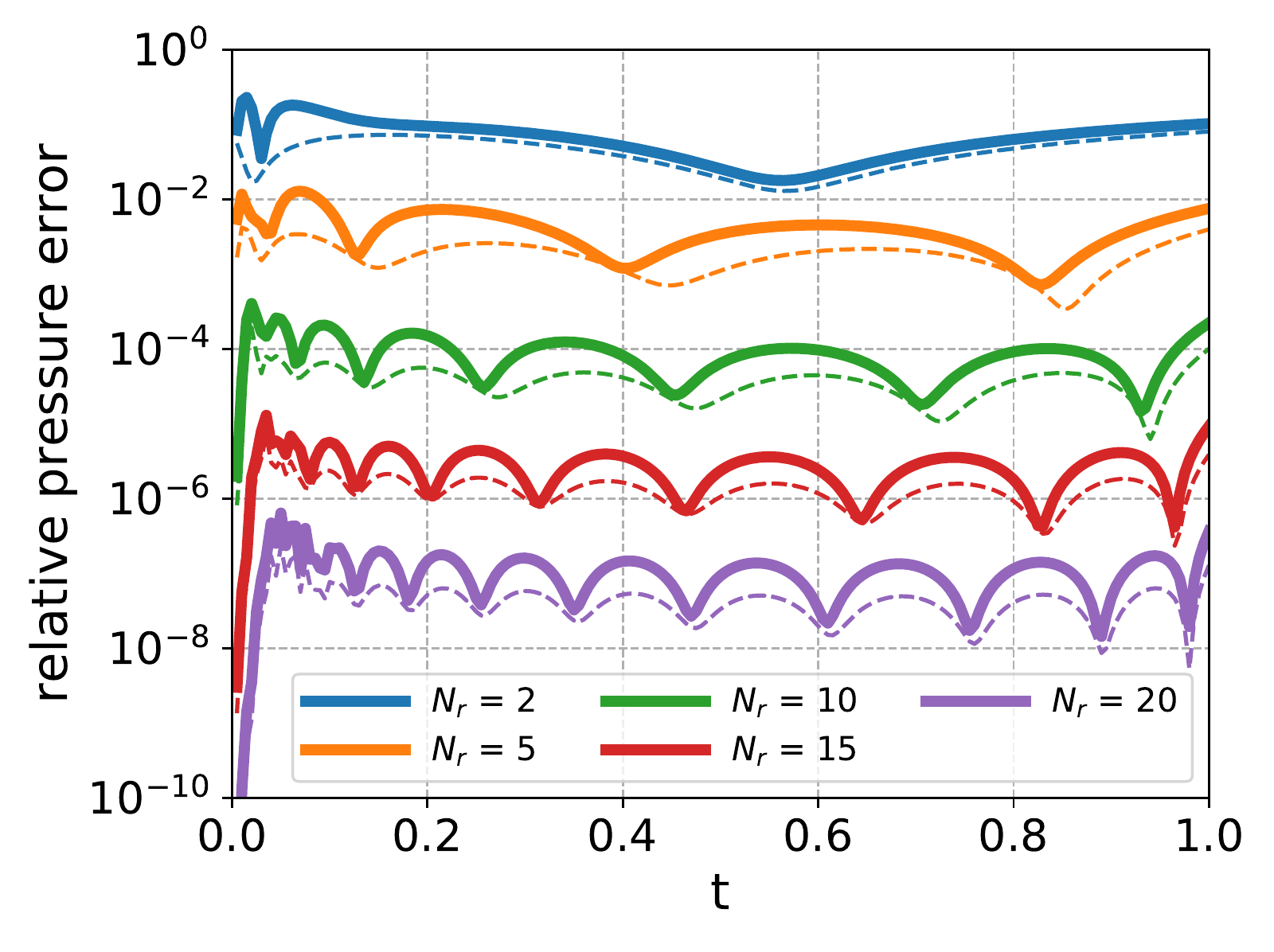}
	\end{subfigure}
	\caption{Relative pressure error as a function of time for different number of modes for the lid driven cavity flow case: (left) inconsistent flux method; (right) consistent flux method. Dashed lines: basis projection error (projecting snapshots onto truncated basis).}
	\label{fig:L2error_P_LDC}
\end{figure}

The relative prediction and basis projection errors are plotted in Figure~\ref{fig:L2error_U_LDC} for velocity and Figure~\ref{fig:L2error_P_LDC} for pressure. We clearly see how the accuracy increases when increasing the number of modes. The relative error for a certain number of modes appears to be almost the same for both projection methods. This means that both ROMs are consistent with the FOMs used for the snapshot collection~\cite{grimberg2020stability}.

\begin{figure}[h!]
	\centering
	\begin{subfigure}{.5\linewidth}
		\centering
		\includegraphics[width=1.\linewidth]{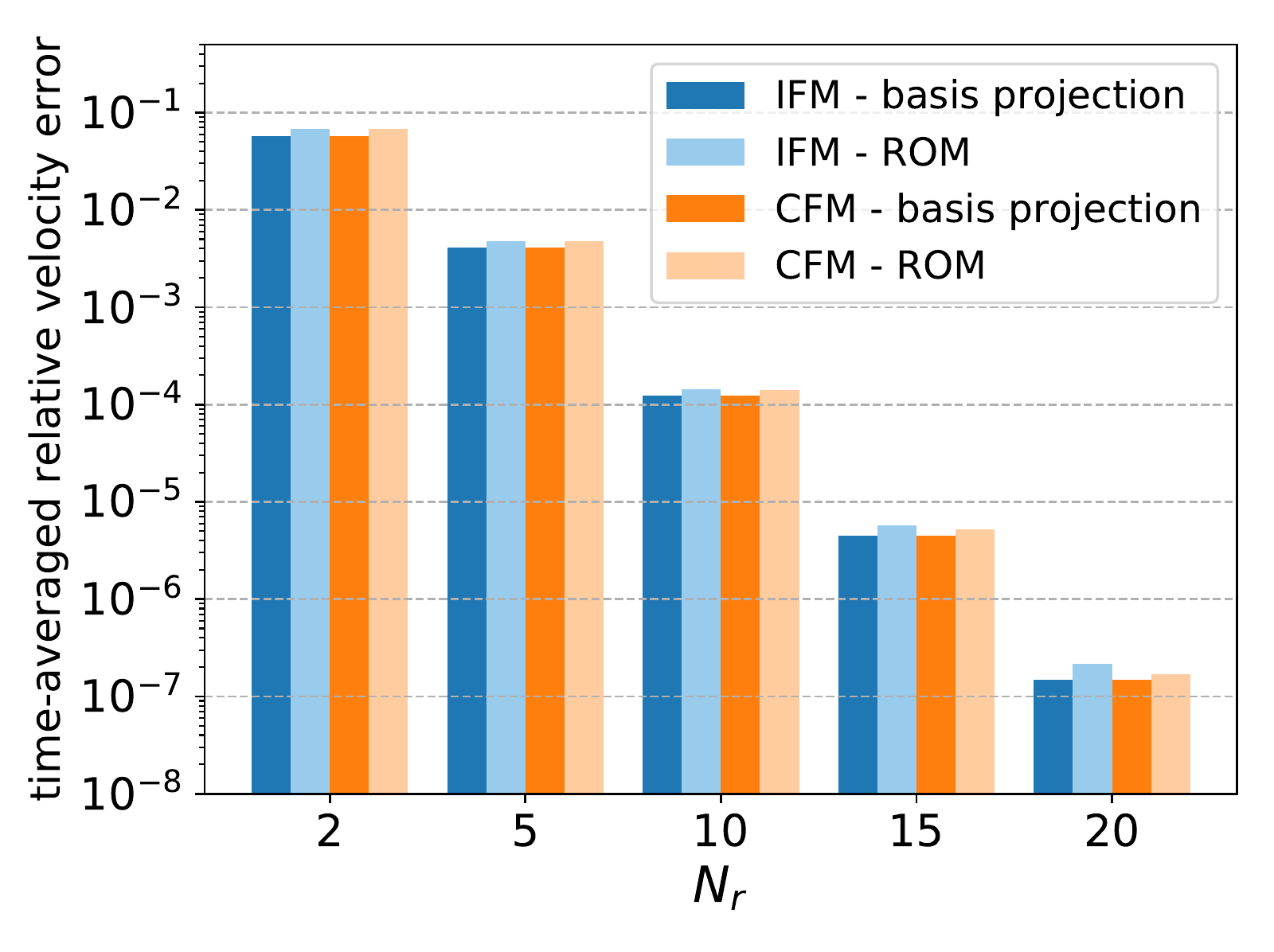}
	\end{subfigure}%
	\begin{subfigure}{.5\linewidth}
		\centering
		\includegraphics[width=1.\linewidth]{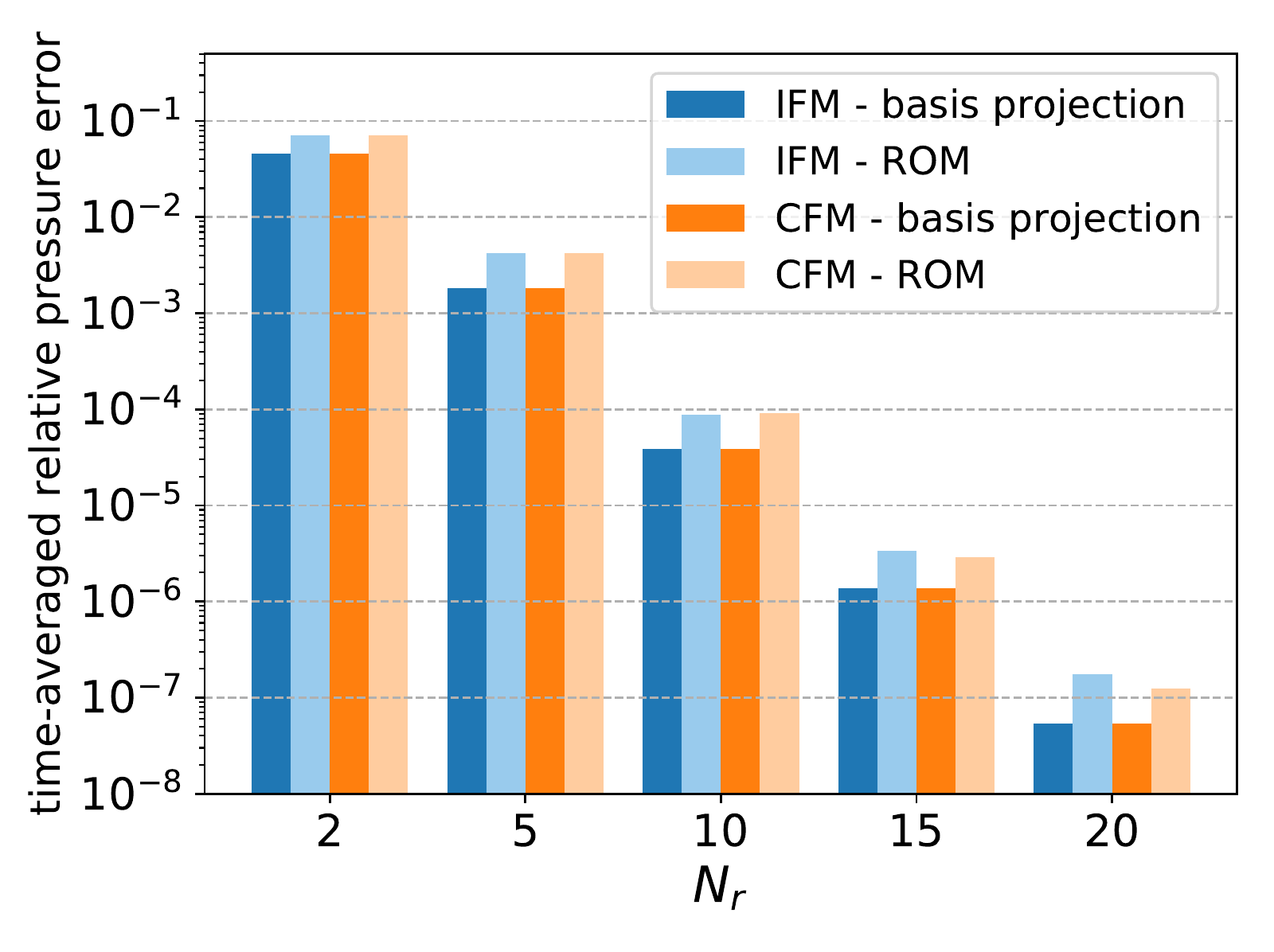}
	\end{subfigure}
	\caption{Time-averaged relative basis projection and prediction errors of the lid driven cavity flow problem: (left) velocity; (right) pressure.}
	\label{fig:LDC_time_ave_L2_errors}
\end{figure}

Furthermore, for both ROM methods, the relative velocity errors (Equation~\ref{eq:l2_prediction}) are very close to the relative basis projection errors (Equation~\ref{eq:l2_projection}) as they are almost overlapping. In addition, the relative pressure errors are of the same order as the velocity errors for the same number of modes. This is also shown in Figure~\ref{fig:LDC_time_ave_L2_errors} in which we plotted the time-averaged basis projection errors (Equation~\ref{eq:l2_projection}) and time-averaged ROM prediction errors (Equation~\ref{eq:l2_prediction}) for velocity and pressure, respectively. 

In all cases, \replaced{accurate}{stable} ROM results were obtained with the proposed explicit projection method. This indicates that additional pressure stabilization methods, such as the supremizer enrichment technique, the exploitation of a pressure Poisson equation during the projection stage or the novel local projection stabilization methods~\cite{rubino2020numerical}, are not required. Moreover, in this test case a relative error of about $\mathcal{O}$(\num{e-4}), which is accurate enough for many engineering applications, is obtained with only 10 velocity and 10 pressure modes (plus 10 face velocity modes in the case of the consistent flux method).

Furthermore, the local continuity errors computed for the IFM-POD velocity modes and the IFM-ROM are of the order \num{e-6} (regardless the number of modes used). On the other hand, the local continuity errors are of the order \num{e-16} for the CFM, which is of the order of the machine precision. They are of the same order as for the corresponding FOMs. Thus, the discrete face velocity is only approximately discretely divergence free in the case of the IFM, whereas it is discretely divergence free with the consistent flux method.

\begin{figure}[h!]
	\centering
	\begin{subfigure}{.5\linewidth}
		\centering
		\includegraphics[width=1.\linewidth]{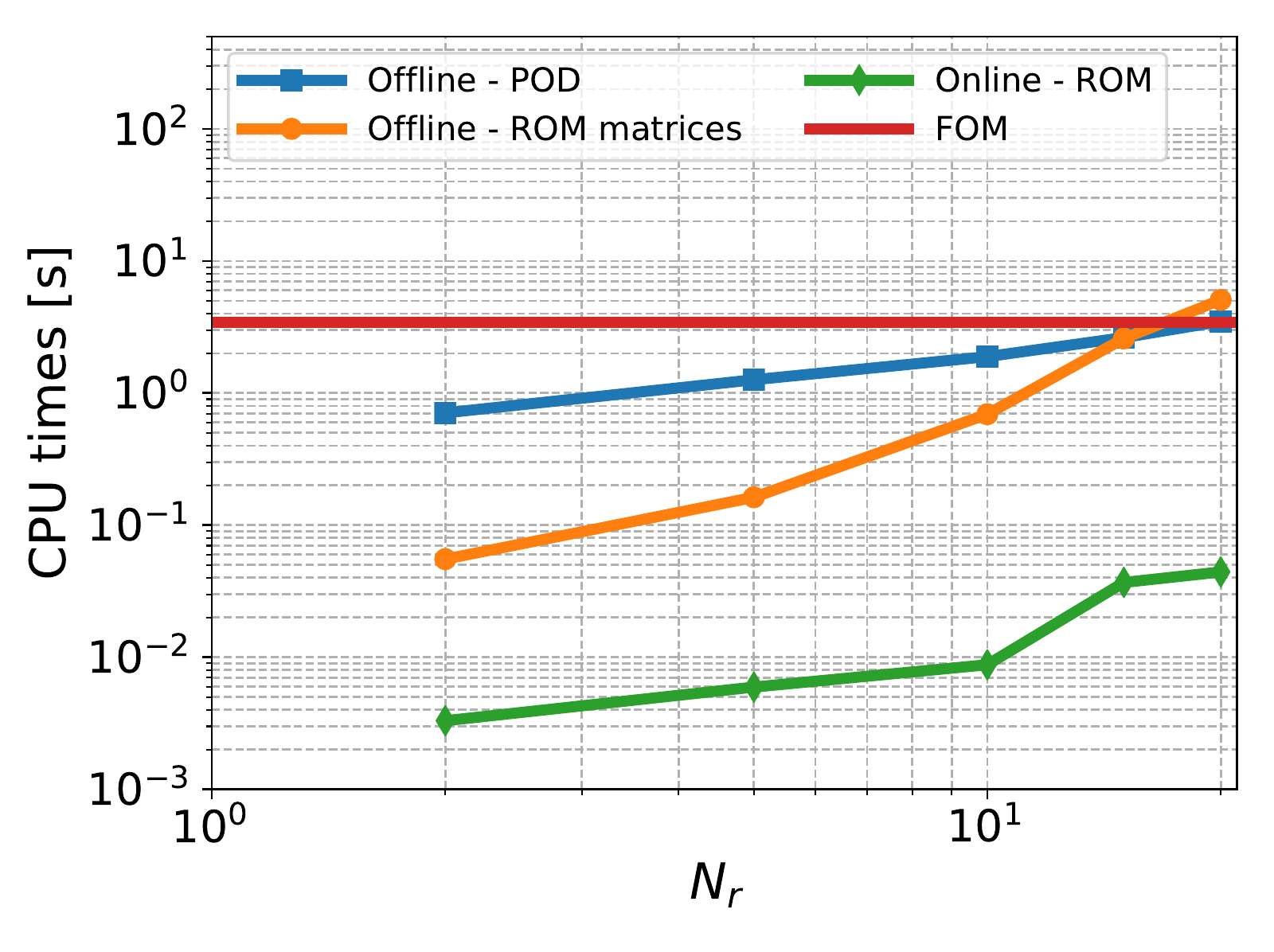}
	\end{subfigure}%
	\begin{subfigure}{.5\linewidth}
		\centering
		\includegraphics[width=1.\linewidth]{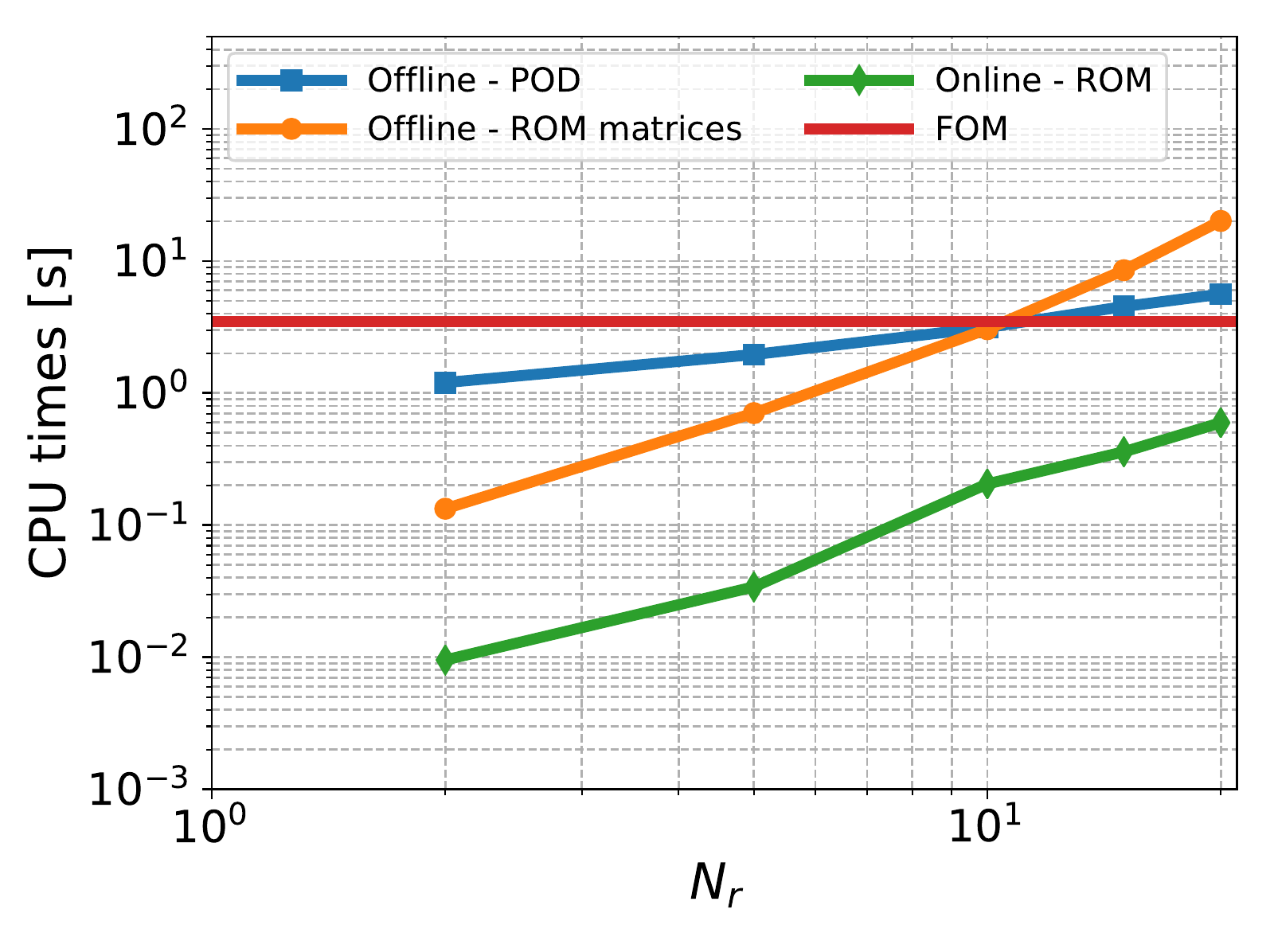}
	\end{subfigure}
	\caption{Computational times in seconds as function of number of modes for the lid driven cavity flow case: (left) inconsistent flux method; (right) consistent flux method.}
	\label{fig:CPU_IFM_LDC}
\end{figure}

\begin{figure}[h!]
	\centering
	\includegraphics[width=0.5\linewidth]{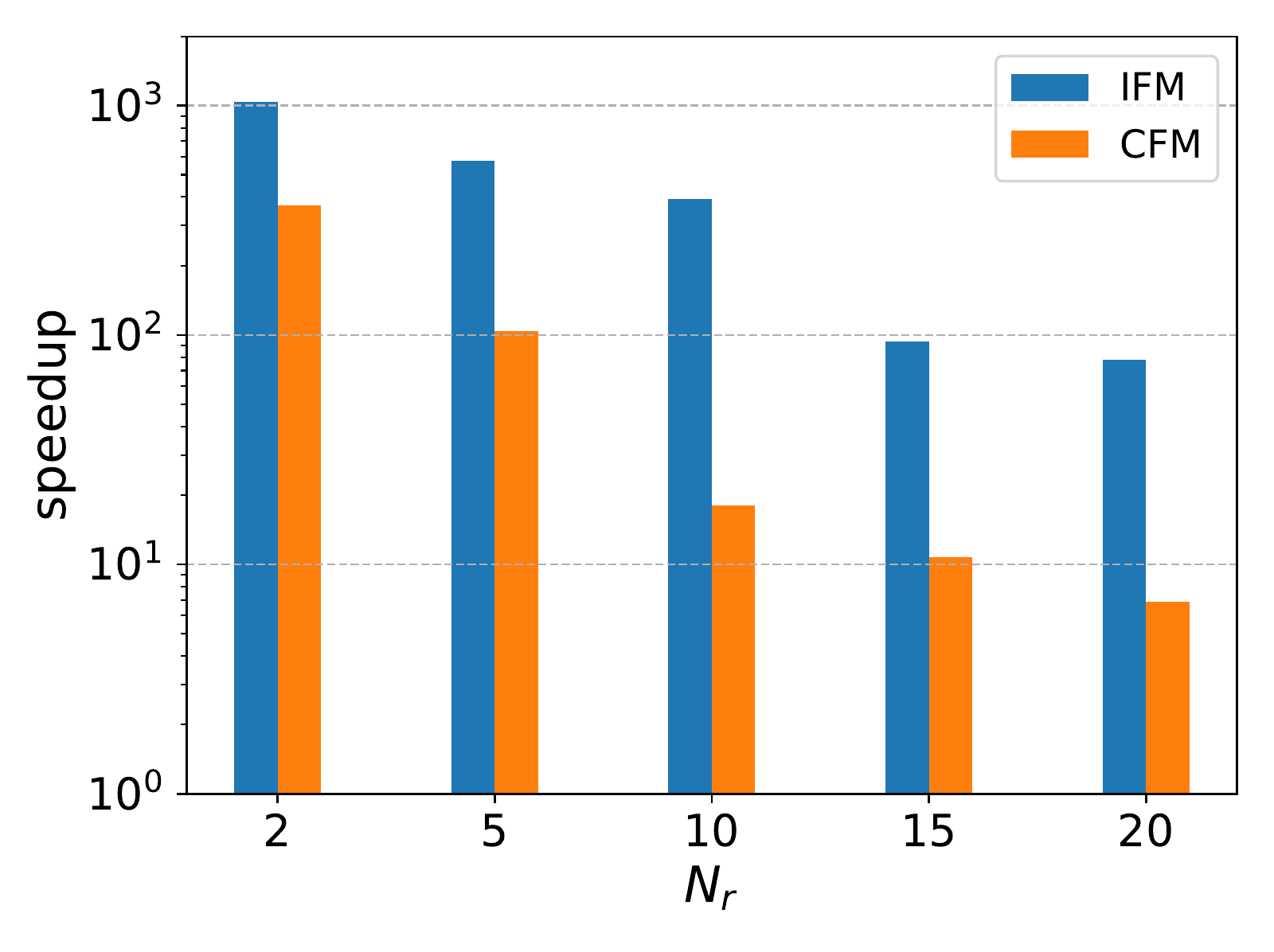}
	\caption{Speedup in computational time of the ROM compared to the FOM in seconds as function of number of modes for the lid driven cavity flow case.}
	\label{fig:speedup_LDC}
\end{figure}

\newpage
Finally, the computational time required by the ROMs is compared to the FOM CPU times in Figure~\ref{fig:CPU_IFM_LDC}. The plotted computational times are the average times of two simulations. For both methods, the speedup ratio between the ROM and the FOM is shown in Figure~\ref{fig:speedup_LDC}, which depend strongly on the number of modes used for the ROMs. In the case of the CFM, an additional equation for the face velocity (Equation~\ref{eq:cfm_phi2}) needs to be solved at the reduced order level, which explains the lower speedup compared to the IFM. Moreover, the larger the number of modes, the more time it takes to precompute the reduced matrices. This especially applies to those related to the convection operators as the dimension of the tensors increases with the cube of the number of POD modes. The cost is higher for the CFM\added{-ROM} than the IFM\added{-ROM} as more matrices need to be precomputed due to the additional equation for the face velocity (Equation~\ref{eq:cfm_phi2}). Therefore, the time to compute the POD modes is also higher for the consistent flux method. The POD is relatively expensive compared to the ROM simulation time. However, the POD modes only need to be determined once during the offline phase. 

\clearpage 
\subsection{Open cavity flow problem}\label{sec:resultsOC}
\begin{figure}[h!]
	\centering
	\begin{subfigure}{.49\linewidth}
		\centering
		\includegraphics[width=1.\linewidth]{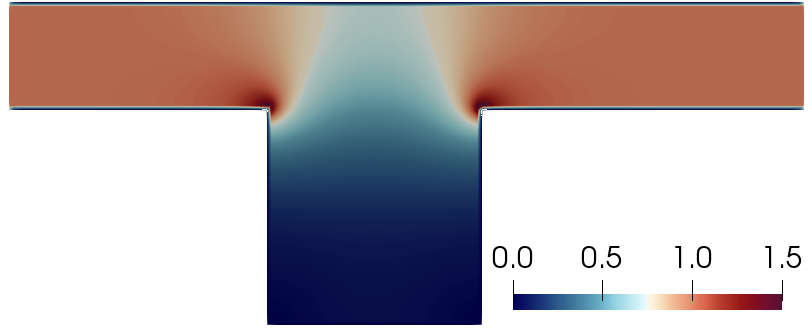}
	\end{subfigure}
	\begin{subfigure}{.49\linewidth}
		\centering
		\includegraphics[width=1.\linewidth]{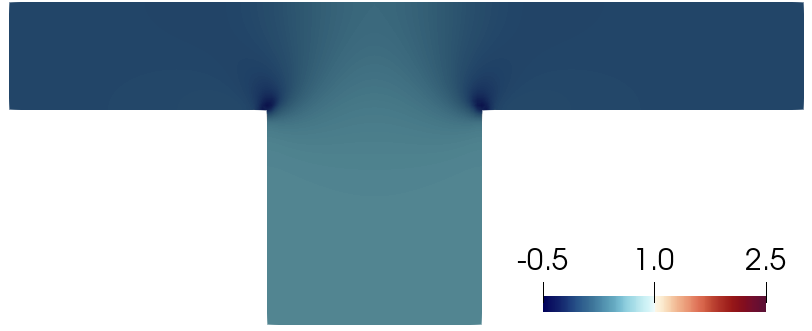}
	\end{subfigure}
	\begin{subfigure}{.49\linewidth}
		\centering
		\includegraphics[width=1.\linewidth]{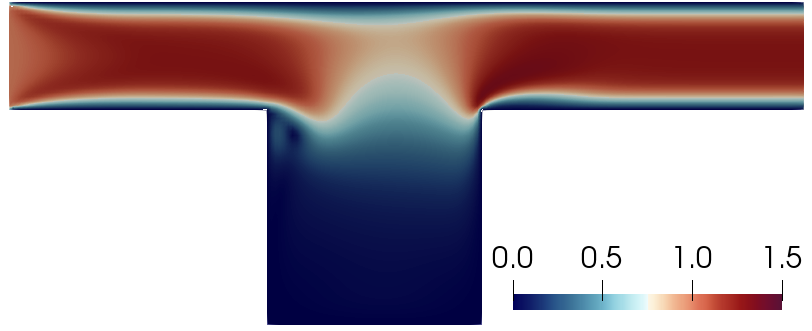}
	\end{subfigure}
	\begin{subfigure}{.49\linewidth}
		\centering
		\includegraphics[width=1.\linewidth]{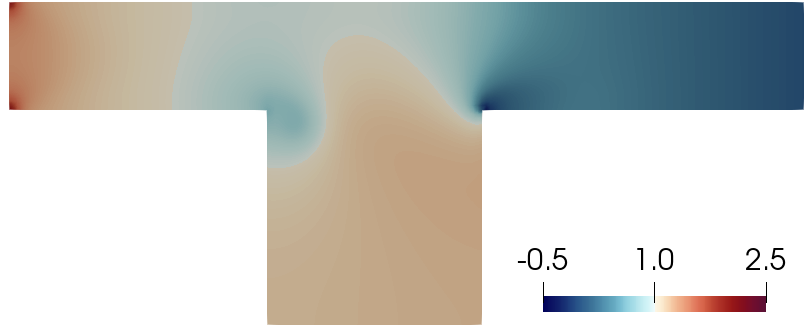}
	\end{subfigure}
	\begin{subfigure}{.49\linewidth}
		\centering
		\includegraphics[width=1.\linewidth]{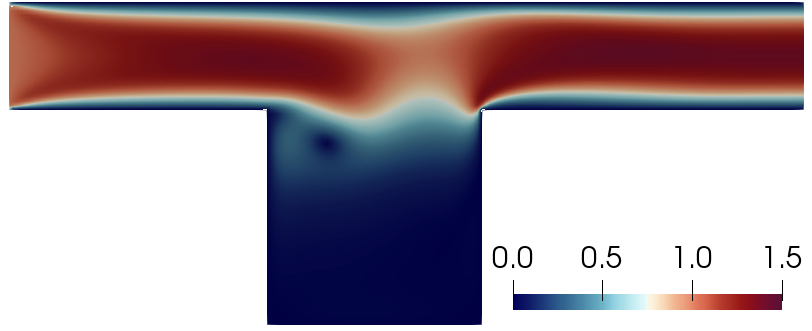}
	\end{subfigure}
	\begin{subfigure}{.49\linewidth}
		\centering
		\includegraphics[width=1.\linewidth]{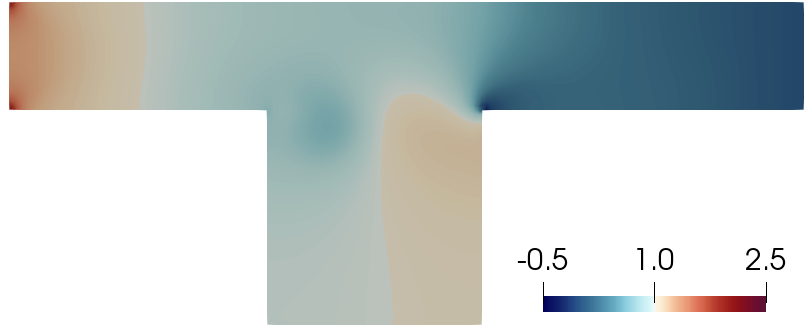}
	\end{subfigure}
	\begin{subfigure}{.49\linewidth}
		\centering
		\includegraphics[width=1.\linewidth]{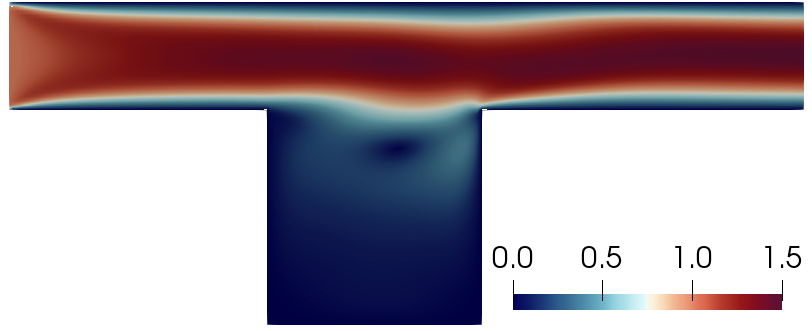}
	\end{subfigure}
	\begin{subfigure}{.49\linewidth}
		\centering
		\includegraphics[width=1.\linewidth]{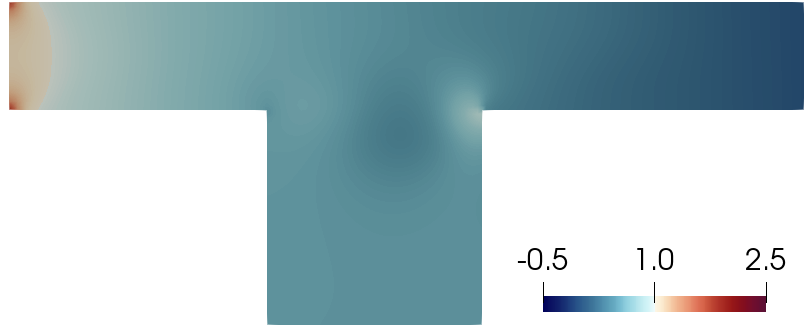}
	\end{subfigure}
	
	\caption{From top to bottom: Snapshots obtained at $t$ = 0; 0.5; 1.0; 2.0 s with the consistent flux method: (left) cell-centered velocity magnitude in m/s; (right) pressure in Pa.}
	\label{fig:OC_snapshots_CFM}
\end{figure}

Full order simulations are performed for the open cavity problem according to Section~\ref{sec:setupOC}. The cell-centered velocity (magnitude) and pressure snapshots at $t$ = 0; 0.5; 1.0; 2.0 s that are obtained with the consistent flux method are shown in Figure~\ref{fig:OC_snapshots_CFM}; these snapshots look similar for the IFM. This figure shows that the problem is unsteady for the simulated time span.

\begin{figure}[h!]
	\centering
	\begin{subfigure}{.5\linewidth}
		\centering
		\includegraphics[width=1.\linewidth]{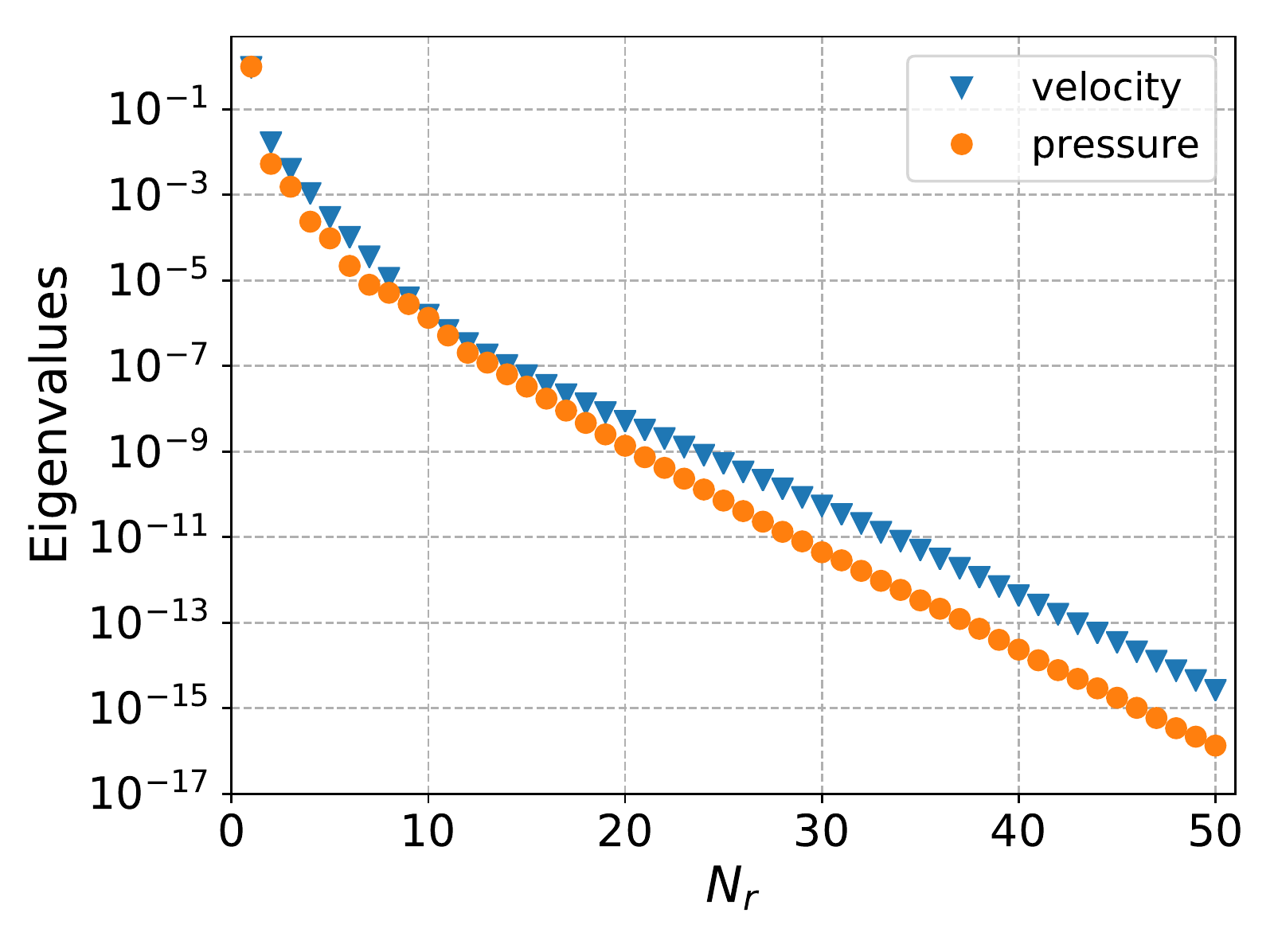}
	\end{subfigure}%
	\begin{subfigure}{.5\linewidth}
		\centering
		\includegraphics[width=1.\linewidth]{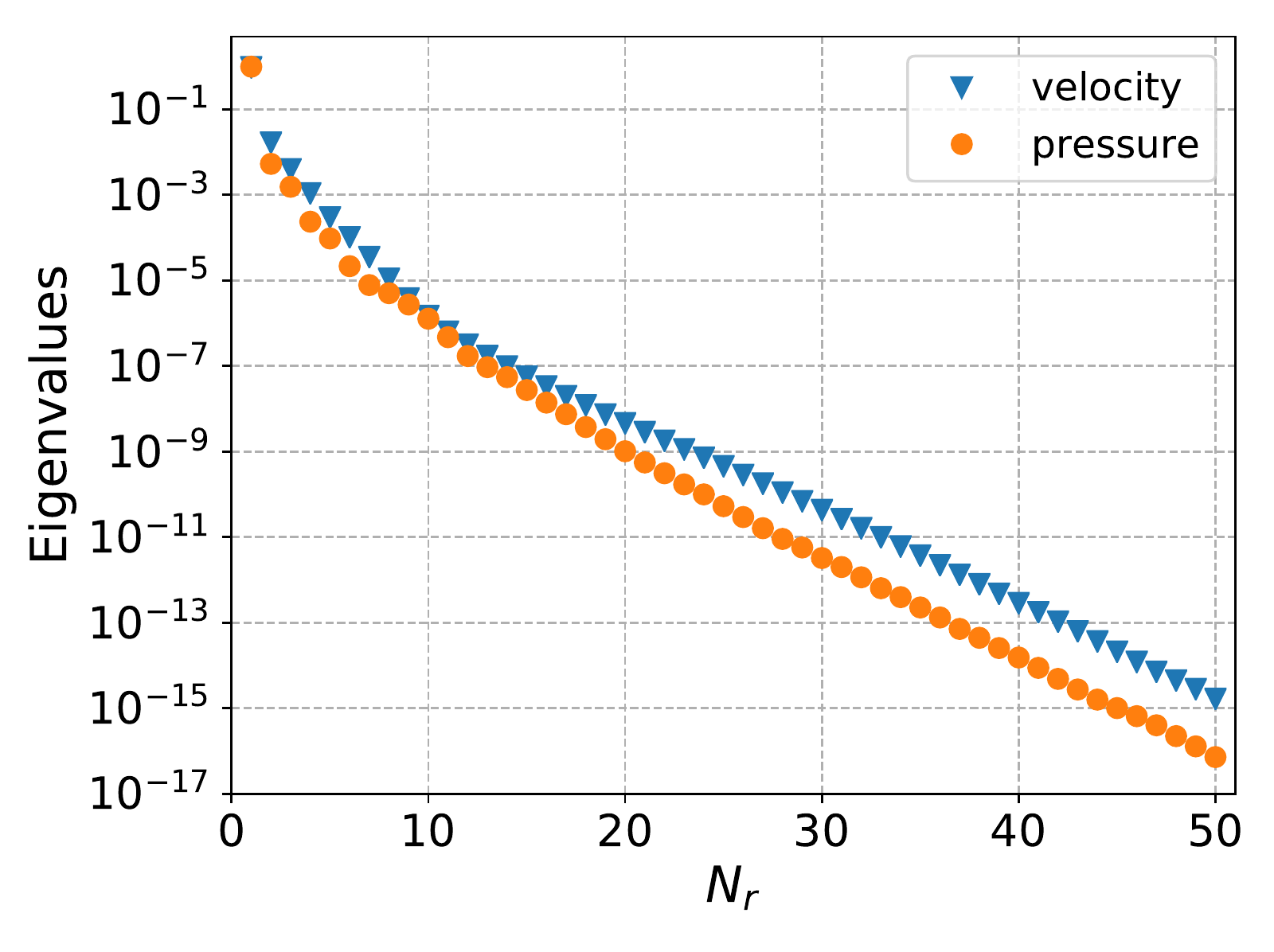}
	\end{subfigure}
	\caption{Eigenvalues for the open cavity flow case: (left) inconsistent flux method; (right) consistent flux method.}
	\label{fig:ev_OC}
\end{figure}

The POD eigenvalues of the cell-centered velocity and pressure modes are shown in Figure~\ref{fig:ev_OC}. The eigenvalues are approximately the same for both projection methods. For both velocity and pressure, the rate of decay of the first ten modes is steeper than the rate of decay of the higher modes. The eigenvalues also decay less rapidly for increasing number of modes compared to the lid driven cavity case (Figure~\ref{fig:ev_LDC}), which indicates that more POD modes are needed to approximate the FOM solutions accurately. As the slope of eigenvalue decay is more or less the same for pressure and velocity, we take equal numbers of modes $N_r$ = 2; 5; 10; 15; 20 for the reduced pressure basis and reduced velocity basis. 

\newpage

\begin{figure}[h!]
	\centering
	\begin{subfigure}{.5\linewidth}
		\centering
		\includegraphics[width=1.\linewidth]{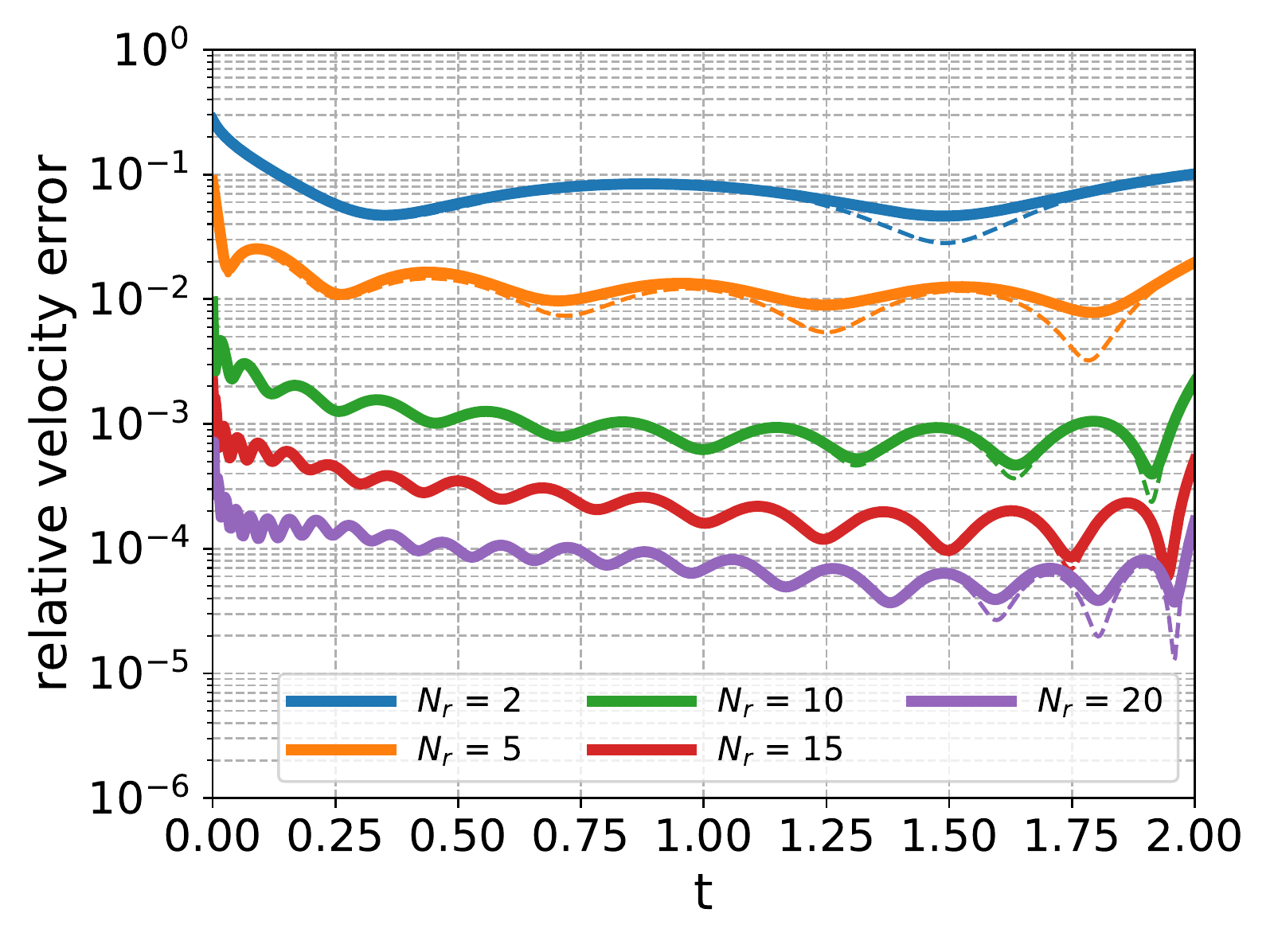}
	\end{subfigure}%
	\begin{subfigure}{.5\linewidth}
		\centering
		\includegraphics[width=1.\linewidth]{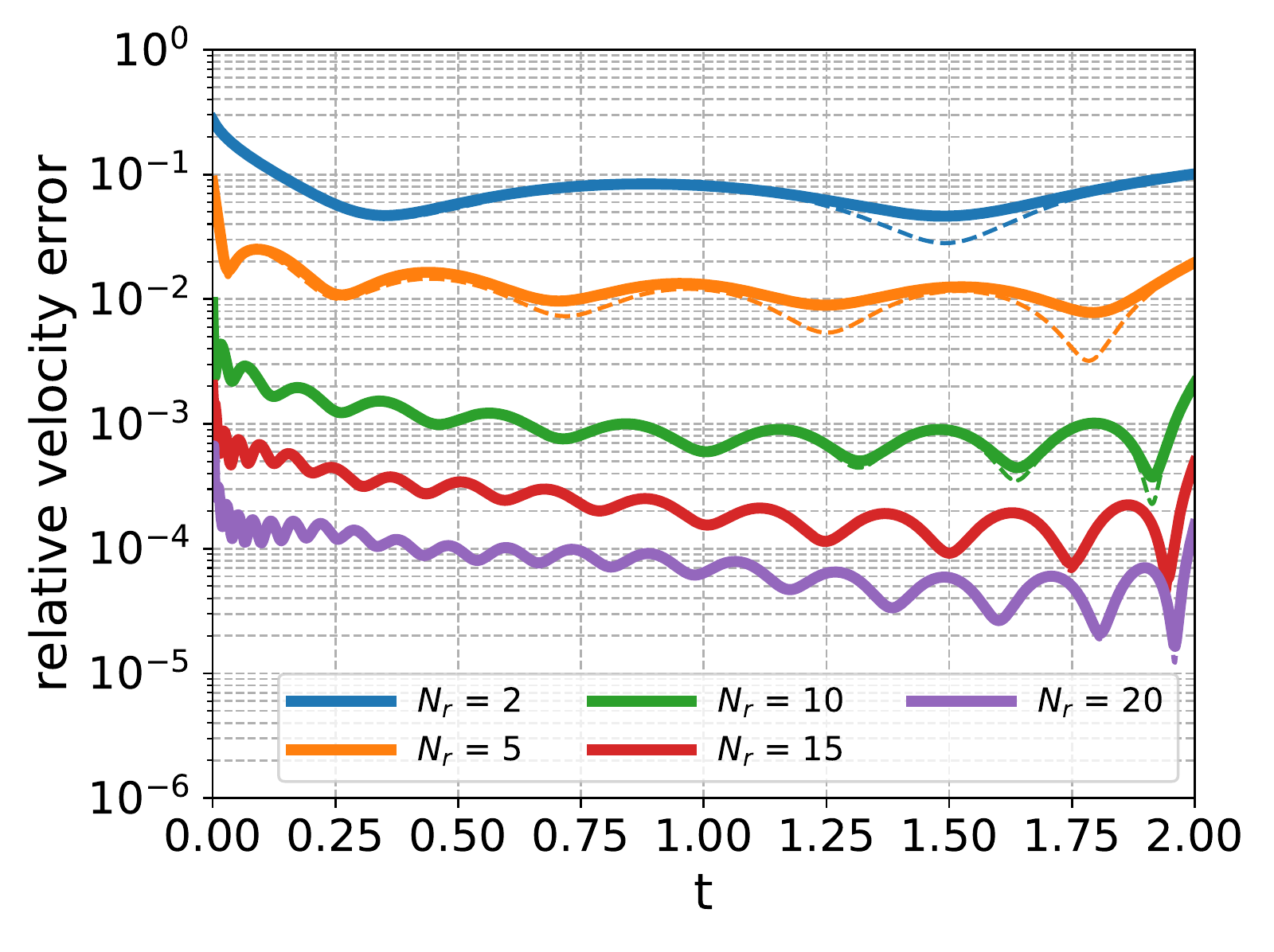}
	\end{subfigure}
	\caption{Relative cell-centered velocity error as a function of time for different number of modes for the open cavity flow case: (left) inconsistent flux method; (right) consistent flux method. Dashed lines: basis projection error (projecting snapshots onto truncated basis).}
	\label{fig:L2error_U_OC}
\end{figure}

\begin{figure}[h!]
	\centering
	\begin{subfigure}{.5\linewidth}
		\centering
		\includegraphics[width=1.\linewidth]{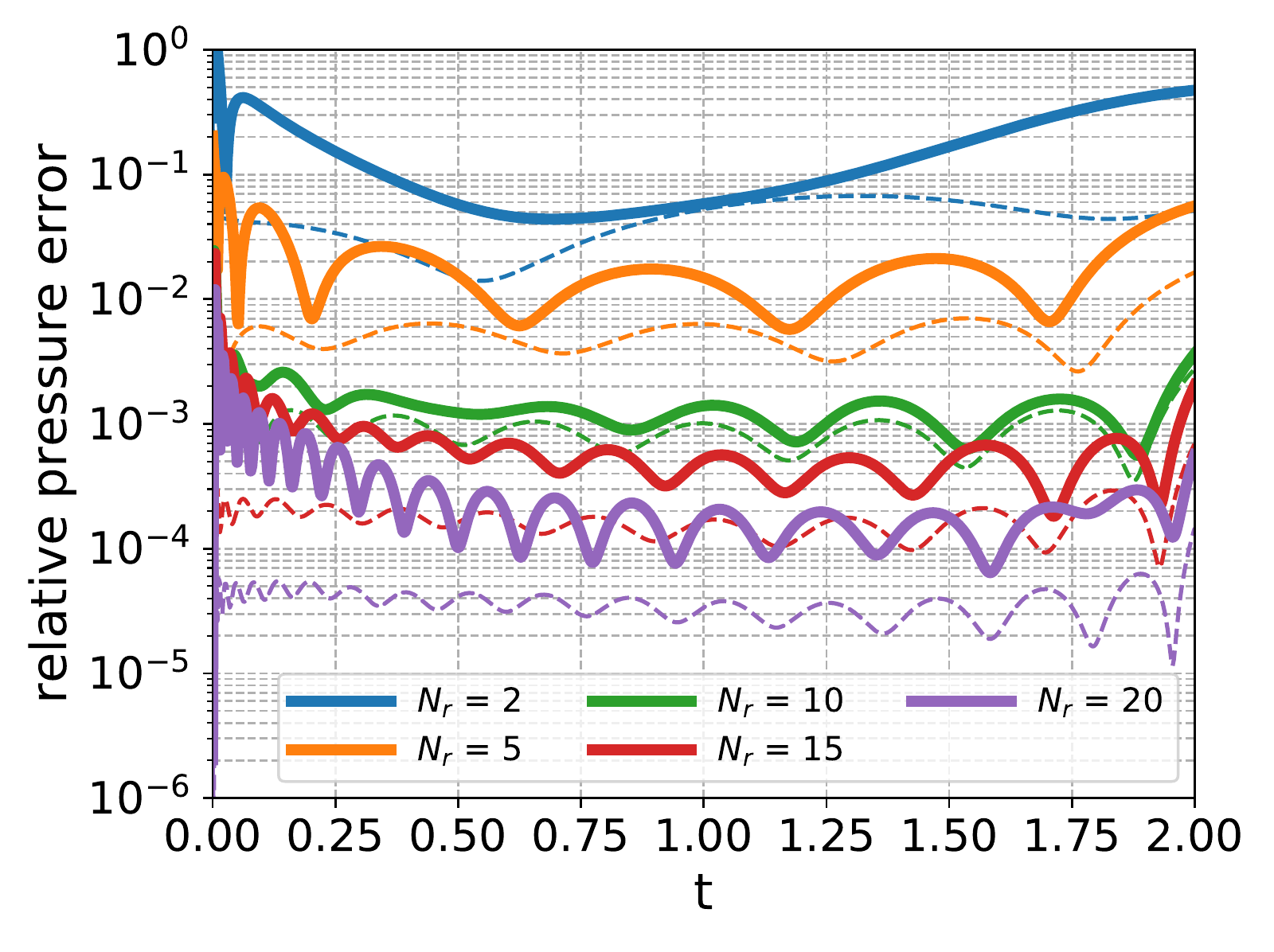}
	\end{subfigure}%
	\begin{subfigure}{.5\linewidth}
		\centering
		\includegraphics[width=1.\linewidth]{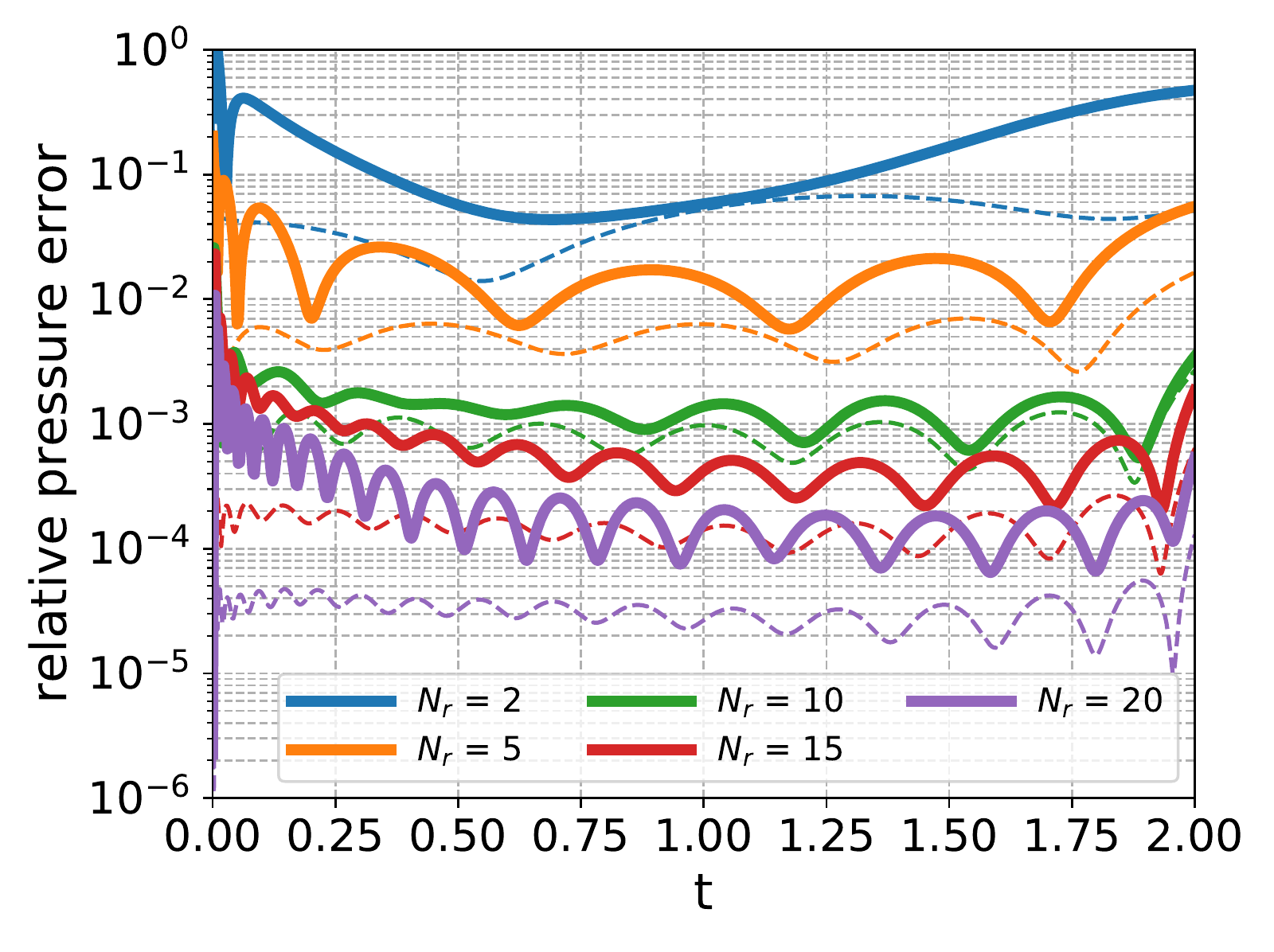}
	\end{subfigure}
	\caption{Relative pressure error as a function of time for different number of modes for the open cavity flow case: (left) inconsistent flux method; (right) consistent flux method. Dashed lines: basis projection error (projecting snapshots onto truncated basis).}
	\label{fig:L2error_P_OC}
\end{figure}

The relative prediction and basis projection errors (Equations~\ref{eq:l2_projection} and~\ref{eq:l2_prediction}) are plotted in Figure~\ref{fig:L2error_U_OC} for velocity and Figure~\ref{fig:L2error_P_OC} for pressure. These figures show that the errors decrease when increasing the number of modes for both the IFM-ROM and CFM-ROM. Figure~\ref{fig:L2error_U_OC} shows that the relative velocity errors (Equation~\ref{eq:l2_prediction}) are very close to the basis projection errors (Equation~\ref{eq:l2_projection}) as they are almost overlapping. However, after about 1.5 seconds of simulation time the prediction error for 20 modes start slightly deviating from the projection error for the same number of modes in the case of the inconsistent flux methods, while the errors are almost overlapping in the case of the consistent flux method. 

\newpage 
The relative pressure errors plotted in Figure~\ref{fig:L2error_P_OC} are of the same order as the velocity errors in Figure~\replaced{\ref{fig:L2error_U_OC} for a certain number of modes}{\ref{fig:L2error_P_OC} for the same number of modes}. \added{However, the relative pressure error is larger than the basis projection error for pressure, especially at the beginning of the simulation. The prediction error decreases as the simulation progresses. Nevertheless, the larger error remains present throughout the simulation. A possible explanation is that the reduced order model does not accurately reproduce the behavior of the flow at the beginning of the simulation. The initial pressure field is obtained by solving an inviscid potential flow problem at the full order level. When the viscous full order problem is solved, a pressure jump occurs at the initial time step. Projection-based reduced order model often do not reproduce strong changes in the flow fields well due to the truncation of the low energy modes that are, however, important for representing such flow behavior~\cite{balajewicz2016minimal}.}

\begin{figure}[h!]
	\centering
	\begin{subfigure}{.5\linewidth}
		\centering
		\includegraphics[width=1.\linewidth]{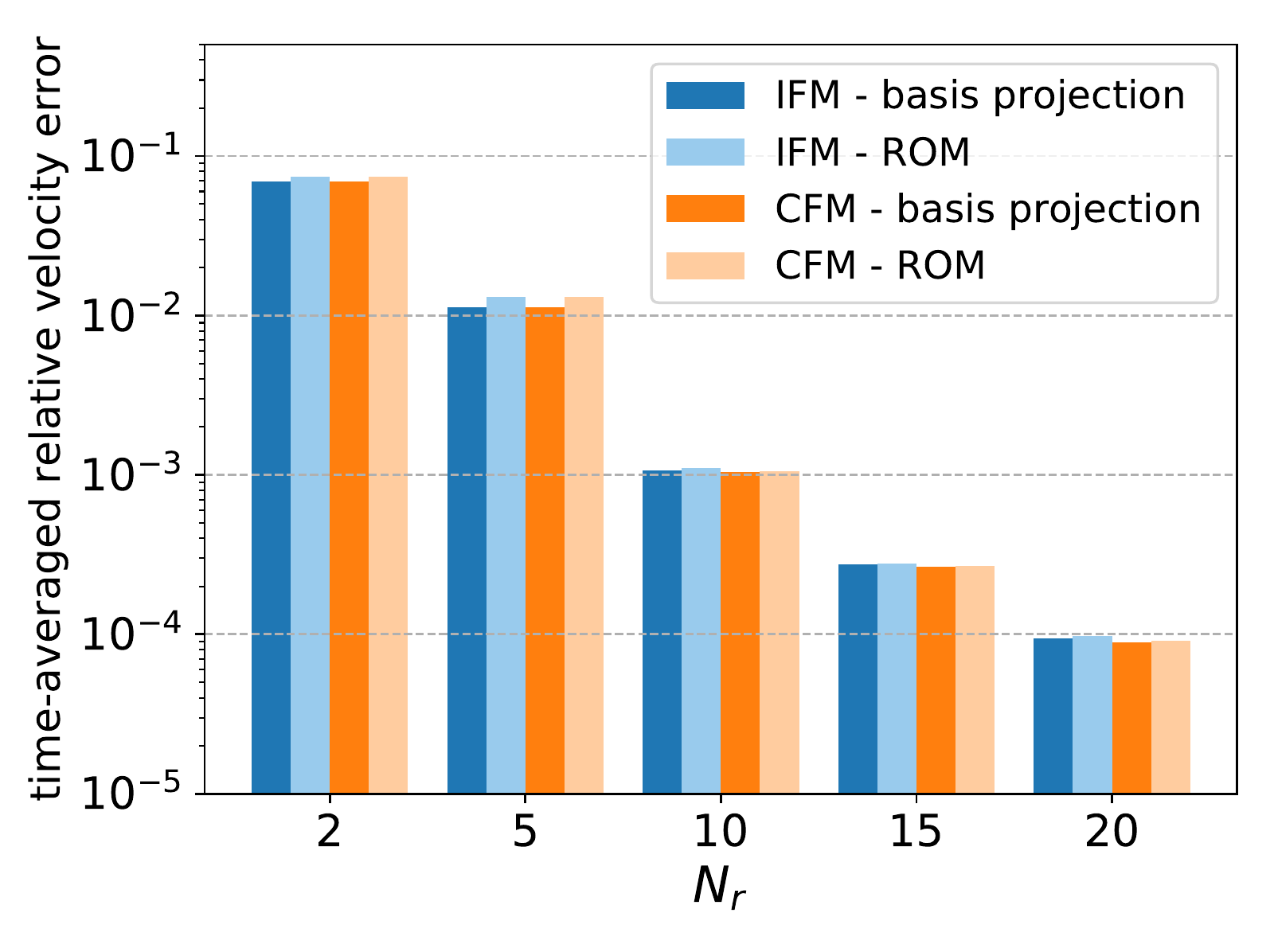}
	\end{subfigure}%
	\begin{subfigure}{.5\linewidth}
		\centering
		\includegraphics[width=1.\linewidth]{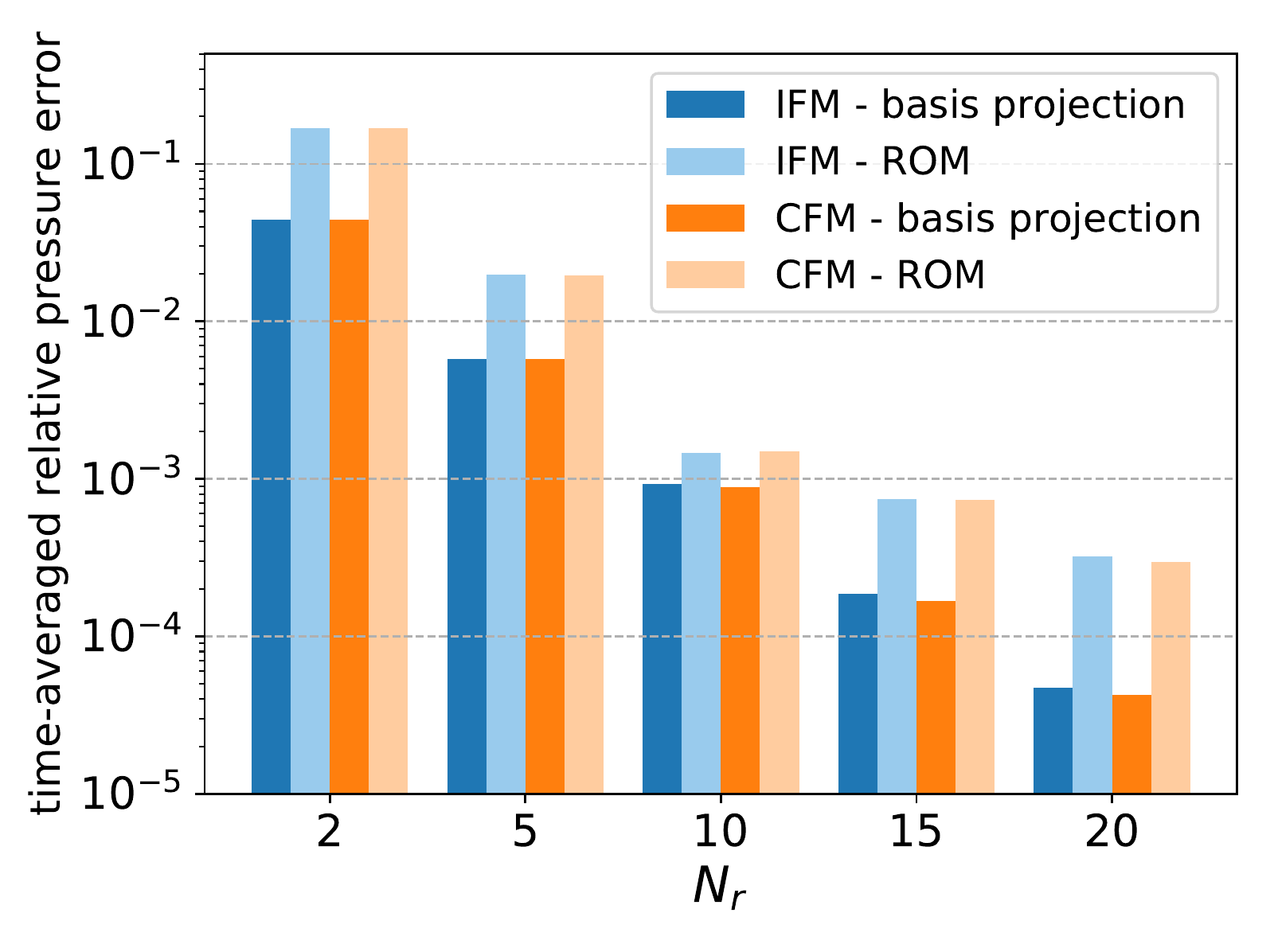}
	\end{subfigure}
	\caption{Time-averaged relative basis projection and prediction errors of the open cavity flow problem: (left) velocity; (right) pressure.}
	\label{fig:OC_time_ave_L2_errors}
\end{figure}

Also for pressure, the prediction error at around 1.8 s of simulation time is higher for the IFM-ROM compared to the CFM-ROM for 20 modes. Moreover, the difference between the prediction and projection errors is the smallest for 10 modes as is also shown in Figure~\ref{fig:OC_time_ave_L2_errors} in which we plotted the time-averaged basis projection errors (Equation~\ref{eq:l2_projection}) and time-averaged ROM prediction errors (Equation~\ref{eq:l2_prediction}) for velocity and pressure, respectively.

In all cases, \replaced{accurate}{stable} ROM results were obtained with the proposed explicit projection method, indicating that additional pressure stabilization methods are not required (as discussed in the Introduction). Moreover, a relative error of about $\mathcal{O}$(\num{e-3}) is obtained with only 10 velocity and 10 pressure modes (plus 10 face velocity modes in the case of the consistent flux method) in this test case.

Furthermore, the local continuity errors (Equation~\ref{local_continuity}) of the IFM-FOM is of the order \num{e-5}. Also the local continuity errors computed for the POD velocity modes and the IFM-ROM are of the order \num{e-5} (regardless the number of modes used). On the other hand, the local continuity errors are of the order \num{e-16} for the CFM, which is of the order of the machine precision. Thus, the discrete face velocity is only approximately discretely divergence free in the case of the IFM, whereas the constraint is fully satisfied with the CFM.

\begin{figure}[h!]
	\centering
	\begin{subfigure}{.49\textwidth}
		\centering
		\includegraphics[width=1.\linewidth]{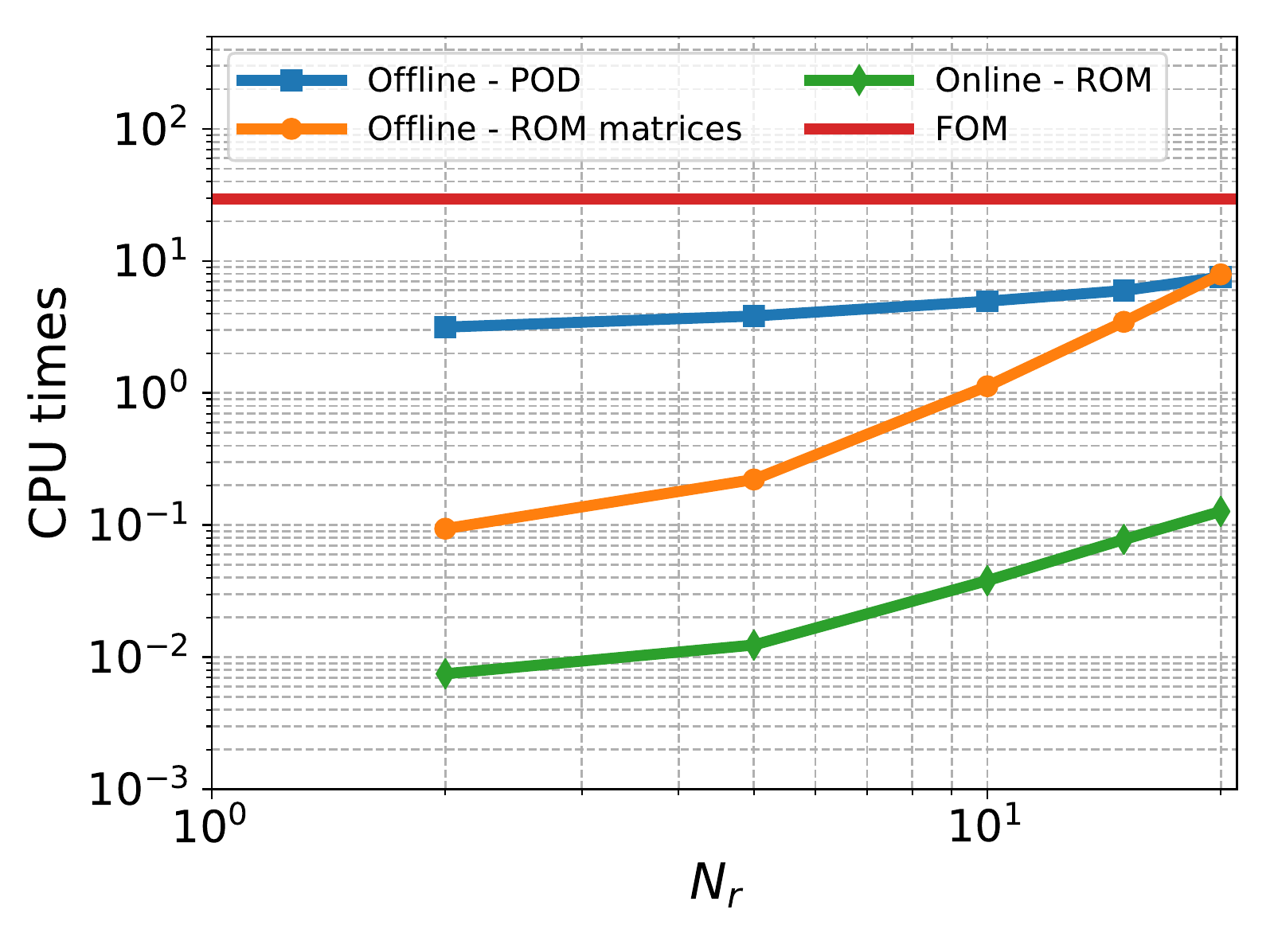}
	\end{subfigure}%
	\begin{subfigure}{.49\textwidth}
		\centering
		\includegraphics[width=1.\linewidth]{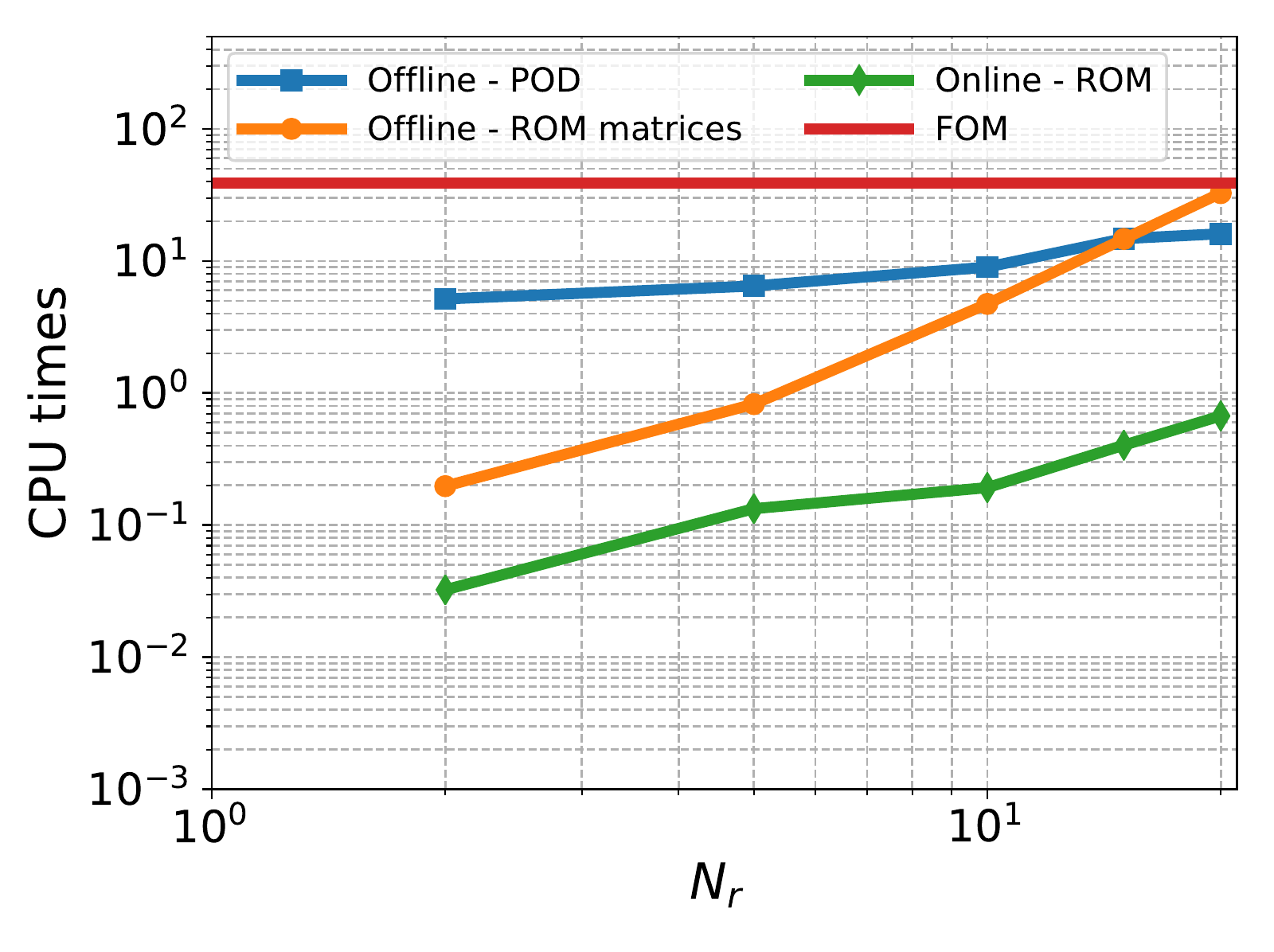}
	\end{subfigure}
	\caption{Computational times in seconds as function of number of modes the open cavity flow case: (left) inconsistent flux method; (right) consistent flux method.}
	\label{fig:CPU_OC}
\end{figure}

\begin{figure}[h!]
	\centering
	\includegraphics[width=0.5\linewidth]{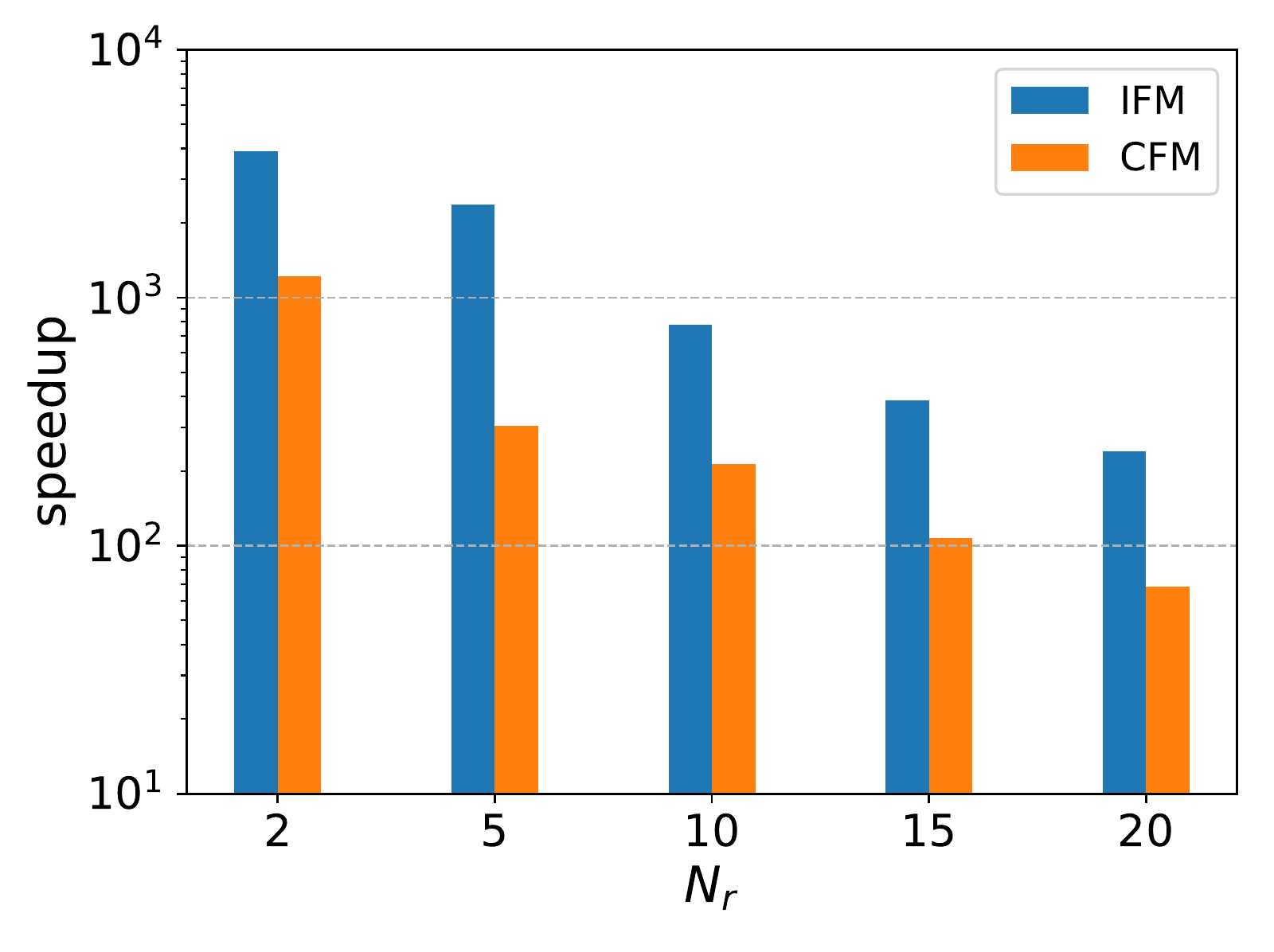}
	\caption{Speedup in computational time of the ROM compared to the FOM in seconds as function of number of modes for the open cavity flow case.}
	\label{fig:speedup_OC}
\end{figure}

Finally, the computational times required by the ROMs is compared to the FOM CPU times in Figure~\ref{fig:CPU_OC}. The plotted computational times are the average times of two simulations. The speedup is plotted in Figure~\ref{fig:speedup_OC} and is between about \num{2e2} and \num{4e3}, depending on the number of modes used for the IFM-ROM, while the speedup is between about \num{6e1} and \num{1e3} for the CFM-ROM. This is according to expectations as an additional equation for the face velocity (Equation~\ref{eq:cfm_phi2}) needs to be determined at \added{the} ROM level. For the same reason, more matrices need to be precomputed for the CFM, which explains the higher cost. Moreover, the larger the number of modes, the more time it takes to precompute the reduced matrices. 

\newpage

\section{Discussion} \label{sec:discussion}

The two main advantages of the `discretize-then-project' approach compared to existing approaches is that it requires no pressure stabilization technique, even though the pressure term is present in the ROM, nor a boundary control technique to impose the boundary conditions at the ROM level.

\replaced{We showed in Section~\ref{sec:NS} that velocity and pressure are coupled in the continuous domain and in Section~\ref{sec:SD} that they are also coupled in the semi-discrete domain. Since we project the equations for the pressure and for the cell-centered velocity at the next time step onto the reduced basis spaces of pressure and velocity, respectively, velocity and pressure are also coupled at the reduced order level. In that way, the ROM and the FOM formulations are consistent with each other and pressure is included in the ROM formulation for the incompressible NS equations. In contrast to `velocity-only' ROMs, pressure is thus not recovered in a post-processing step}{However, in both methods the cell-centered velocity fields are only approximately discretely divergence free. As a consequence, velocity and pressure are also coupled at the reduced order level. Therefore, pressure needs to be included in the	ROM formulation for the incompressible NS equations and cannot be simply recovered in a post-processing step, in contrast to `velocity-only' ROMs}~\cite{caiazzo2014numerical,kean2020error,fonn2019fast}. 

\added{That the ROM formulation is fully corresponding with the FOM formulation makes the `discretize-then-project' approach more straightforward than other popular approaches that recover pressure for stability purposes. As highlighted in the Introduction, a disadvantage of the pressure recovery by, for example, exploitation of a pressure Poisson equation during the projection stage is that it is often not clear how to treat the boundary conditions in the pressure Poisson equation~\cite{Stabile2017CAF,gresho1998incompressible,liu2010stable}. The disadvantage of a supremizer stabilized velocity basis is that it is hard to determine how many supremizer modes need to be used~\cite{Stabile2017CAF,ballarin2015supremizer,kean2020error}. The `discretize-then-project' approach does not encounter such difficulties. Furthermore, the relative ROM error of the velocity and pressure fields using the `discretize-then-project' approach is of the same order as the basis projection error as shown in Figures~\ref{fig:LDC_time_ave_L2_errors} and~\ref{fig:OC_time_ave_L2_errors}, while the ROM velocity and pressure fields using the standard approaches are often about one or two orders less accurate than the fields obtained by projecting the full order solutions onto the POD basis spaces~\cite{Stabile2017CAF,ballarin2015supremizer,kean2020error,busto2020pod}.}

\added{The results have shown that with the current approach of projecting the boundary vectors onto the reduced basis spaces, it is not needed to use additional boundary control methods, such as the penalty method or the lifting function method, to impose the boundary conditions in the ROM. This approach is easier to implement and more generic than other approaches since it does not rely on parameter tuning.} \replaced{Moreover, the approach of projecting the boundary vectors}{Furthermore, the results have shown that with the current approach of projecting the boundary vectors onto the reduced bases, it is not needed to use a penalty method or a lifting function method to enforce the boundary conditions in the ROM. This approach} can also be implemented for PISO or PIMPLE~\cite{ferziger2002computational} algorithms for collocated grids that are more frequently used in engineering applications as the implicit time discretization is, generally, more stable than explicit schemes. 

The main difference between the two projection methods, the inconsistent flux method and the consistent flux method, is \replaced{that}{the divergence freeness of the fluxes. Whereas} the fluxes are discretely divergence free in the case of the CFM, \added{while} they are only approximately discretely divergence free in the case of the IFM. Nevertheless, the difference in the cell-centered velocity and pressure solutions can be considered negligible in our test cases. 

\added{In this work, it has not been investigated whether momentum is globally conserved in the ROM with the `discretize-then-project' approach since both test cases, the lid driven cavity case and the open cavity flow case, do not globally conserve momentum. Nevertheless, there exist techniques to enforce global momentum conservation for problems that conserve momentum. Carlberg et al.~\cite{carlberg2018conservative} considered conservative model reduction in a finite-volume context by solving a constrained optimization problem at each time step. In that way, the resulting reduced order model is globally conservative over sub-domains in a decomposed mesh. Other techniques to conserve momentum in the reduced order model are exploiting the discrete skew-symmetric structure of the full order system at the level of the reduced system as presented by Afkham et al.~\cite{afkham2020conservative} and using a constrained singular value decomposition approach approach to enforce global momentum conservation on periodic domains as presented by Sanderse~\cite{sanderse2020non}.}

The CFM-FOM and CFM-ROM simulation\added{s} take more computational time than the equivalent models with the IFM as shown in Figures~\ref{fig:CPU_IFM_LDC} and~\ref{fig:CPU_OC}, for the lid driven cavity and open cavity test cases, respectively. This is mostly due to the additional equation that needs to be solved for the fluxes at \replaced{at the}{FOM and} ROM level as well as computing the reduced POD basis space for the face-centered velocity. Therefore, it is plausible to prefer the IFM method despite the fact that the velocity fields are only approximately discretely free. On the other hand, as observed for the open cavity case, the IFM-ROM is slightly less accurate than the CFM-ROM towards the end of the ROM simulation when a large number of modes is used for the construction of the reduced \replaced{basis spaces}{bases}. For different test cases than the cases studied in this work, the divergence error of the IFM could be potentially much larger, leading to possibly non-physical or inaccurate results. Another possible cause of the slight difference, which only occurred for large number of modes, is that the modes with smaller eigenvalues are dominated by numerical noise. Therefore, the drop in eigenvalue magnitude does not always provide a reliable identification of a reduced basis of high quality~\cite{lee2020importance}. 

On the other hand, as observed for the open cavity case, the IFM-ROM is slightly less accurate than the CFM-ROM towards the end of the ROM simulation when a large number of modes is used for the construction of the reduced \replaced{basis spaces}{bases}. For different test cases than the cases studied in this work, the divergence error of the IFM could be potentially much larger, leading to possibly non-physical or inaccurate results.

\begin{figure}[h!]
	\centering
	\includegraphics[width=0.5\linewidth]{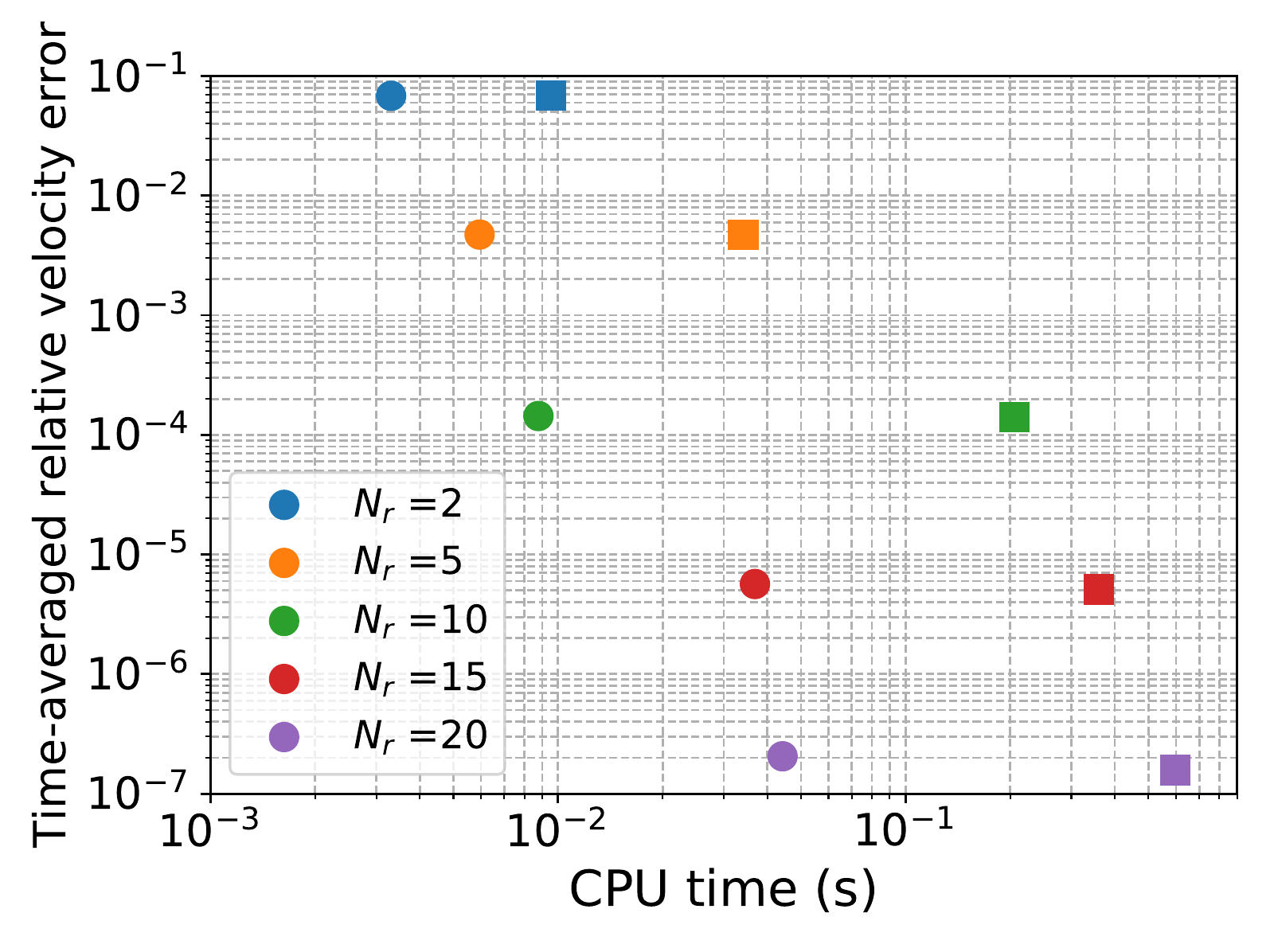}
	\caption{Computational times in seconds versus the time-averaged relative velocity error of the lid driven cavity flow problem for different numbers of modes. Circles: inconsistent flux method; squares: consistent flux method.}
	\label{fig:ParatoLDC}
\end{figure}

\begin{figure}[h!]
	\centering
	\includegraphics[width=0.5\linewidth]{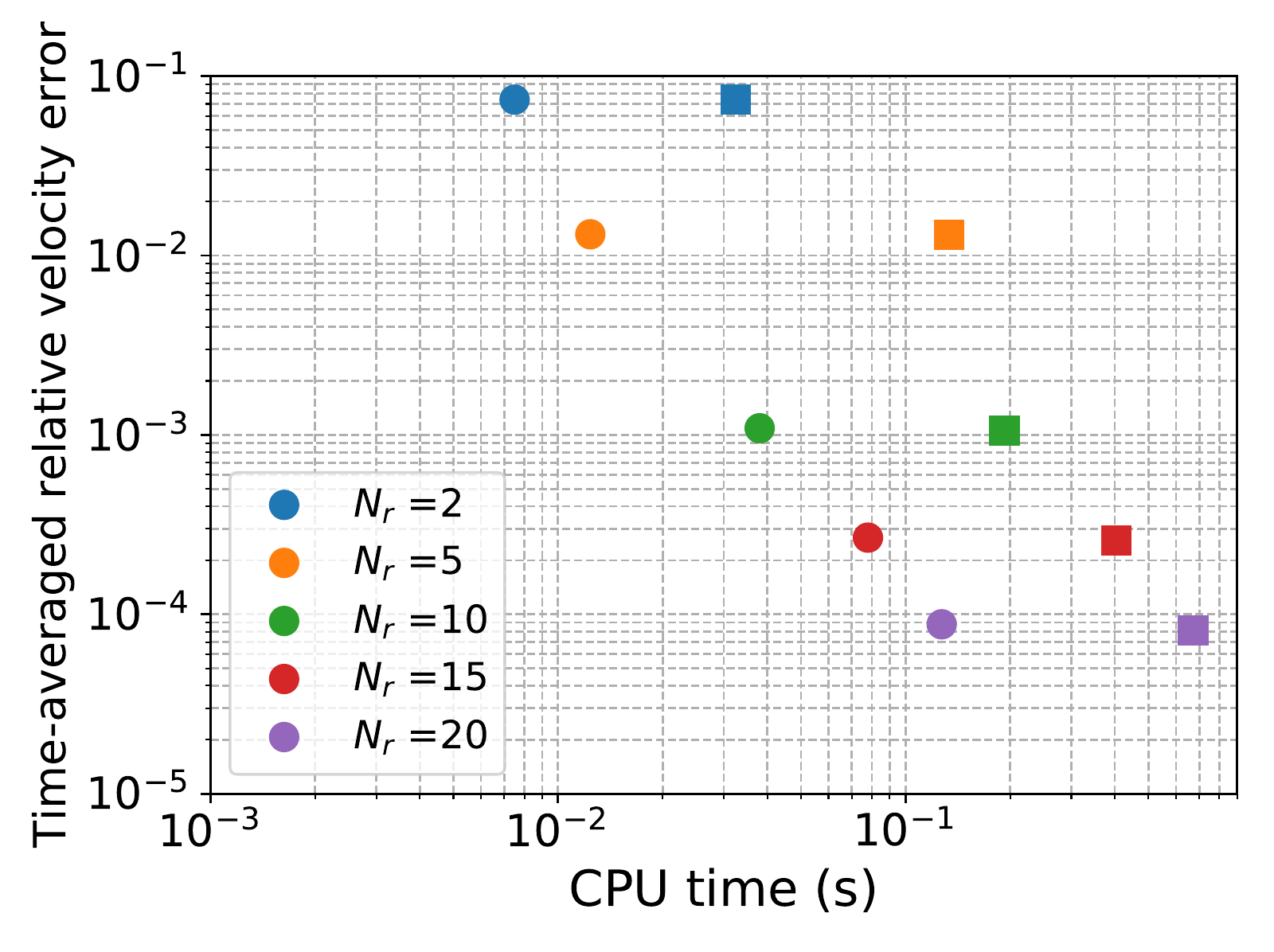}
	\caption{Computational times in seconds versus the time-averaged relative velocity error of the open cavity flow problem for different numbers of modes. Circles: inconsistent flux method; squares: consistent flux method.}
	\label{fig:ParatoOC}
\end{figure}

\added{Figures~\ref{fig:ParatoLDC} and~\ref{fig:ParatoOC} depict the computational times in seconds versus the time-averaged relative velocity error of the lid driven cavity flow problem and open cavity flow problem, respectively. These plot show that the gain in accuracy by using the CFM over the IFM is almost neglicible in comparison with the gain in computational time. Therefore, the IFM is the preferred method for the cases tested in this work.}

\newpage
We have only investigated first order explicit temporal discretization. Moreover, we first discretize in space and in time before performing the Galerkin projection. Therefore, the ROM formulations are fully corresponding with the FOM formulations. Higher-order explicit (Runge-Kutta) methods, such as those analyzed by Komen et al.~\cite{komen2020analysis}, are generally more accurate than the Forward Euler scheme used in this work. However, to keep the ROM and the FOM \added{formulations} consistent with each other, higher-order methods would require the implementation of the different stages also at \added{the} reduced order level. This is, in contrast to the FOM level, not straightforward at the ROM level. 

Moreover, the disadvantage of explicit schemes is that the systems become unstable for Courant numbers larger than unity. This can form a severe limitation for the time step~\cite{Jasak}. The standard OpenFOAM method is PISO, which is an implicit pressure-based scheme for the NS equations. The segregated nature of PISO induces a decoupling between mass and momentum equations. The PISO algorithm has similarities with the consistent flux method presented in this work. Therefore, it would be an asset to extend the CFM to implicit schemes. However, a number of corrections of the pressure and velocity fields are needed to enforce the pressure-velocity coupling at each time step and to minimize the errors. Therefore, the same challenge as for higher order (explicit) Runge-Kutta schemes applies, namely keeping the ROM and the FOM consistent with each other.

The methodology can be extended to parametric problems as the ROM formulations are already written in such a way that viscosity is not part of the diffusion operator and the associated boundary vector (when projecting the boundary vectors, $\boldsymbol{r}_p^C$ and $\boldsymbol{r}_p^D$, onto the reduced basis spaces separately).

Finally, the speedup is higher for the open cavity case compared to the lid driven cavity case as the FOM contains a larger number degrees of freedom. With an increasing number of modes, the precomputing phase (in particular assembling the reduced convection operator) becomes the dominant factor in the ROM execution. In our test cases this is not a concern, as the number of modes is typically sufficient before the precomputing phase becomes a dominant factor. Nevertheless, one could reduce the complexity of the convection operator (a third order tensor) by using hyper-reduction techniques such as the discrete empirical interpolation method~\cite{chaturantabut2010nonlinear}.

\section{Conclusions and outlook} \label{sec:conclusion}
The novel reduced order models are developed using a `discretize-then-project' approach. The ROM formulations are fully corresponding to the discrete FOM formulations of the incompressible NS equations on collocated grids. No pressure stabilization method is needed, even though the pressure term is present in the ROM. Moreover, the boundary conditions at the ROM level are imposed via the projection of the boundary vectors that are specified at the discrete FOM level. Therefore, it is not needed to use a boundary control method such as the penalty method or lifting function method. 

We considered two variants of a forward Euler time discretization: the inconsistent flux method, for which the velocity at the cell centers are considered only approximately discretely divergence free and the consistent flux method, for which the face velocities are discretely divergence free.

The ROMs predict well the underlying FOMs as \deleted{stable and }accurate results are obtained with the proposed methods for the lid driven cavity and open cavity flow cases. The ROMs obtained with the consistent flux method, having divergence-free velocity fields, are slightly more accurate compared to the inconsistent flux method. 

However, the speedup of the ROM compared to the FOM is lower for the consistent flux method due to the additional equation for the face velocity that also needs to be solved at the ROM level. Furthermore, the speedup strongly depends on the number of modes used for the reduced basis spaces. For any number of modes, the speedup is the highest for the open cavity test case with the inconsistent flux method as it contains more degrees of freedom than the lid driven cavity case at \added{the} full order level. 

In future work, the methodology can be extended to higher-order explicit (Runge-Kutta) methods. \added{Also, the ROM needs to be tested for time evolutions that are different from those of the full order simulations and for long time integration. Therefore, we are planning to make the time step adaptive, for example by estimating the eigenvalues of the ROM operators using a linear stability theory~\cite{sanderse2020non}. Furthermore, it needs to be investigated whether momentum is globally conserved in the ROM with the `discretize-then-project' approach~\cite{carlberg2018conservative,sanderse2020non,afkham2020conservative}, for example for problems with periodic boundary conditions.} Moreover, an analogy of the consistent flux method can be constructed for the PISO algorithm, which is a widely used implicit time discretization method. The approach of projecting the full order boundary vectors containing the contributions of the boundary conditions can still be applied in the context of implicit models. Finally, we plan to extend our approaches to parametric (time-dependent) boundary conditions as well as physical parametrization, such as parameterizing the value of the viscosity.

\section*{CRediT authorship contribution statement}
\textbf{S.K. Star}:  Conceptualization, Methodology, Software, Formal analysis, Data curation, Writing - original draft. \textbf{B. Sanderse}: Conceptualization, Methodology,  Writing - review \& editing, Supervision. \textbf{G. Stabile}: Methodology, Software,  Writing - review \& editing. \textbf{G. Rozza}: Project administration, Funding acquisition. \textbf{J. Degroote}: Writing - review \& editing, Supervision, Funding acquisition.

\section*{Acknowledgment}
We acknowledge the support provided by the European Research Council Executive Agency by the Consolidator Grant project AROMA-CFD "Advanced Reduced Order Methods with Applications in Computational Fluid Dynamics" - GA 681447, H2020-ERC CoG 2015 AROMA-CFD, by the Italian Ministry of Education (MIUR) through the FARE-X-AROMA-CFD and NA-FROM-PDES PRIN projects. The authors thank E.M.J. Komen and E.M.A. Frederix from NRG for discussing and explaining the implementation of a selection of explicit Runge-Kutta schemes.

\bibliographystyle{ieeetr}
\bibliography{mybibfile}  
\end{document}